\documentclass[useAMS,usenatbib]{mn2e}

\usepackage{natbib}

\usepackage{latexsym,graphicx}
\usepackage{color}
\usepackage{fixltx2e}
\usepackage{verbatim} \usepackage{float}
\usepackage{amsmath,amssymb}
\usepackage{times}

\usepackage{setspace}
\usepackage{rotating}
\usepackage{pdflscape}
\usepackage{subfig}

\usepackage[squaren, Gray, cdot]{SIunits}

\newcommand\hii{H\,{\sc ii} \,}

\def\apgt{\ {\raise-.5ex\hbox{$\buildrel>\over\sim$}}\ }
\def\aplt{\ {\raise-.5ex\hbox{$\buildrel<\over\sim$}}\ }

\let\oldhat\hat
\renewcommand{\hat}[1]{\oldhat{\mathbf{#1}}}

\def\HII{\rm{H}{\textsc {ii}}\, }

\title[Forming spectroscopic massive proto-binaries by disk fragmentation.]{Forming spectroscopic massive proto-binaries by disk fragmentation}  
   
\author[D. M.-A.~Meyer et al.]
       {D. M.-A.~Meyer,$^{1}$\thanks{E-mail: dominique.meyer@uni-tuebingen.de} R.~Kuiper,$^{1}$ W.~Kley,$^{1}$ \textcolor{black}{K.~G.~Johnston}$^{2}$ and E. Vorobyov$^{3,4,5}$\\
       $^{1}$Institut f\" ur Astronomie und Astrophysik, Universit\" at T\" ubingen,  Auf der Morgenstelle 10, 72076 T\" ubingen, Germany\\
       $^{2}$School of Physics and Astronomy, E.C. Stoner Building, The University of Leeds, Leeds LS2 9JT, UK\\
       $^{3}$\textcolor{black}{Institute of Fluid Mechanics and Heat Transfer, TU Wien, Vienna, 1060, Austria}\\ 
       $^{4}$Department of Astrophysics, The University of Vienna, Vienna, A-1180, Austria\\
       $^{5}$Research Institute of Physics, Southern Federal University, Stachki 194, Rostov-on-Don, 344090, Russia\\       
       }

\begin{document}

\date{Received Month Day Year; accepted Month Day Year}

\maketitle
   
\label{firstpage}

\begin{abstract} 
The surroundings of massive protostars \textcolor{black}{constitute} \textcolor{black}{an 
accretion disc which has numerically been shown} to be subject to fragmentation 
and responsible for luminous accretion-driven outbursts. Moreover, it is 
suspected to produce close binary companions which will later strongly 
influence \textcolor{black}{the star's} future evolution in the Hertzsprung-Russel diagram. 
We present three-dimensional gravitation-radiation-hydrodynamic numerical 
simulations of $100\, \rm M_{\odot}$ pre-stellar cores.  
We \textcolor{black}{find} that accretion discs of young massive stars violently fragment without 
preventing the (highly variable) accretion of \textcolor{black}{gaseous clumps} onto the protostars.  
\textcolor{black}{
While acquiring the characteristics of a nascent low-mass companion, 
some disc fragments migrate onto the central massive protostar with dynamical properties 
showing that its final Keplerian orbit is close enough to constitute a close massive proto-binary 
system, having a young high-mass and a low-mass component. 
}
We conclude on the viability of the disc fragmentation channel for the formation of such short-period binaries, 
and that both processes --close massive binary formation and accretion bursts-- may happen at the same time. 
\textcolor{black}{
FU-Orionis-type bursts, such as observed in the young high-mass star S255IR-NIRS3,
may not only indicate ongoing disc fragmentation, but also be considered as a tracer for
the formation of close massive binaries -- progenitors
of the subsequent massive spectroscopic binaries -- once the high-mass component of
the system will enter the main-sequence phase of its evolution. 
} 
\textcolor{black}{Finally, we investigate the ALMA-observability of the disc fragments.} 
\end{abstract}



\section{Introduction}
\label{sect:introduction}

Massive star formation is a process happening in large, cold 
and opaque molecular clouds~\citep{zinnecker_araa_45_2007,langer_araa_50_2012} 
but its repercussions \textcolor{black}{upon} the functioning of the interstellar medium are 
huge~\citep{vink_asp_353_2006,vink_aarv_20_2012}. 
\textcolor{black}{
Understanding the formation of these high-mass stars in detail would allow us to 
predict - given a set of initial properties of the parent molecular cloud in which 
these stars form - their final fate, e.g. as core-collapse supernovae or 
gamma-ray bursts~\citep{woosley_rvmp_74_2002}. 
}
%
It has been shown that most of the massive stars experience \textcolor{black}{(close)} multiplicity even 
during their early main-sequence phase~\citep{vanbeveren_ssrv_56_1991,2013A&A...550A..27M}, which in turn strongly impacts 
their future evolutionary path in the \textcolor{black}{Hertzsprung}-Russel 
diagram~\citep{dessart_aa_404_2003,petrovic_phdt_2004,sana_sci_337_2012}. Additionally, while rapidly leaving their 
pre-main-sequence tracks, high accretion rates of circumstellar material $\sim 
10^{-3}\, \rm M_{\odot}\, \rm yr^{-1}$ can already affect the evolution of 
massive protostars before entering the main-sequence 
phase~\citep{hosokawa_apj_691_2009,haemmerle_585_aa_2016,haemmerle_aa_602_2017}.

\textcolor{black}{
Close massive binaries are believed to \textcolor{black}{have} formed in the \textcolor{black}{surroundings} of 
pre-main-sequence massive stars, by analogy with low-mass star 
formation~\citep{bonnell_mnras_271_1994}. Consequently, the 
detailed study of the proto-circumstellar medium of young high-mass stars is a 
preponderant step towards a complete stellar evolution of massive stars. 
}
Amongst the several formation scenarios of massive binaries that have been 
proposed, the direct formation of low-mass companion(s) from the fragmentation 
of the accretion disc of a still growing massive protostar is one of the most 
natural~\citep{cesaroni_natur_444_2006}. It consists in 
assuming that high-mass star formation is a scaled-up version \textcolor{black}{of 
the formation of low-mass stellar systems~\citep{fuente_aa_366_2001,testi_2003} and the formation of companions 
to massive stars} might be seen as a scaled up version of gas giant planet formation via 
gravitational instability. This is strengthened by more and 
more unambiguous observations of collapsing high-mass pre-stellar 
cores~\citep{beuther_aa_584_2015}, 
\textcolor{black}{jets~\citep{purser_mnras_460_2016}} and accretion flows at their 
center~\citep{keto_apj_637_2006}. Several circumstellar mechanisms associated 
to disc accretion have been observed, such as ionized \HII regions channeled into 
cavities shaped by bipolar outflows~\citep{cesaroni_aa_509_2010}. 

Direct imaging of discs are only beginning to be resolvable by modern 
facilities~\citep{beuther_aa_543_2012}. 
Particularly, the {\it Atacama Large (sub-)Millimeter Array (ALMA)} recently 
revealed the existence of a Keplerian accretion disc around the \textcolor{black}{forming O-type} 
star AFL 4176~\citep{johnston_apj_813_2015} and in the surroundings of the young 
early massive star G11.92$−$0.61 MM1~\citep{ilee_mnras_462_2016}. 
\textcolor{black}{
More recently, observations of the high-mass proto-binary IRAS17216-3801 in~\citet{kraus_apj_835_2017} 
revealved together the circumstellar accretion discs of each massive components of a system similar to the 
numerical model of~\citet{Krumholz_sci_2009}, themselves surrounded by a circumbinary disc. 
}

The physics of collapsing 
gaseous clouds has been extensively studied \textcolor{black}{~\citep[][and 
references therein]{1999ARA&A..37..311E,2007ARA&A..45..339B,2007ARA&A..45..565M}, 
and angular momentum conservation may lead to the 
formation of an accretion disc.} 
Many mechanisms have been proposed to explain 
angular momentum transport \textcolor{black}{in accretion discs around low-mass stars such as baroclinic 
instabilities~\citep{klahr_apj_582_2003,klahr_apj_606_2004}, 
convective instabilities~\citep{lin_apj_416_1993}, magneto-rotational 
instabilities~\citep{ruden_apj_375_1991,hawley_apj_400_1992,flaig_mnras_420_2012} and 
vertical shear instabilities~\citep{nelson_mnras_435_2013,stoll_aa_572_2014}. Nonetheless, gravitational torques arising from the 
innermost to the outer part of self-gravitational discs is the most efficient momentum 
transport phenomenon in discs around high-mass stars~\citep{kuiper_apj_732_2011}}. 
\textcolor{black}{After the gravitational collapse of} pre-stellar cores, 
spiral arms developing in accretion discs may become unstable \textcolor{black}{and} 
experience local gravitational collapse leading to \textcolor{black}{the formation of 
dense gaseous clumps - progenitors of massive giant planets, brown 
dwarfs \textcolor{black}{or stellar companions}~\citep{stamatellos_mnras_392_2009,vorobyov_aa_552_2013,boss_2017arXiv170104892B}}. The migration of 
disc \textcolor{black}{sub-structures (clumps, spiral arcs)} onto their central protostar 
is responsible for violent accretion-driven 
outbursts broadly studied in the low-mass~\citep{vorobyov_apj_723_2010}, 
\textcolor{black}{the high-mass~\citep{meyer_mnras_464_2017}} and the 
primordial~\citep{vorobyov_aa_552_2013,hosokawa_2015} regimes of star formation.


\textcolor{black}{
The surroundings of young stars has been investigated with sophisticated numerical 
models including \textcolor{black}{appropriate} physical processes such as the presence of 
the cloud magnetic fields~\citep{banerjee_mnras_355_2004,banerjee_apj_641_2006,
commercon_aa_510_2010,fendt_apj_737_2011, 
fendt_apj_774_2013,machida_mnras_431_2013, 
seifried_mnras_432_2013,fendt_apj_796_2014, 
seifried_mnras_446_2015} or (external) stellar radiation 
feedback~\citep{yorke_aa_315_1996,richling_aa_327_1997,kessel_aa_337_1998, 
richling_aa_340_1998,richling_apj_539_2000}. 
However, only a small number of studies tackled the problem of star formation in its high-mass 
regime~\textcolor{black}{\citep{1998MNRAS.298...93B,2002ApJ...569..846Y,benerjee_apj_660_2007,
krumholz_apj_656_2007,kuiper_apj_722_2010,
peters_apj_711_2010,commercon_apj_742_2011, 
kuiper_apj_732_2011,seifried_mnras_417_2011,seifried_mnras_422_2012,kuiper_apj_800_2015,harries_mnras_448_2015,
kuiper_apj_832_2016,klassen_apj_823_2016,harries_2017}}. 
The outcomes of 
those studies have been post-processed with radiative transfer methods in order 
to obtain synthetic observations, e.g. as seen from the \textcolor{black}{{\it ALMA } 
telescope~\citep{krumholz_apj_665_2007,seifried_mnras_571_2016}}. 
In addition to different spatial resolution that limits the 
proper capture of local gravitational collapse in a disc, those studies devoted 
to massive star formation differ by (i) the level of realism taken into 
account in order to model the proto-stellar feedback, (ii) the spatial resolution 
allowed by the scheme and (iii) the criteria 
applied to consider a dense region of the disc as a fragment, e.g. introducing 
sink particles~\textcolor{black}{\citep[see][for a description of such technique]{federrath_apj_713_2010}}. 
}

The characterisation of the fragmentation of self-gravitating discs is not a 
trivial task and several criteria necessary but not sufficient have been derived so far. 
The coupling of N-body solvers to hydrodynamical codes via 
the introduction of sink particles \textcolor{black}{alleviate the computation 
of longer timescales, because subgrid models save computing time since they do 
not have to resolve the smallest scales~\citep{federrath_apj_713_2010}.}
%
%
However, the required conditions of their creation throughout a simulation is subject to debate and 
may lead to artificial \textcolor{black}{fragmentation} which can, for instance, affects their 
host spiral arm dynamics. The two main of those 
criteria are the so-called Toomre and Gammie criteria and they compare the local 
\textcolor{black}{effects} of gravitational instability~\citep{toomre_apj_138_1963} and 
cooling~\citep{gammie_apj_462_1996} against the combined effects of the disc 
thermodynamics with the gas rotation shear, respectively. A large literature 
confronts these criteria with numerical simulations of 
protostellar and low-mass self-gravitating discs and highlights their sensitivity 
to initial conditions, viscosity, resolution and 
numerics~\citep{stamatellos_mnras_400_2009,forgen_mntas_410_2011,lodato_mnras_413_2011,meru_mnras_411_2011,
forgan_mnras_420_2012,rice_mnras_420_2012,paardekooper_mnras_421_2012, rice_mnras_438_2014,tsukamoto_mnras_446_2015}. 
Interestingly, in~\citet{klassen_apj_823_2016}, numerical simulations of massive protostellar discs 
that fulfill those criteria may not necessarily \textcolor{black}{form fragments} while this apparent 
stability is consistent with the so-called Hill criterion~\citep{roger_mnras_423_2012}. 

\begin{figure*}
	\centering
	\begin{minipage}[b]{ 0.33\textwidth}
		\includegraphics[width=1.0\textwidth]{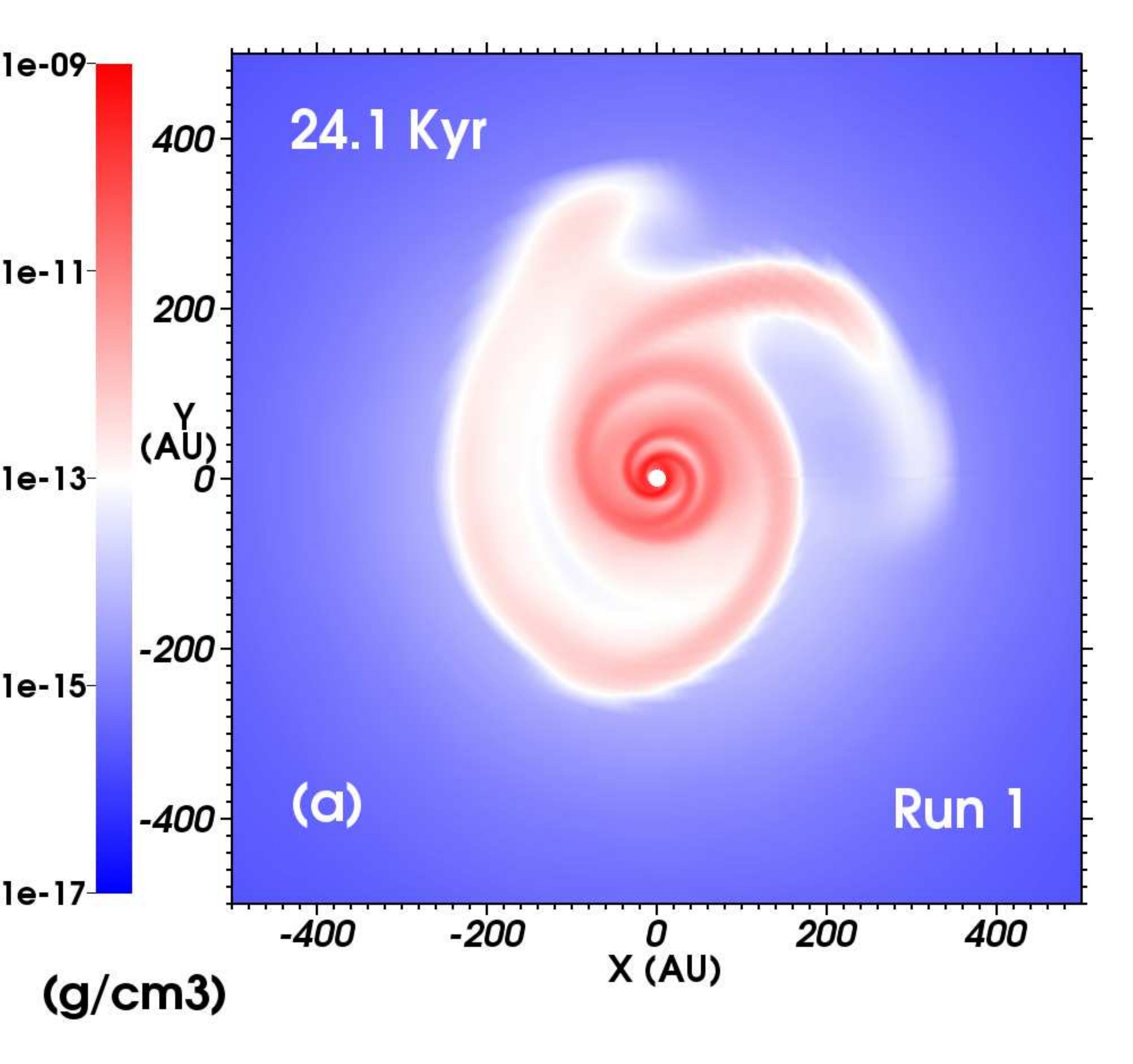}
	\end{minipage}	
	\begin{minipage}[b]{ 0.33\textwidth}
		\includegraphics[width=1.0\textwidth]{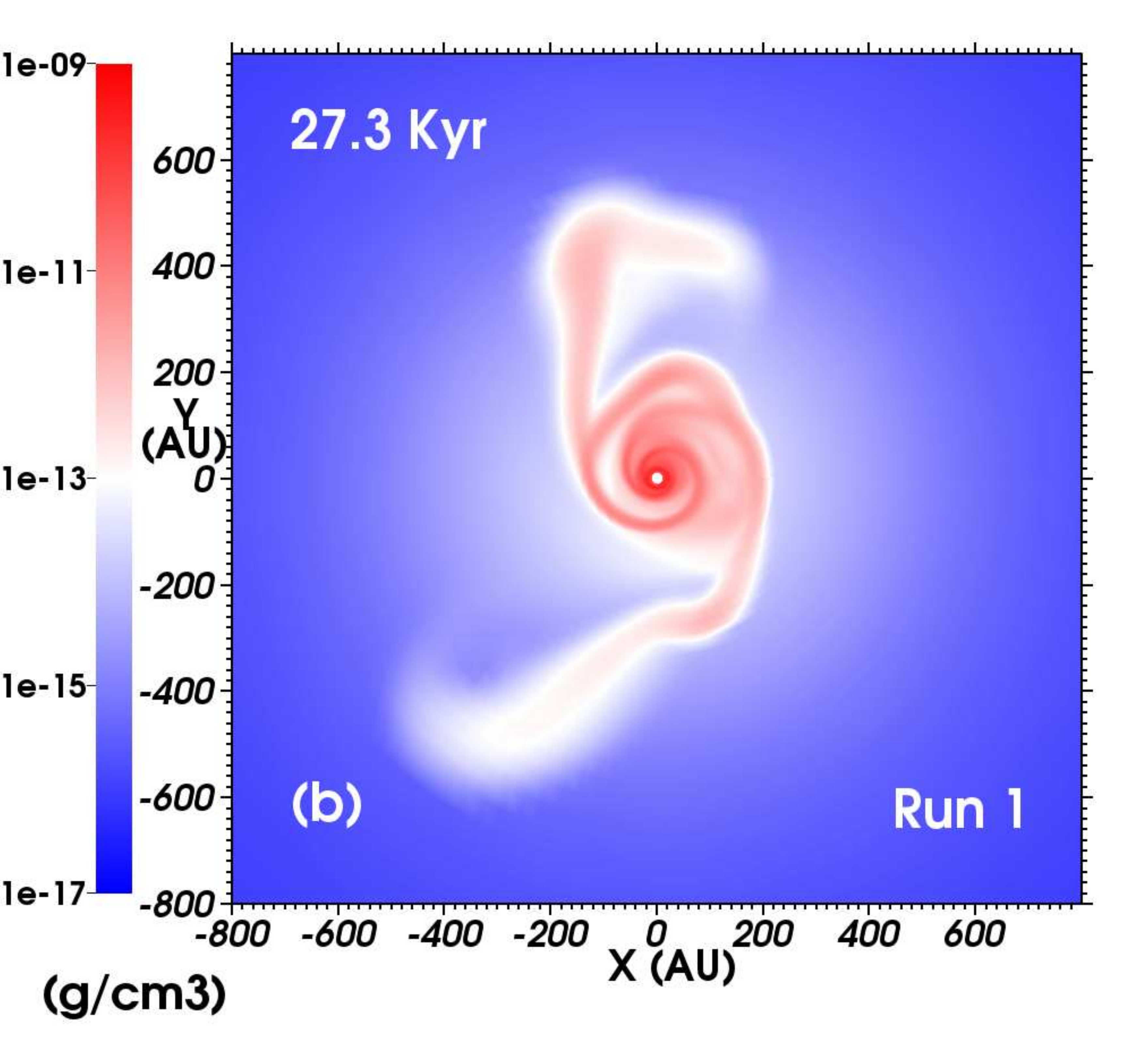}
	\end{minipage} 
	\begin{minipage}[b]{ 0.33\textwidth}
		\includegraphics[width=1.0\textwidth]{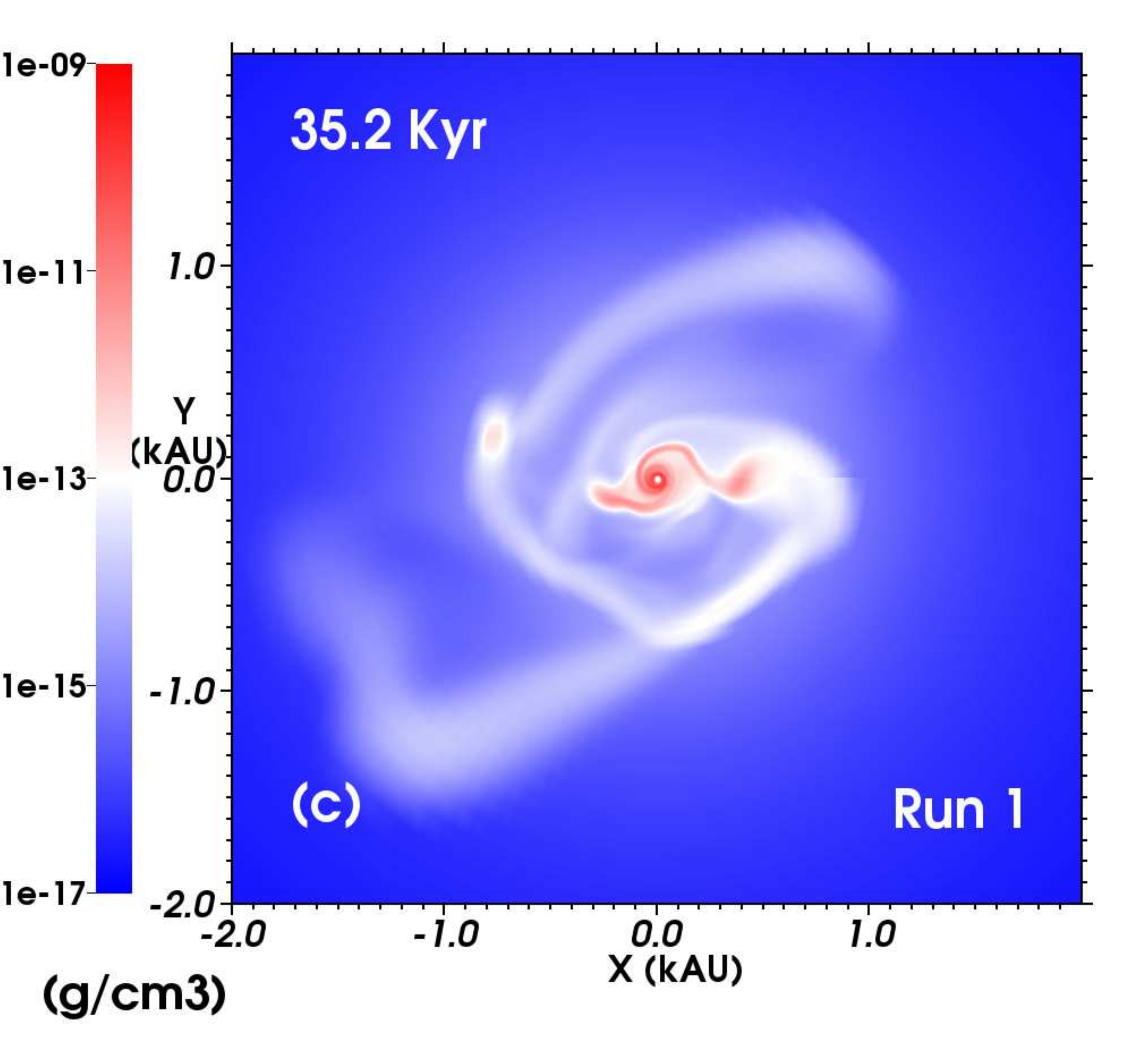}
	\end{minipage}	\\
	\begin{minipage}[b]{ 0.33\textwidth}
		\includegraphics[width=1.0\textwidth]{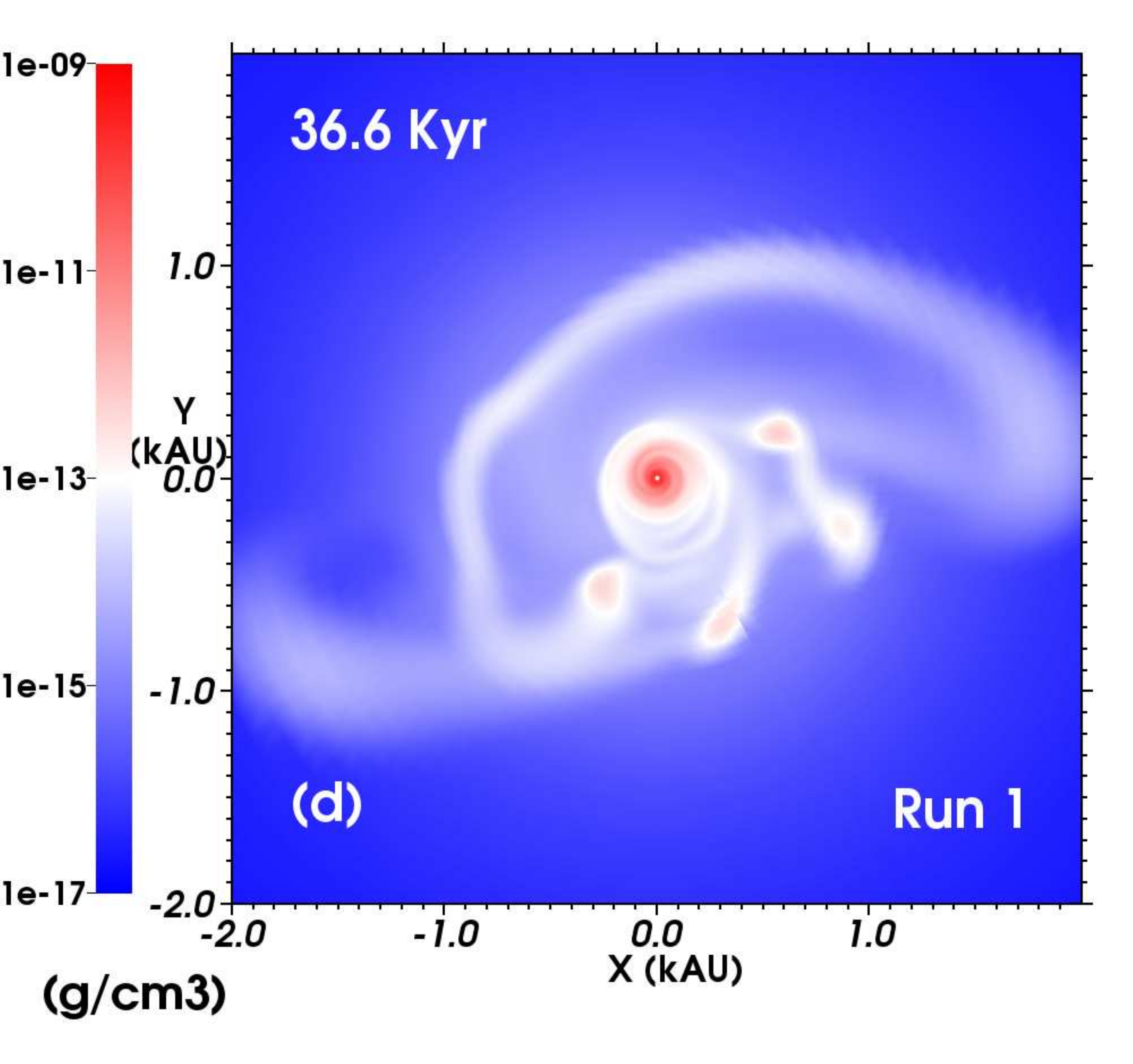}
	\end{minipage}	
	\begin{minipage}[b]{ 0.33\textwidth}
		\includegraphics[width=1.0\textwidth]{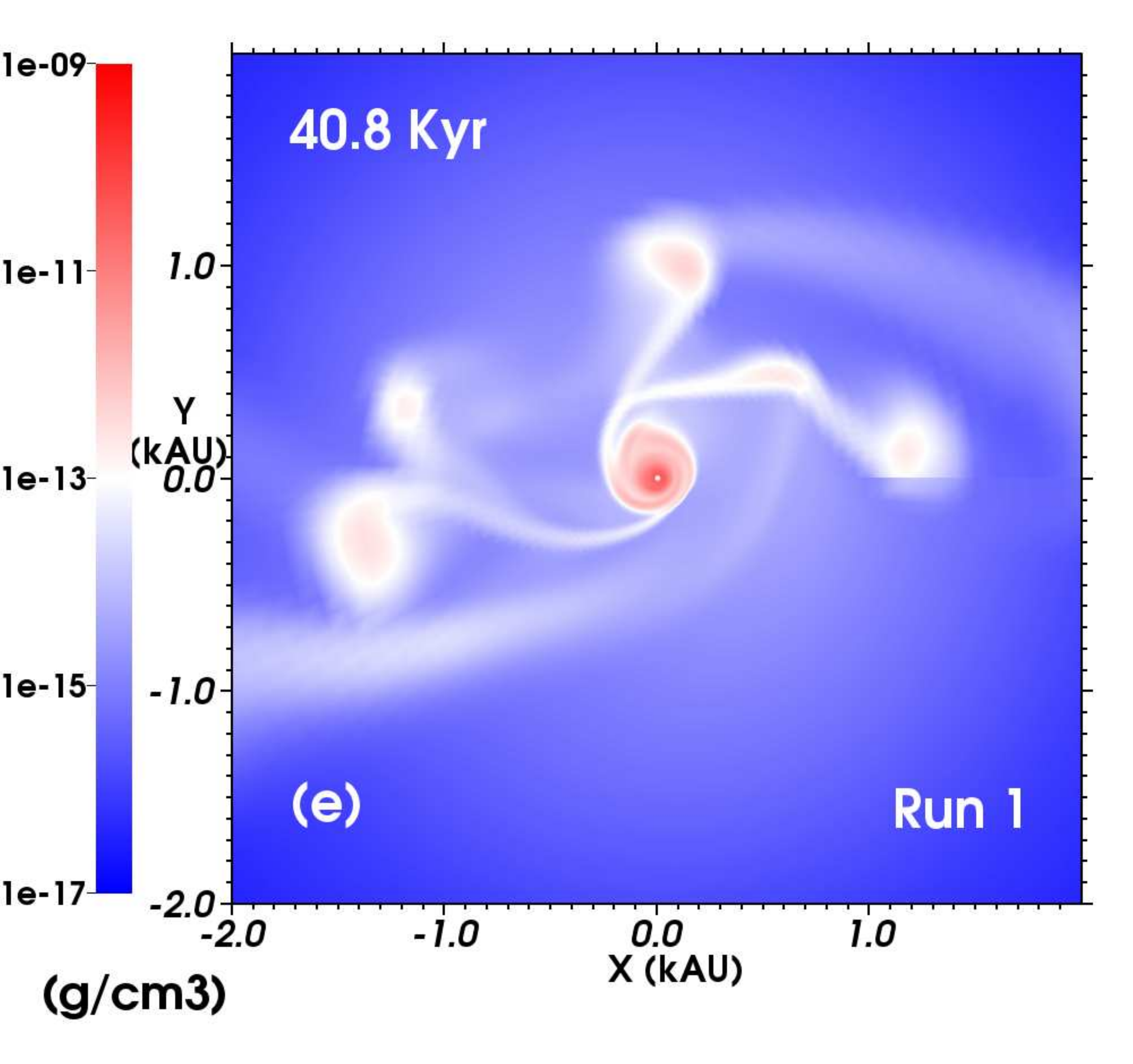}
	\end{minipage}
	\begin{minipage}[b]{ 0.33\textwidth}
		\includegraphics[width=1.0\textwidth]{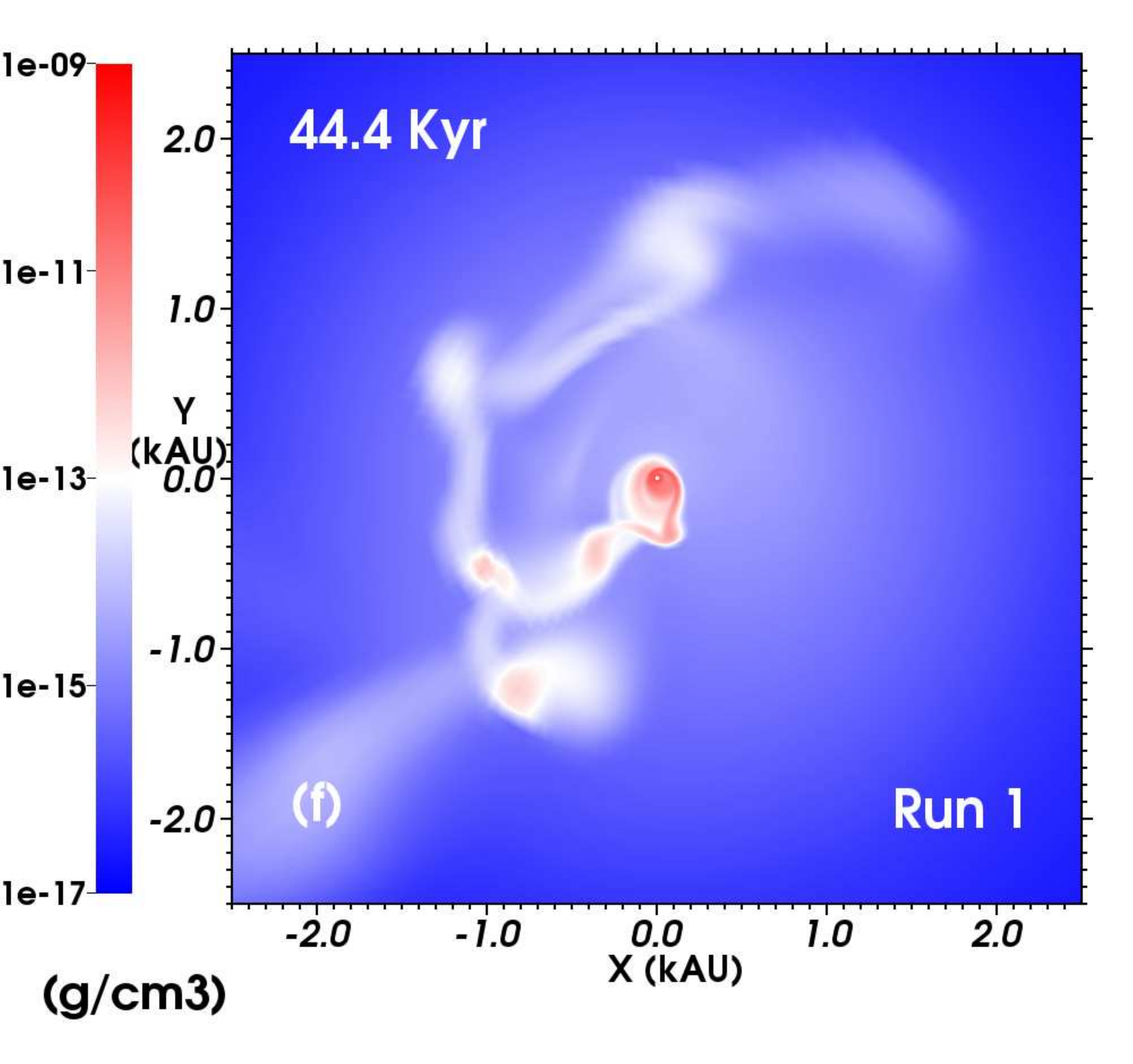}
	\end{minipage}
	\caption{ 
		 Midplane density (in $\rm g\, \rm cm^{-3}$) of the accretion disc generated in Run~1, 
		 showed at different characteristic times of its evolution (in $\rm kyr$).  
		 The density is plotted on a logarithmic scale and the size of the panels is 
		 in AU \textcolor{black}{(top left and middle panel) or kAU (\textcolor{black}{elsewhere})}. 
		 }	
	\label{fig:disc_evol1}  
\end{figure*}

The consistency of a FU-Orionis-like scenario producing strong accretion 
outbursts by fast migration of disc debris has just been shown for massive star 
formation~\citep{meyer_mnras_464_2017} and is in accordance with the outflows  
observed in the young massive stellar system 
S255IR-SMA1~\citep{burns_mnras_460_2016,caratti_nature_2016} 
and NGC6334I-MM1~\citep{2017arXiv170108637H}. In the present study, we extend our 
numerical investigation of the stability of self-gravitating accretion discs 
around massive protostars and its connection to the accretion-driven 
flares \textcolor{black}{as well as multiplicity}. 
While the scientific literature already covered the question of the 
gravitational collapse of massive ($\ge\, 30\, \rm M_{\odot}$) pre-stellar cores 
generating a massive protostar surrounded by a growing accretion disc, those 
studies generally assumed a simplified treatment of the protostellar 
feedback, \textcolor{black}{
i.e. within the flux-limited diffusion approximation~\citep{krumholz_apj_656_2007,Krumholz_sci_2009} 
or suffer from a lack of resolution~\citep{kuiper_apj_732_2011}, especially in the region close to 
the star~\citep{krumholz_apj_656_2007,Krumholz_sci_2009,rosen_mnras_463_2016,klassen_apj_823_2016}
}. 
\textcolor{black}{
\textcolor{black}{The studies above presented evidence for} (i) multiplicity in the 
high-mass regime by generating large-separation low-mass companions in spiral arms of massive protostellar accretion 
discs~\citep{krumholz_apj_656_2007,rosen_mnras_463_2016}, (ii) massive binaries made of two high-mass 
protostars~\citep{Krumholz_sci_2009}, (iii) close massive binaries of two high-mass components~\citep{bonnell_mnras_271_1994}, 
while (iv)~\citet{klassen_apj_823_2016} does not conclude on the formation of stars around young massive protostellar objects. 
Nevertheless, no satisfactory theoretical models have yet explained the existence of small-separation/short-period low-mass 
companions to young high-mass stars. This is an important question because such systems are believed to evolve towards 
so-called spectroscopic (massive) binary systems, i.e. a particular class of close massive binaries made of \textcolor{black}{an} O-type star 
surrounded \textcolor{black}{by one or more closely-separated low-mass companions~\citep[see][]{2013A&A...550A..27M,2014ApJS..213...34K}}. 
}

\textcolor{black}{ 
The goal of this study consists in testing the disc fragmentation scenario for the formation of those short-period massive 
binaries made of a proto-O star and (at least) a close low-mass companion, 
using high-resolution hydrodynamical simulations of the gravitational collapse of rotating pre-stellar cores. 
We perform three-dimensional numerical simulations including self-gravity, radiative transport and 
a midplane-symmetric spherical computational domain centered at a sink cell in which a protostar is assumed to 
form and evolve. The young star acquires 
mass by disc accretion, as in many current studies devoted to disc fragmentation around low-mass and primordial 
protostars~\citep{vorobyov_mnras_381_2007,voroboyov_aa_557_2013,vorobyov_apj_805_2015,hosokawa_2015}. 
Our highly-resolved models neglect the internal turbulence and magnetization of the collapsing 
pre-stellar cores but allow a sub-AU spatial resolution close to the protostar which is higher 
than in any models previsouly published so far. 
}
We find that all our \textcolor{black}{modeled discs} show clear sign of instability, fragment and generate 
\textcolor{black}{a series of sudden increase of the accretion rate} \textcolor{black}{caused by fragments migrating} towards the central protostar. 
Interestingly, some of those migrating clumps have \textcolor{black}{interior thermodynamic properties} indicating 
that \textcolor{black}{they} may be secondary pre-stellar \textcolor{black}{embryos}, 
and, therefore, we conclude on the viability of disc fragmentation as a channel 
for \textcolor{black}{the formation of close low-mass companions to forming massive stars}.

Our study is organized as follows. We begin in Section~\ref{sect:method} with 
introducing the reader to our numerical setup. In Section~\ref{sect:ci}, we 
describe our particular initial conditions in terms of density and angular 
frequency distribution of the collapsing pre-stellar cores that we consider. 
Their evolution, the formation of accretion discs and their subsequent 
fragmentation are qualitatively described in Section~\ref{section:results_disc}. 
\textcolor{black}{
In Section~\ref{section:numerics}, we investigate the physical and numerical 
processes at work in this study, i.e. the role of the central protostellar irradiation 
and the effects of the spatial resolution in the simulations,}
while we further test our \textcolor{black}{outcomes} for accretion disc against various criteria 
for disc stability in Section~\ref{section:results_fragmentation}.
\textcolor{black}{
In section~\ref{section:closebin} we detail the fate of the disc debris and investigate 
their properties as forming low-mass stars and further discuss our 
findings in the context of the formation of close massive binary systems.
}
In Section~\ref{section:discussion} we compare our 
results to previous theoretical and observational studies.  
%
%
Finally, we \textcolor{black}{give} our conclusions in Section~\ref{section:cc}.


\section{Method}
\label{sect:method}

In the following paragraphs, we introduce the reader to the method used to 
perform our numerical simulations. Our \textcolor{black}{simulations model} the collapse of a 
pre-stellar core forming around a self-gravitating accretion disc that evolves 
as being irradiated by the time-dependent luminosity of the central protostar. 
We detail the various assumptions and prescriptions used to model the 
microphysical processes of the gas and the dust of the pre-stellar core, 
describe the manner we treat the protostellar feedback onto the disc and we 
present the computational grid and the numerical scheme utilised in this study. 
 
\begin{table*}
	\centering
	\caption{
	\textcolor{black}{Initial conditions. Quantity $\beta_{\Omega}$
	is the slope of the \textcolor{black}{angular velocity} \textcolor{black}{index} of the pre-stellar core.  
	Simulations are run until $t_{\rm end}$ (in $\rm kyr$).  } 
	}
	\begin{tabular}{lcccr}
	\hline
	${\rm {Models}}$       &   Angular frequency index $\beta_{\Omega}$         &  $t_{\rm end}$ ($\rm kyr$) & $\mathrm{Motivation}$ \\ \hline   
	{\rm Run 1}            &   $~~0.0 $                                         &  50.0                    & Solid-body \textcolor{black}{rotation}~\citep[cf.~run~M100~in][]{klassen_apj_823_2016} \\  
	{\rm \textcolor{black}{Run 1-noIrr}}      &   $~~0.0 $                                         &  45.0                    & cf. Run 1, without protostellar irradiation \\ 
	{\rm \textcolor{black}{Run 1-LR}}         &   $~~0.0 $                                         &  40.0                    & cf. Run 1, with a coarser spatial resolution \\ 
	{\rm \textcolor{black}{Run 1-HR}}         &   $~~0.0 $                                         &  32.1                    & cf. Run 1, with a finer spatial resolution \\ 	{\rm Run 2}            &   $-0.35$                                          &  40.0                    & Intermediate initial \textcolor{black}{angular velocity} radial distribution  \\     		
	{\rm Run 3}            &   $-0.75$                                          &  35.0                    & Steep initial \textcolor{black}{angular velocity} radial distribution~\citep[see][]{meyer_mnras_464_2017} \\     			
	\hline    
	\end{tabular}
\label{tab:sigma}
\end{table*}

\subsection{Governing equations}
\label{sect:eq}

As other studies devoted to star formation, we work at the interplay 
between gas dynamics, gravity and radiation transport. 
The hydrodynamics is described by the equations of fluid dynamics 
for compressible gas that consist of the relations for mass,
\begin{equation}
	   \frac{\partial \rho}{\partial t}  + 
	   \bmath{\nabla}  \cdot (\rho\bmath{v}) =   0,
\label{eq:euler1}
\end{equation}
momentum,
\begin{equation}
	   \frac{\partial \rho \bmath{v} }{\partial t}  + 
           \bmath{\nabla} \cdot ( \rho  \bmath{v} \otimes \bmath{v} )  + 
           \bmath{\nabla}p 			      =   \bmath{f},
\label{eq:euler2}
\end{equation}
and total energy conservation,
\begin{equation}
	  \frac{\partial E }{\partial t}   + 
	  \bmath{\nabla} \cdot ((E+p) \bmath{v})  =	   
	  \bmath{v} \cdot \bmath{f} 
\label{eq:euler3}
\end{equation}
where $\rho$, $\bmath{v}$, $p$ are the Eulerian variables, i.e. the 
mass density, the vector velocity and the gas pressure, respectively. 
The gas assumes an ideal equation of state $p=(\gamma-1)E_{\rm int}$ 
where $\gamma=5/3$ is the constant adiabatic index. The variable 
$E_{\rm int}$ is the internal energy of the medium such that,
\begin{equation}
	  E  = E_{\rm int} + \textcolor{black}{\frac{1}{2} \rho \bmath{v}^{2}},
\label{eq:etot}
\end{equation}
represents the total energy of the plasma.

The radiation transport scheme is taken \textcolor{black}{from}~\citet{kuiper_aa_511_2010}. 
In Eqs.~(\ref{eq:euler2}) and~(\ref{eq:euler3}) the quantity, 
\begin{equation}
          \textcolor{black}{
	  \bmath{ f } = -\rho \bmath{\nabla} \Phi_{\rm tot} 
	                - \frac{\rho  \kappa_{\rm R}(T_{\star})}{c}    \bmath{F_{\star}}  
			- \frac{\rho  \kappa_{\rm R}(T)}{c}            \bmath{F}          ,
          }
\label{eq:f}
\end{equation}
is the force density vector with \textcolor{black}{$\Phi_{\rm tot}$ total 
gravitational potential, $\kappa_{\rm R}$ the \textcolor{black}{Rossland} opacity, $c$ the speed of light, 
$\bmath{F}$ the radiation flux and $\bmath{F_{\star}}$ the 
stellar radiation flux}. The equations of radiation transport govern the time-evolution 
of the \textcolor{black}{radiation} energy density $E_{\rm R}$, 
\begin{equation}
	  \frac{\partial E_{\rm R} }{\partial t}   + 
	  f_{\rm c} \bmath{\nabla} \cdot (   \bmath{F}  +   \bmath{F_{\star}}  )  =	   
	  0,
\label{eq:rad1}
\end{equation}
where $f_{\rm c}=1/( c_{\rm v} \rho / 4 a T^{3} + 1 )$ with $c_{\rm v}$ the 
specific heat capacity and $a$ the radiation constant. We solve 
Eq.~(\ref{eq:rad1}) in the flux-limited diffusion (FLD) approximation, where 
$\bmath{ F } = -D \bmath{\nabla} E_{\rm R}$. The 
diffusion constant is $D = \lambda c / \rho \kappa_{\rm R}$ where 
\textcolor{black}{$\lambda$ is the flux limiter and $\kappa_{\rm R}$ 
is the mean Rossland opacity}. The relation, 
\begin{equation}
	  \frac{ \bmath{F_{\star}}(r)  }{ \bmath{F_{\star}}( R_{\star})  } = \Big( \frac{ R_{\star} }{ r } \Big)^{2} e^{-\tau(r)},
\label{eq:rad2}
\end{equation}
reports the decrease of the flux during the ray-tracing of the stellar 
incident radiation \textcolor{black}{with the stellar radius $R_{\star}$ and 
$\tau(r) = \kappa_{\rm P}(T_{\star}) \rho(r)$ being} the optical depth of the medium. 
The total opacity takes into account both the gas \textcolor{black}{and} the dust 
attenuation of the radiation, i.e., $\kappa_{\rm P}=\kappa_{\rm P}^{\rm g}+\kappa_{\rm P}^{\rm d}$, 
where $\kappa_{\rm P}^{\rm g}$ and $\kappa_{\rm P}^{\rm d}$ are the gas and dust opacities, respectively. 
To this end, we use constant gas opacity of $\kappa_{\rm P}^{\rm g}=0.01\, 
\textcolor{black}{\rm cm^{2}\, \rm g^{-1}}$ and utilise the tabulated dust opacity 
of~\citet{laor_apj_402_1993}. Gas and dust temperature are calculated assuming 
the equilibrium \textcolor{black}{between the dust temperature and the total radiation field}, 
\begin{equation}
	  a T^{4}  = E_{\rm R} + \frac{\kappa_{\rm P}(T_{\star})}{\kappa_{\rm P}(T)} \frac{ |\bmath{F_{\star}}| }{ c }.  
\label{eq:radfield}
\end{equation}
\textcolor{black}{
We employed the time-dependent treatment of the dust for photoevaporation and sublimation described 
in~\citet{kuiper_apj_732_2011} and used in subsequent works~\citep{hosokawa_2015}. At every \textcolor{black}{timestep} and in each cells 
of the mesh, the dust temperature that is obtained solving Eq.~(\ref{eq:radfield}) is compared 
with the local evaporation temperature of the dust grains  $T_{\rm evap}$. It is initially determined 
in~\citet{1994ApJ...421..615P} and has been reformulated in~\citet{isella_aa_438_2005} as the 
power-law expression $T_{\rm evap}= g \rho^{\beta}$ where $g=2000\, \rm K$ and $\beta=0.0195$. 
It permits to adjust the local dust-to-gas mass ratio by taken into account the evaporation ($T>T_{\rm evap}$) 
and the sublimation ($T<T_{\rm evap}$) of the dust grains component of our mixture treated as a single-fluid. 
It particularly applies to the hot regions of the computational domain such as in the innermost part of 
the disc, see~\citet{kuiper_aa_511_2010}.
}

The stellar luminosity represents the total irradiation, i.e. the sum of the photospheric 
luminosity $L_{\star}$ plus the accretion luminosity \textcolor{black}{$L_{\mathrm{acc}} \propto  
M_{\star}\dot{M}/R_{\star}$} of the central protostar. \textcolor{black}{This} quantity is 
mainly \textcolor{black}{a} function of the stellar radius $R_{\star}$, the effective temperature $T_{\rm 
eff}$ and \textcolor{black}{the accretion rate $\dot{M}$} that are interpolated from the pre-main-sequence stellar evolutionary 
models assuming high accretion of~\citet{hosokawa_apj_691_2009}. \textcolor{black}{The} tracks 
allow us to time-dependently adapt the protostellar luminosity as a function of the 
accretion history onto the forming massive star.

The gravity of the gas is taken into account calculating the total 
gravitational potential felt by a volume element of the gas,  
\begin{equation}
     \Phi_{\rm tot}  = \Phi_{\star} + \Phi_{\rm sg}, 
   \label{eq:potential}    
\end{equation}
where $\Phi_{\star}=-G M_{\star}/r$, with $M_{\star}$ the protostellar mass 
and $G$ the universal gravity constant. The self-gravity of the gas 
\textcolor{black}{$\Phi_{\rm sg}$} is described by the Poisson equation, 
\begin{equation}
	  \bmath{\Delta} \Phi_{\rm sg}   =  4\pi G\rho,
\label{eq:poisson}
\end{equation}
that we solve separately. \textcolor{black}{This allows the simulation to tend towards a solution} which reports 
both the accretion phenomenons onto the central protostar but also the 
gravitational interactions that disc substructures share with each other. No 
additional artificial \textcolor{black}{shear viscosity} is injected into the system, via e.g. an 
$\alpha$-prescription of~\citet{shakura_aa_24_1973}, since three-dimensional 
models produce self-consistently gravitational torques accounting for efficient 
angular momentum transport~\citep{kuiper_apj_732_2011,hosokawa_2015}.

\begin{figure*}
	\centering
	\begin{minipage}[b]{ 0.33\textwidth}
		\includegraphics[width=1.0\textwidth]{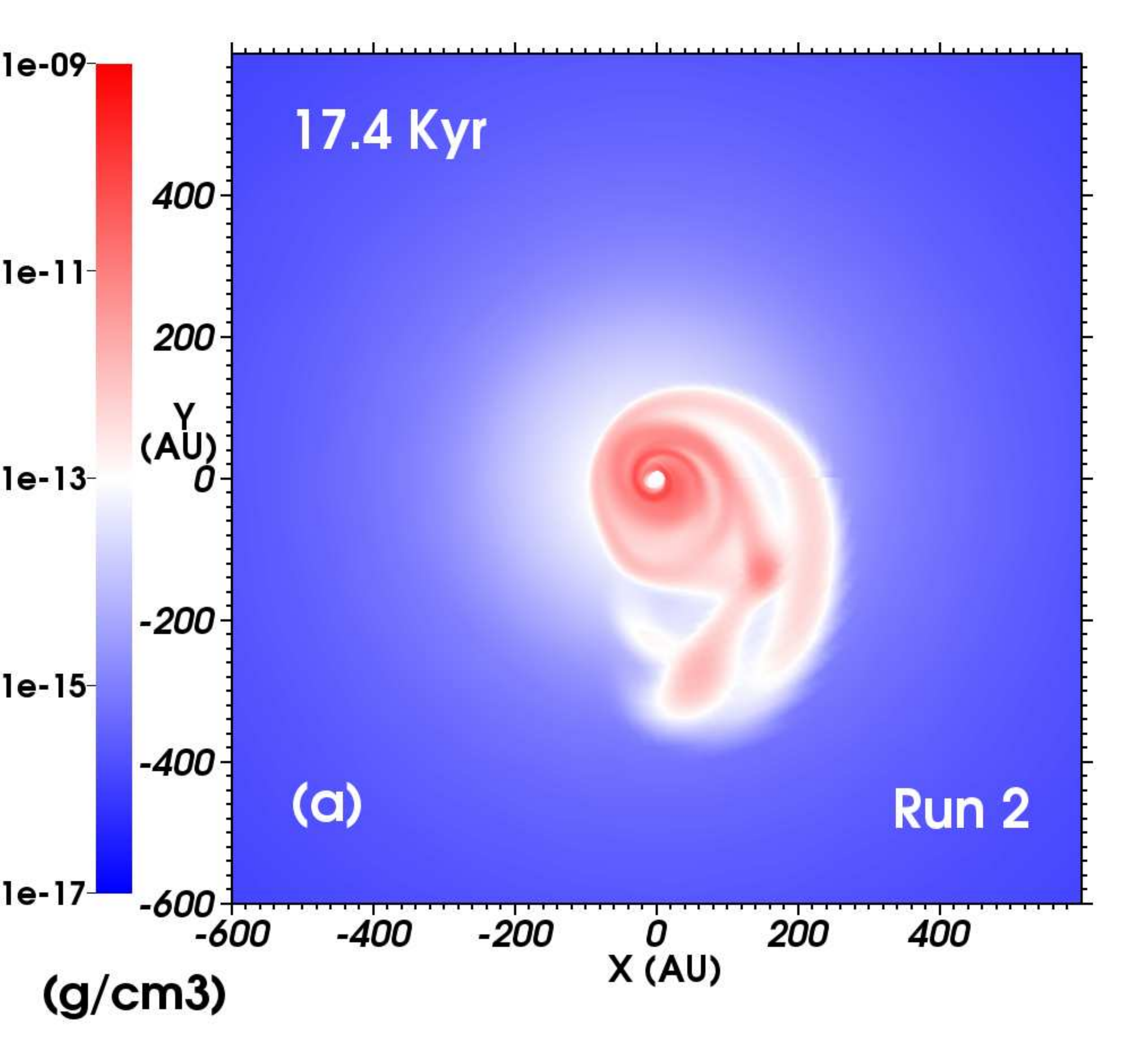}
	\end{minipage}	
	\begin{minipage}[b]{ 0.33\textwidth}
		\includegraphics[width=1.0\textwidth]{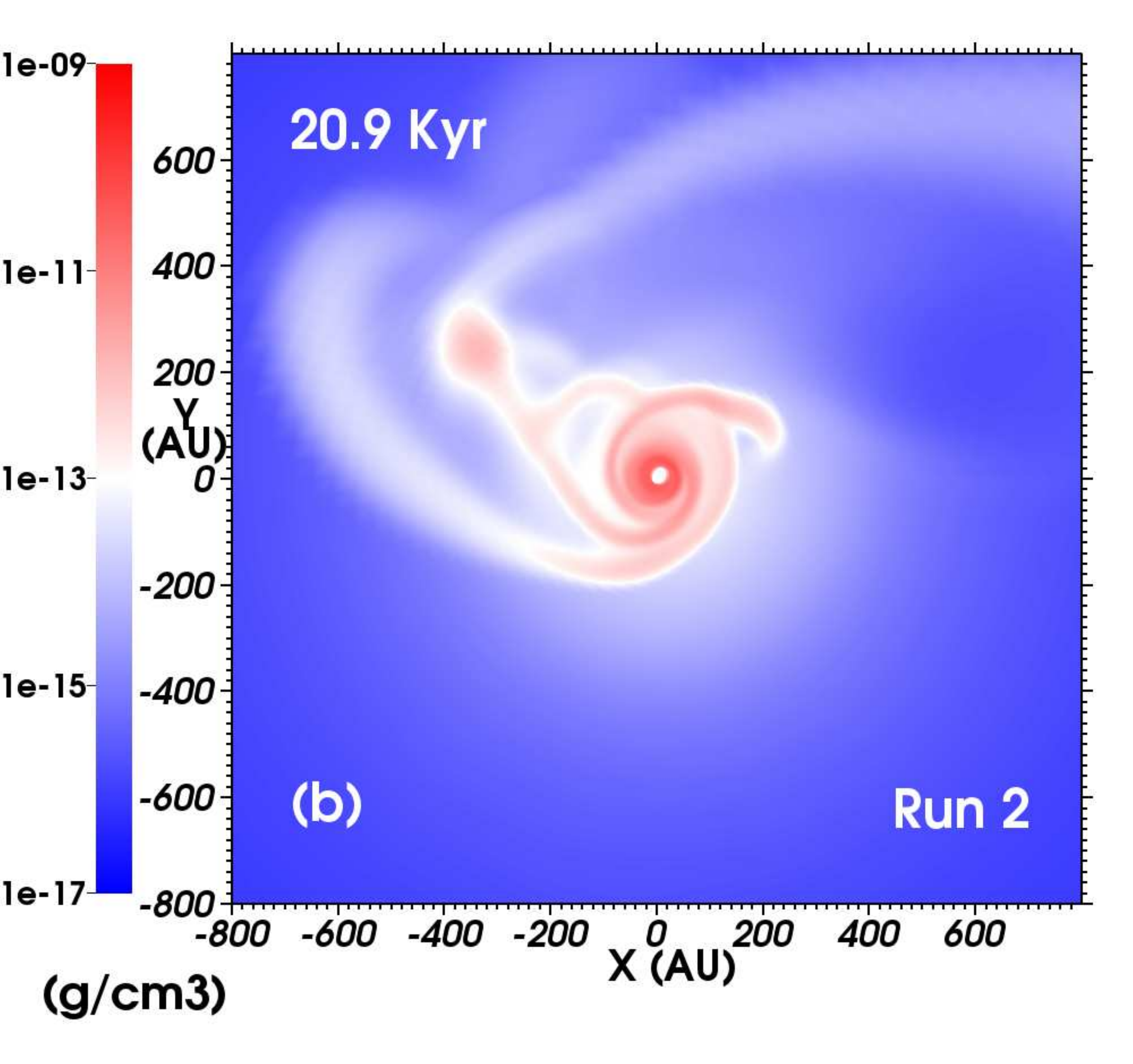}
	\end{minipage} 
	\begin{minipage}[b]{ 0.33\textwidth}
		\includegraphics[width=1.0\textwidth]{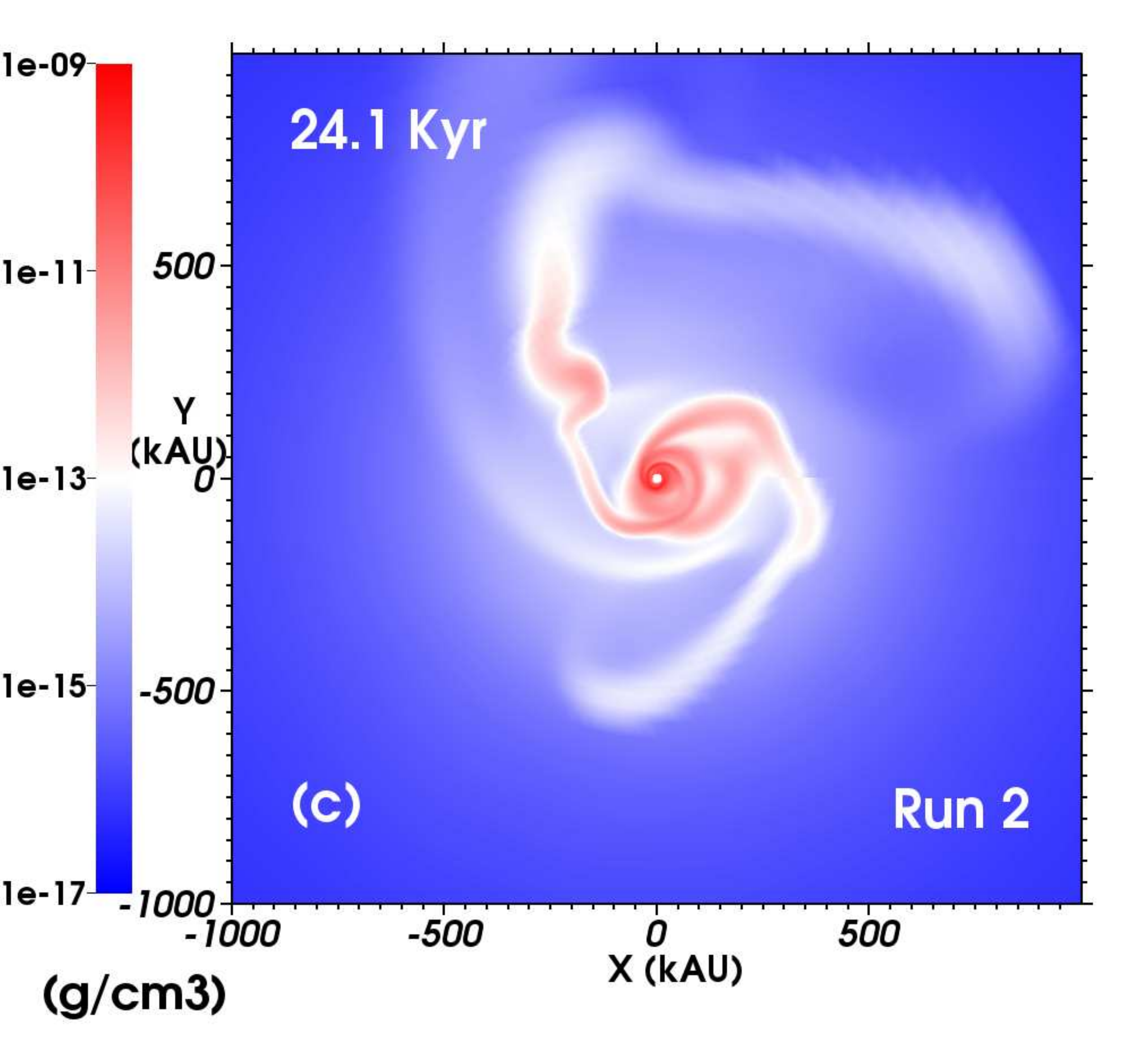}
	\end{minipage}	\\
	\begin{minipage}[b]{ 0.33\textwidth}
		\includegraphics[width=1.0\textwidth]{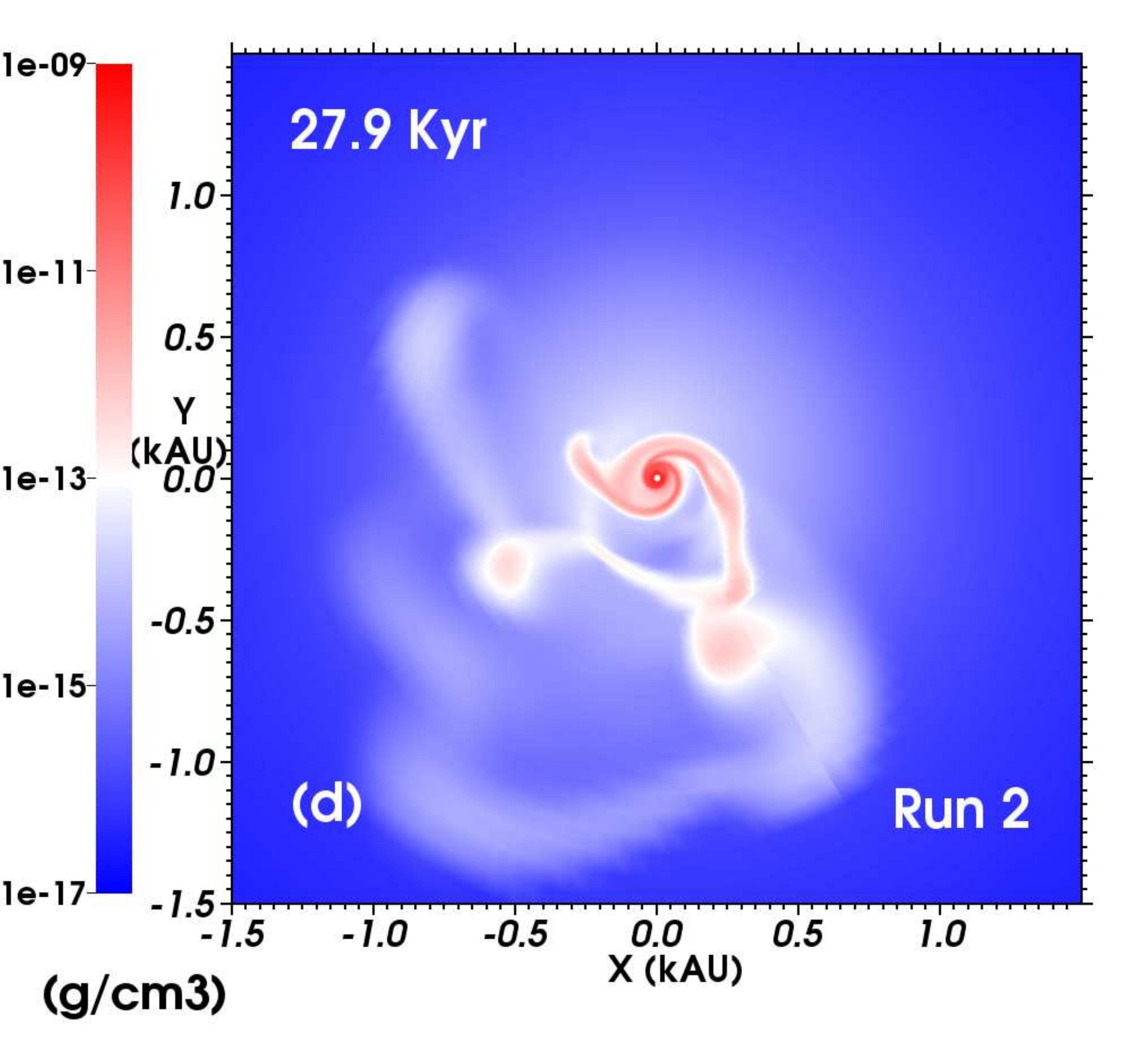}
	\end{minipage}	
	\begin{minipage}[b]{ 0.33\textwidth}
		\includegraphics[width=1.0\textwidth]{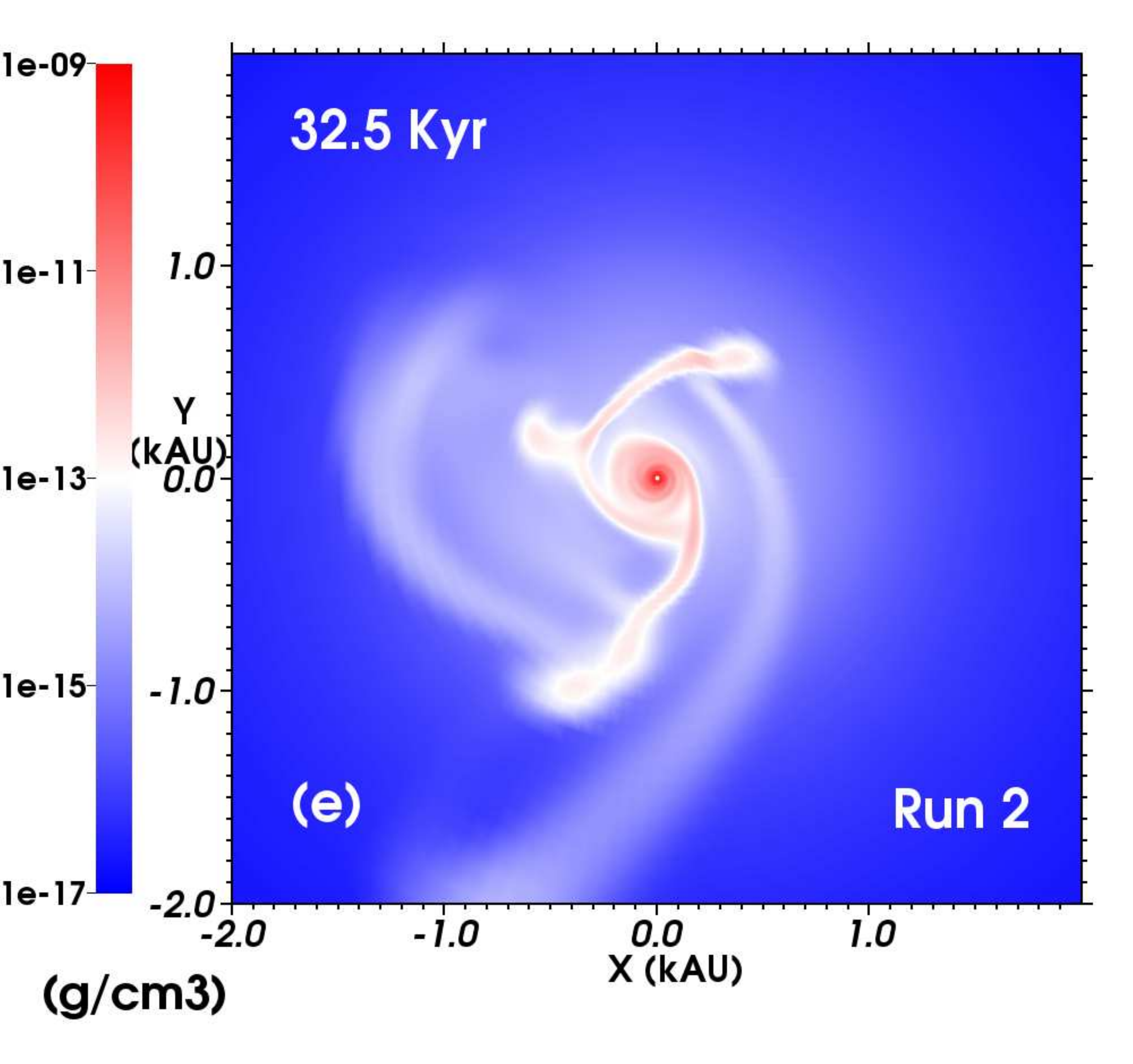}
	\end{minipage}
	\begin{minipage}[b]{ 0.33\textwidth}
		\includegraphics[width=1.0\textwidth]{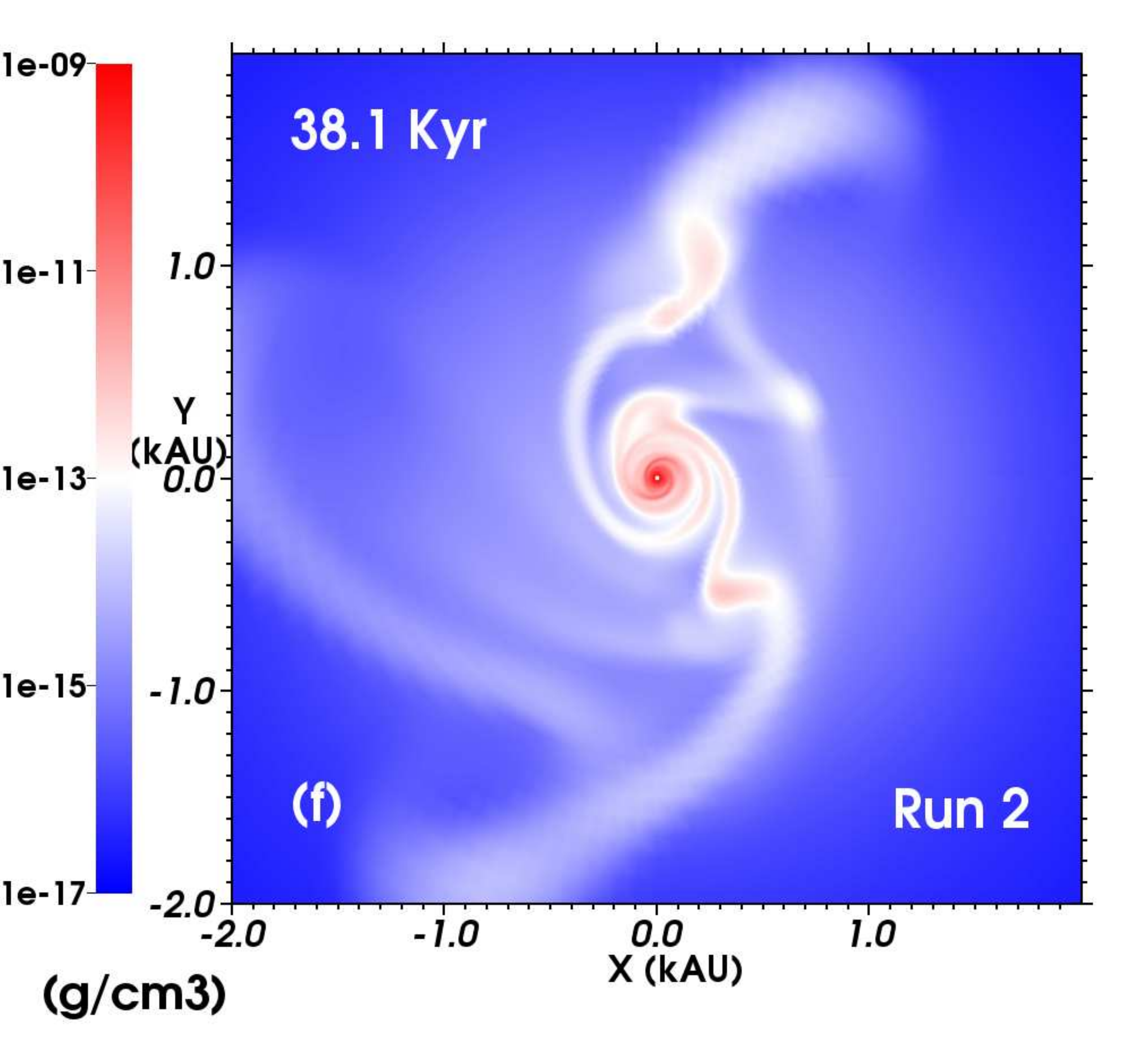}
	\end{minipage}
	\caption{ 
		 \textcolor{black}{Similar to Fig.~\ref{fig:disc_evol1} for Run~2.}   
		 }	
	\label{fig:disc_evol2}
\end{figure*}

\subsection{Grid, boundary conditions and numerical scheme}
\label{sect:setup}

We perform our three-dimensional numerical simulations using a static, 
non-uniform grid mapped onto a spherical coordinate system 
\textcolor{black}{$(r,\theta,\phi)$}. Assuming mid-plane symmetry of the computational domain 
with respect to $\theta=\pi/2$, the grid size \textcolor{black}{is} $[r_{\rm in},r_{\rm 
max}]\times[0,\pi/2]\times[0,2\pi]$ along the radial, polar and azimuthal 
directions, respectively. The grid is made of $128\times21\times128$ cells so 
that we use \textcolor{black}{squared} cells in inner region of the midplane. The mesh is 
logarithmically stretched in the radial direction, i.e. their size radially 
increases as \textcolor{black}{$r(10^{f}-1)$} with $f=\log(r_{\rm max}/r_{\rm in})$, 
and it expands as $\cos(\theta)$ in the polar 
direction~\citep{ormel_mnras_447_2015} whereas it is uniform along the azimuthal 
direction. This grid choice allows us to save computing time in reducing the 
total number of cells while having both a high spatial resolution of up to 
$\Delta r < 1\rm AU$ in the inner region of the midplane. Particularly, the grid 
resolution \textcolor{black}{is at} below $10\, \rm AU$ in the $\approx\, 500\rm AU$ region of the domain. 
The inner boundary in $r_{\rm in}=10\, \rm AU$ whereas the outer one is $R_{\rm 
c}=0.1\, \rm pc$. Outflow conditions are assigned at both boundaries of the 
radial directions so that we can on-the-fly estimate, e.g. $\dot{M}$ as 
the mass of the material crossing $r_{\rm in}$.

We solve the above described equations using the {\sc pluto} 
code~\citep{mignone_apj_170_2007,migmone_apjs_198_2012} that has been augmented 
with several modules for self-gravity, radiation transport and stellar evolution 
that are described 
\textcolor{black}{in~\citet{kuiper_aa_511_2010,kuiper_apj_732_2011,kuiper_apj_772_2013}}. The 
\textcolor{black}{distinctive feature} of this method consists in carefully treating the stellar radiation, 
first ray-tracing the photons from the stellar atmosphere to the disc and then 
\textcolor{black}{mimicking} their propagation into the disc by flux-limited diffusion~\citep[see 
also][]{kuiper_aa_537_2012,kuiper_aa_555_2013}. 
\textcolor{black}{This algorithm has shown the existence of a dust-free front in the 
massive stars' surroundings~\citep{kuiper_apj_722_2010} and this has been strengthened by semi-analytical 
calculations~\citep{vaidya_apj_742_2011}}.

Our method uses a Godunov-type 
solver made of a shock-capturing Riemann solver embedded in a conservative 
finite-volume scheme. We use the hllc solver for fluid dynamics that is 
ruled by the Courant-Friedrich-Levy parameter \textcolor{black}{set to 
$C_{\rm cfl }$ = 0.1 (Run~3, see Section~\ref{sect:ci}) and to higher 
initial values ($C_{\rm cfl }$ = 0.2$-$0.3) for the other runs}. Furthermore, we use the minmod flux limiter and the 
WENO3 interpolation scheme with the \textcolor{black}{third order Runge-Kutta (RK3) time integrator. 
\textcolor{black}{
Additionally, we use the FARGO (Fast Advection in Rotating Gaseous Objects) method~\citep{masset_aa_141_2000} which 
permits larger timesteps than if exclusively controlled by a strict application of the Courant-Friedrich-Levy rule. 
FARGO has been designed to be utilised in the context of problems with a background orbital motion like our simulations 
and it is available in {\sc pluto}~\citep{migmone_apjs_198_2012}. 
}
Therefore, our overall scheme is third} order in space and time. To 
reduce the huge computing time of such calculations, the radiation transport is 
performed within the gray approximation.
The self-gravity of the gas is calculated up to reaching 
\textcolor{black}{$\bmath{\Delta} \Phi_{\rm sg}/\Phi_{\rm sg}\le 10^{-5}$ \textcolor{black}{as in}~\citet{kuiper_apj_732_2011}}. 
\textcolor{black}{
Finally, note that the seed perturbations for the non-axisymmetric modes are 
numerical~\citep[cf.][]{kuiper_apj_732_2011,hosokawa_2015}.  
}


\begin{figure*}
	\centering
	\begin{minipage}[b]{ 0.33\textwidth}
		\includegraphics[width=1.0\textwidth]{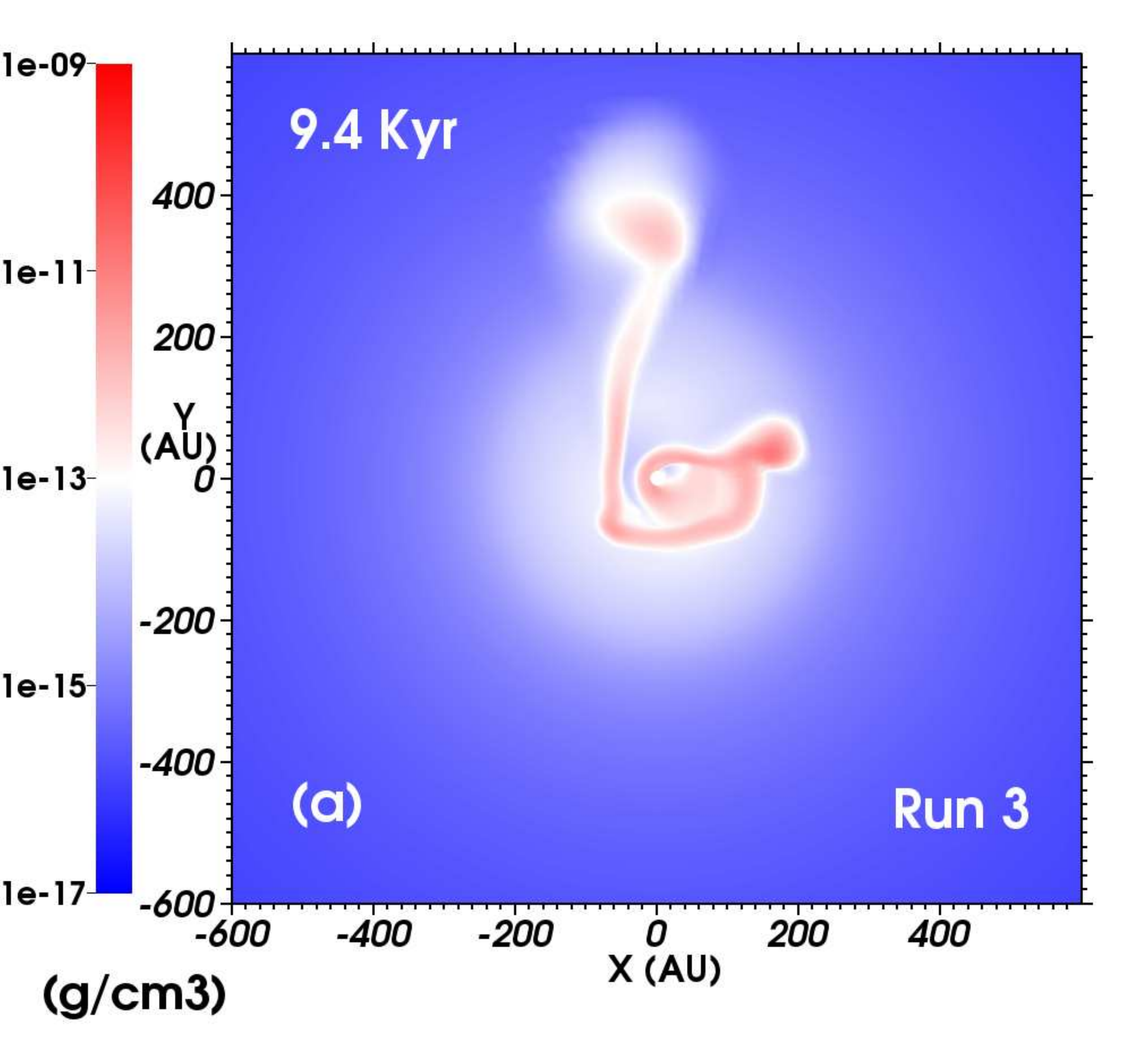}
	\end{minipage}	
	\begin{minipage}[b]{ 0.33\textwidth}
		\includegraphics[width=1.0\textwidth]{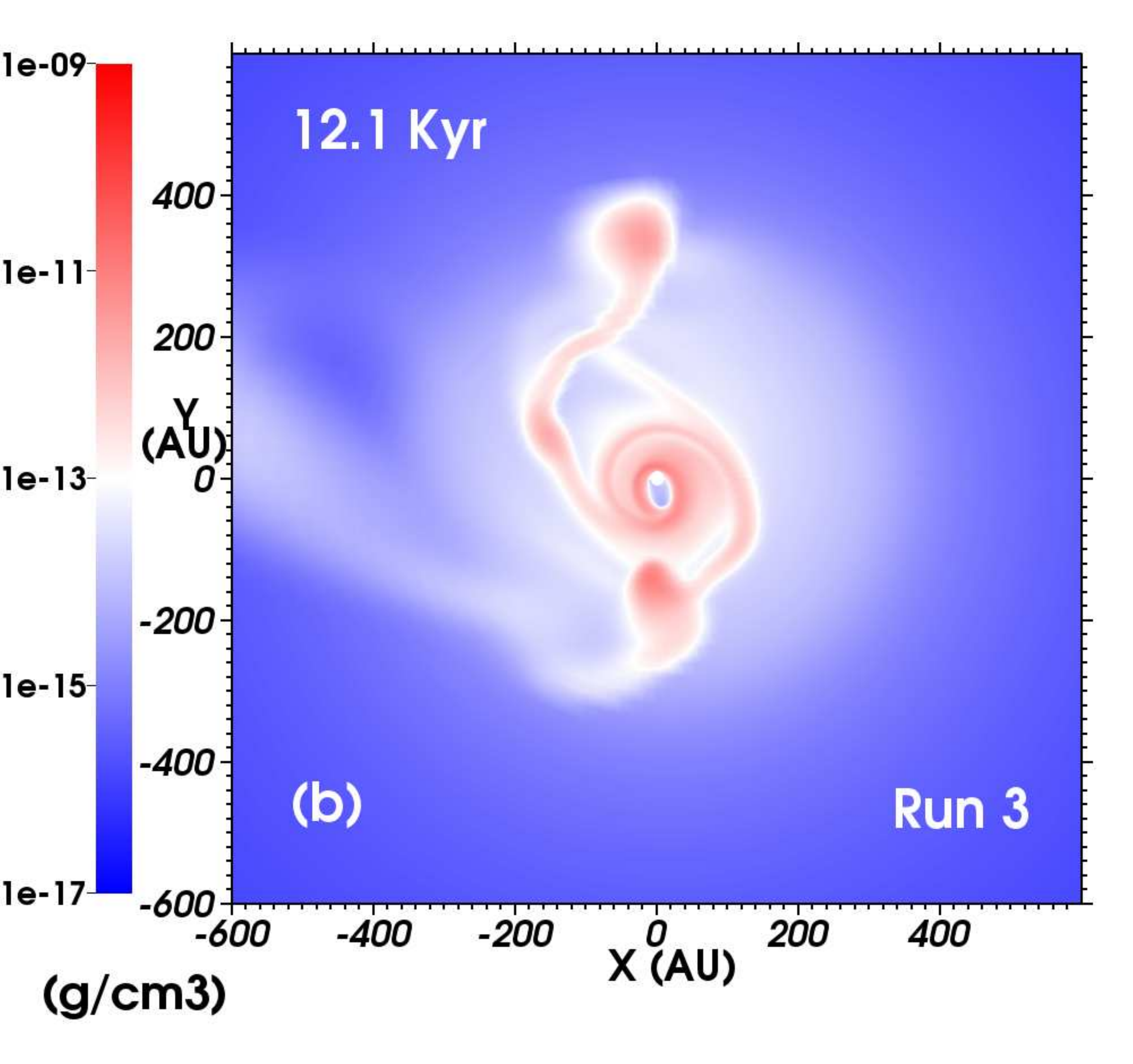}
	\end{minipage} 
	\begin{minipage}[b]{ 0.33\textwidth}
		\includegraphics[width=1.0\textwidth]{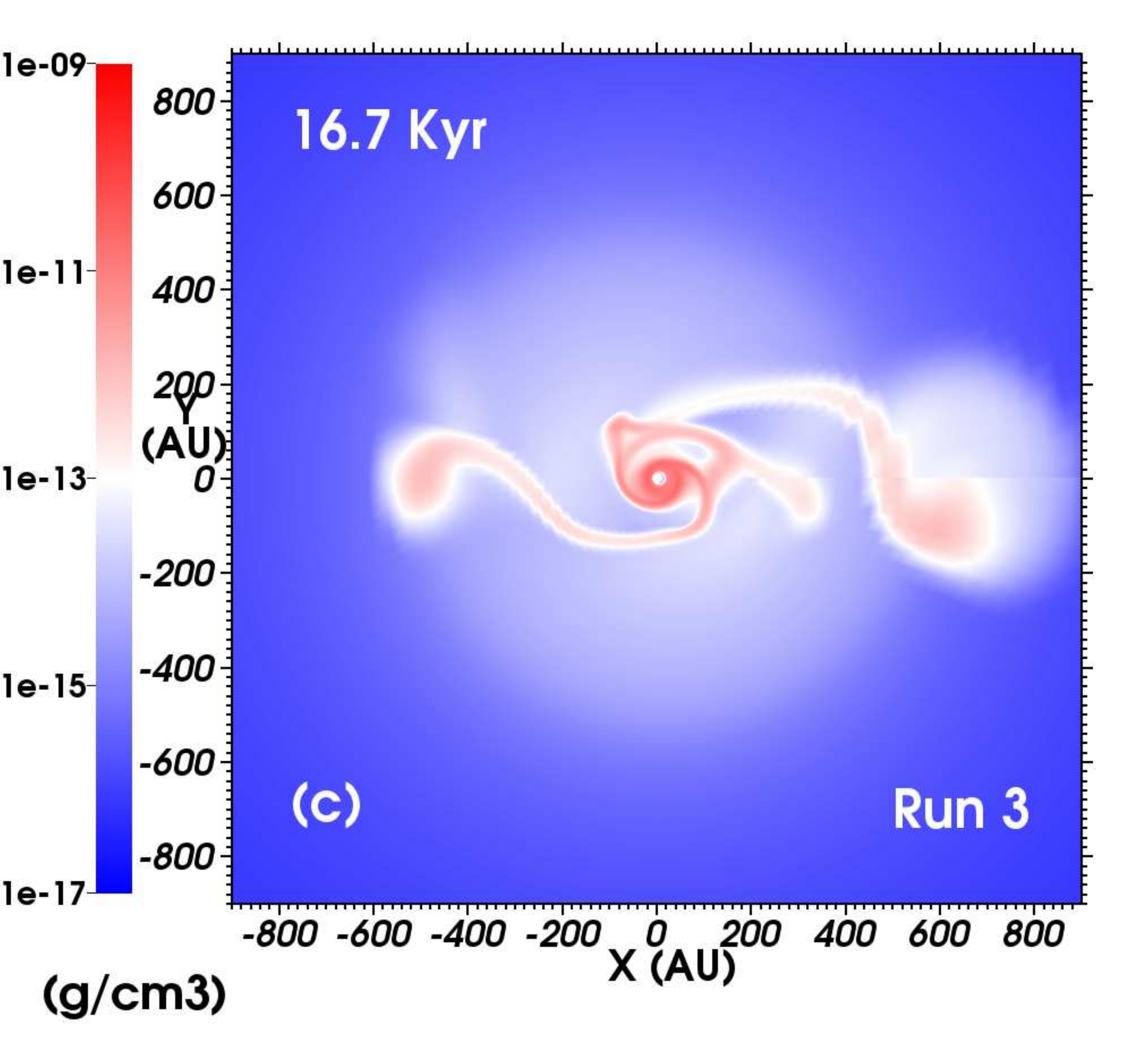}
	\end{minipage}	\\
	\begin{minipage}[b]{ 0.33\textwidth}
		\includegraphics[width=1.0\textwidth]{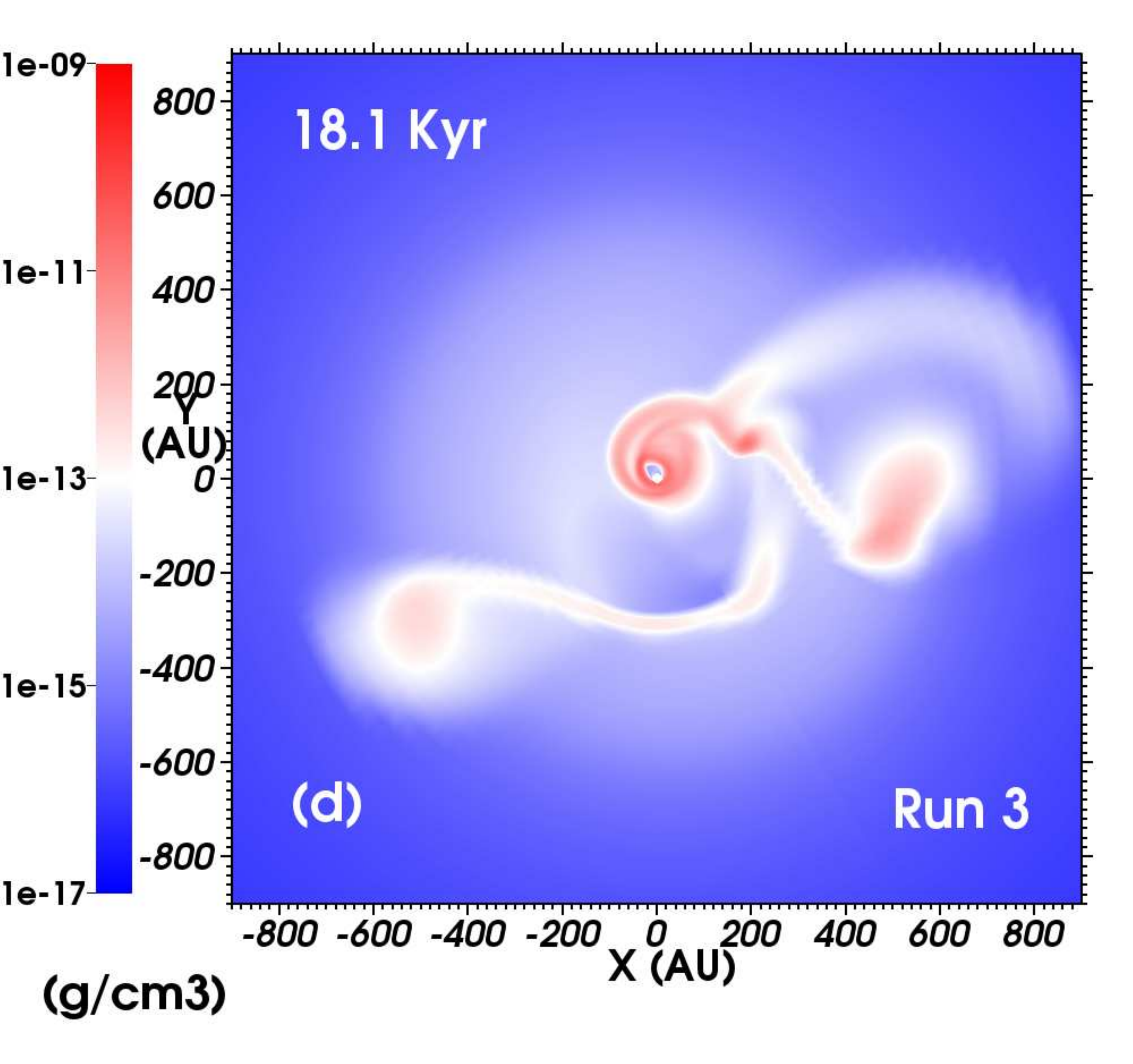}
	\end{minipage}	
	\begin{minipage}[b]{ 0.33\textwidth}
		\includegraphics[width=1.0\textwidth]{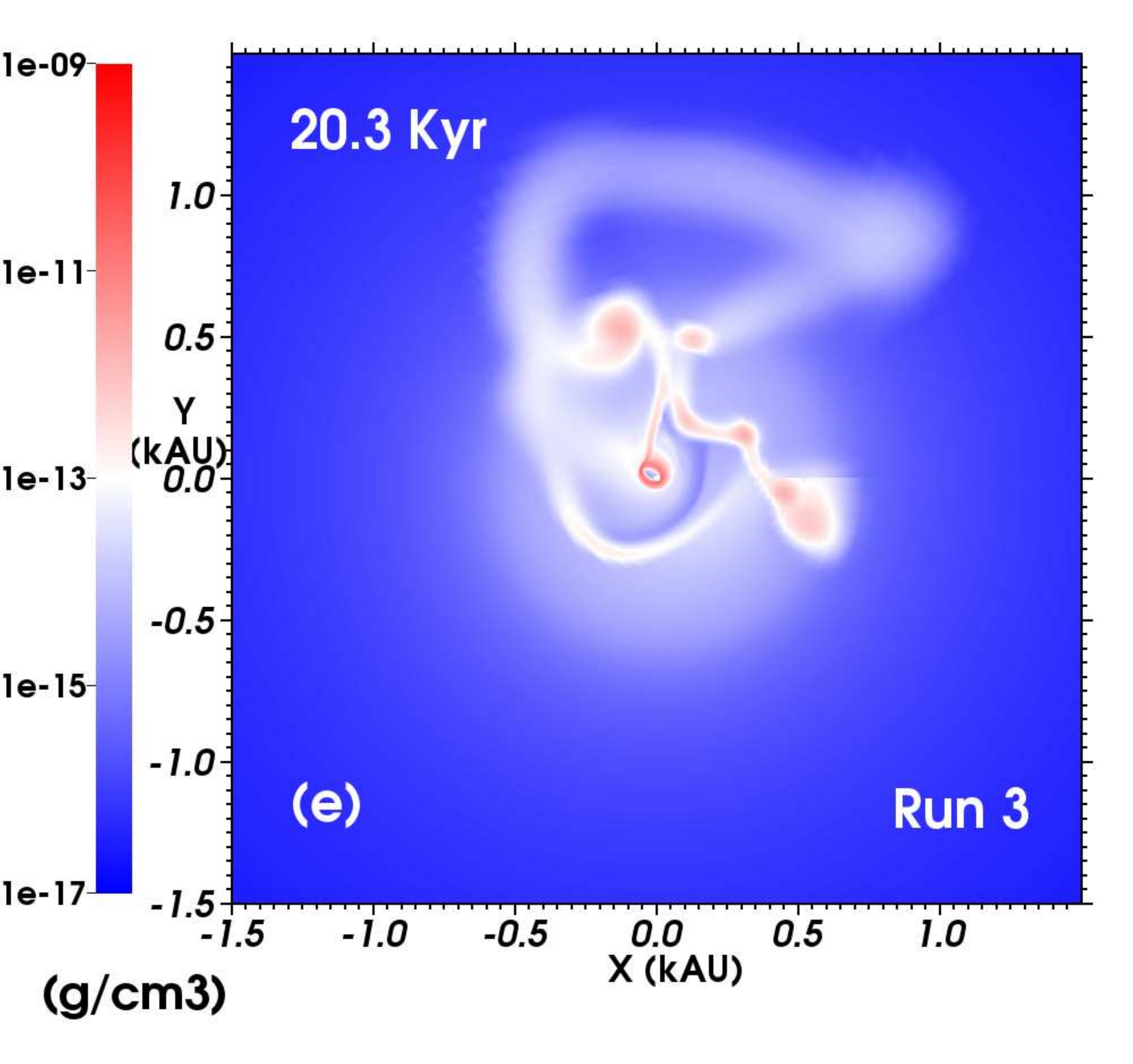}
	\end{minipage}
	\begin{minipage}[b]{ 0.33\textwidth}
		\includegraphics[width=1.0\textwidth]{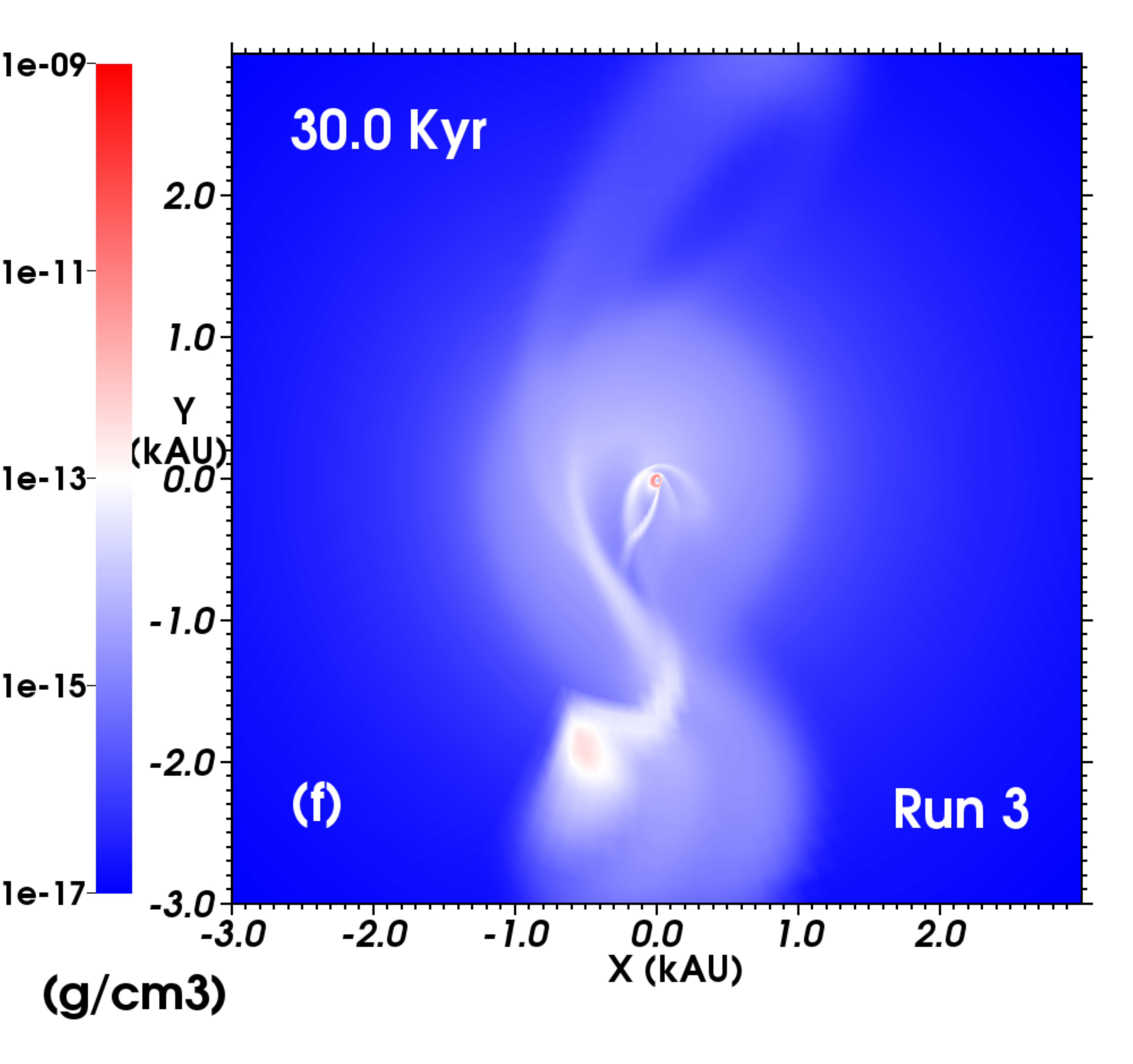}
	\end{minipage}
	\caption{ 
		 \textcolor{black}{Similar to Fig.~\ref{fig:disc_evol1} for Run~3.}  
		 }	
	\label{fig:disc_evol3}  
\end{figure*}

\begin{figure}
	\includegraphics[width=0.48\textwidth]{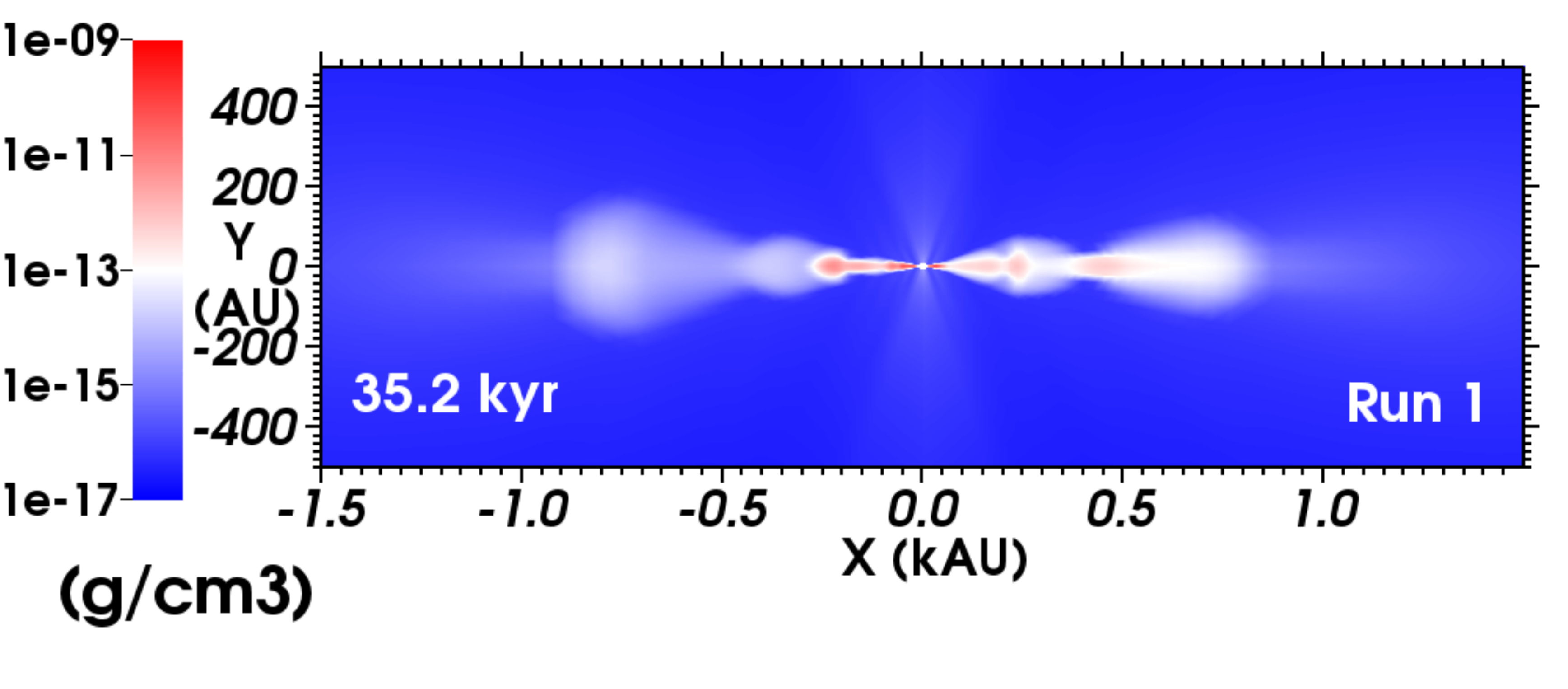}
	\caption{ 
	         \textcolor{black}{
		 Vertical density structure (in $\rm g\, \rm cm^{-3}$) of the accretion disc generated in Run~1 
		 at a time $35.2\, \rm kyr$ \textcolor{black}{and shown with $\phi=0$}. The density is plotted on a logaritmic scale 
		 and the size of the \textcolor{black}{figure} is in (k)AU. 
		 }
		 }	
	\label{fig:cylinder}  
\end{figure}

\section{Initial conditions}
\label{sect:ci}

In this section, we present the \textcolor{black}{initial} internal structure of the pre-stellar core of 
our numerical calculations. We first describe the gas density and velocity 
distribution laws that we consider, before \textcolor{black}{presenting} our models and their 
respective initial characteristics.

\subsection{Gas density distribution}
\label{sect:dens}

We consider a rotating pre-stellar core that has a radial mass density distribution 
$\rho(r)$ represented by the following standard power-law,
\begin{equation}
    \rho(r) = K_{\rho} r^{ \beta_{\rho} },
    \label{eq:density_profile}
\end{equation}
where $K_{\rho}$ is a function of the core size and mass and where $\beta_{\rho}$ 
is a negative exponent characterizing the steepness of the density profile, 
respectively. The total mass of the pre-stellar core that is included between 
the origin of the computational domain and its outermost boundary 
$r_{\rm max}$ is $M_{\rm c} = M(r=r_{\rm max})$ where,
\begin{equation}
   M(r) = M_{\rm c} \Big( \frac{r}{R_{\rm c}} \Big)^{ \beta_{\rho} + 3 },
    \label{eq:mass2}   
\end{equation}
and one can therefore \textcolor{black}{determine} the constant $K_{\rho}$. The density distribution 
of the pre-stellar core is, 
\begin{equation}
    \rho(r) = \frac{ ( \beta_{\rho} + 3 ) }{ 4\pi  } \frac{ M_{\rm c} }{  R_{\rm c}^{ \beta_{\rho} + 3 } } r^{ \beta_{\rho} },
    \label{eq:density_prof_tot}
\end{equation}
with $r$ the distance to the central protostar. 

\subsection{Gas velocity distribution}
\label{sect:momentum}

Similarly, an angular momentum distribution is initially imposed in the 
pre-stellar core via a particular choice in \textcolor{black}{the initial velocity field 
in} the $\phi$ direction. The \textcolor{black}{angular velocity} profile is given by,
\begin{equation}
    \Omega(R) = \Omega_{0} \Big( \frac{ R }{ r_{0} } \Big)^{ \beta_{ \Omega } },
    \label{eq:momentum_distribution1}    
\end{equation}
where $R = r \sin(\theta)$ is the cylindrical radius and where \textcolor{black}{$\beta_{ \Omega }$} 
is an exponent that describes the profile such that \textcolor{black}{$\beta_{ \Omega }=0$} corresponds 
to a \textcolor{black}{core in solid-body rotation}. The parameter $\Omega_{0}$ normalizes the distribution and  
depends on the ratio of kinetic energy with respect to the gravitational energy $\beta = 
E_{\rm rot}/E_{\rm grav}$ that initially characterizes the system. Considering both 
density and momentum profiles in Eqs.~(\ref{eq:density_profile}) 
and~(\ref{eq:momentum_distribution1}), the gravitational 
energy of the \textcolor{black}{pre-stellar} cloud is,
\begin{equation}
    E_{\rm grav} =  \frac{ \beta_{\rho} + 3 }{ 2\beta_{\rho} + 5 }  \frac{G M_{\rm c}^{2}}{R_{\rm c}}, 
   \label{eq:Egrav}    
\end{equation}
whereas its rotational kinetic energy is, 
\begin{equation}
    E_{\rm rot} = \frac{ ( \beta_{\rho} + 3 )  }{ ( \beta_{\rho} + 2\beta_{\Omega} + 5 )  }
		  \frac{ \Omega_{0}^{2} M_{\rm c} R_{\rm c}^{ 2(\beta_{\Omega} + 1 ) }  }{  4 r_{0}^{ 2\beta_{\Omega} }  } 
                  \int_{ 0 }^{ \pi } d\theta \sin( \theta )^{ 3+2\beta_{\Omega} },
   \label{eq:Erot}                  
\end{equation}
%
%
which must be integrated numerically. Finally, one can, for a given molecular cloud 
characterized by a choice of $M_{\rm c}$, $\beta_{\rho}$ and $\beta_{\Omega}$ and 
fixing the desired ratio $\beta$, find the corresponding $\Omega_{0}$.

\subsection{Models presentation}
\label{sect:ic}

The momentum distribution is initially 
implemented into the code via the $\phi$ component of the velocity 
$v_{\phi}(R)=R\Omega(R)$ while the other components of the velocity field are 
$v_{\rm r}=v_{\theta}=0$. We assume that the percentage of kinetic energy with 
respect to the gravitational energy is $\beta=4\%$ \textcolor{black}{for all three models}. 
%
%
Additionally, we assume that the dust is \textcolor{black}{coupled} to the gas with a gas-to-dust 
mass ratio of 100. The thermal pressure is set to $p=R \rho T_{\rm c}/\mu$ with 
$R$ the ideal gas constant, $\mu$ the mean molecular \textcolor{black}{weight} and $T_{\rm c}=10\, \rm K$ 
the temperature of the pre-stellar core. The dust temperature $T_{\rm d}$ is 
considered as equal to the gas temperature $T_{\rm g}=T_{\rm d}=T_{\rm c}$. 
We run three simulations (our Table~\ref{tab:sigma}) initially setting the 
density distribution $\beta_{\rho}=-3/2$ that is a typical value for a 
collapsing \textcolor{black}{pre-stellar core that generates} a present-day massive 
protostar~\textcolor{black}{\citep{2000ApJ...537..283V,2002ApJS..143..469M}}. \textcolor{black}{The} difference between simulations 
concern the initial \textcolor{black}{angular velocity} distribution of the pre-stellar 
cloud, \textcolor{black}{which} is $\beta_{\Omega}=0$~\citep[Run 1, cf.][]{klassen_apj_823_2016}, 
$\beta_{\Omega}=-0.35$ (Run 2) and $\beta_{\Omega}=-3/4$ \textcolor{black}{(Run 3)}. 
\textcolor{black}{Run~3 is the same simulation \textcolor{black}{as} in~\citet{meyer_mnras_464_2017}, but with a longer run-time.} 
Our simulations explore the parameter space in terms of initial angular 
frequency distribution $\Omega(R)$ at fixed $M_{\rm c}$, $R_{\rm c}$, $T_{\rm c}$, $\beta$ and $\beta_{\rho}$. 
\textcolor{black}{
Additionally, we perform three other models with the same initial conditions as Run 1, but 
without protostellar irradiation (Run 1-noIrr), with a lower spatial resolution using a grid 
of $64\times11\times64$ cells (Run 1-LR) and with a higher spatial resolution using a grid 
of $256\times41\times256$ cells (Run 1-HR) (see Section~\ref{section:numerics}). 
}

\begin{figure}
	\centering
	\begin{minipage}[b]{ 0.48\textwidth}
		\includegraphics[width=1.0\textwidth]{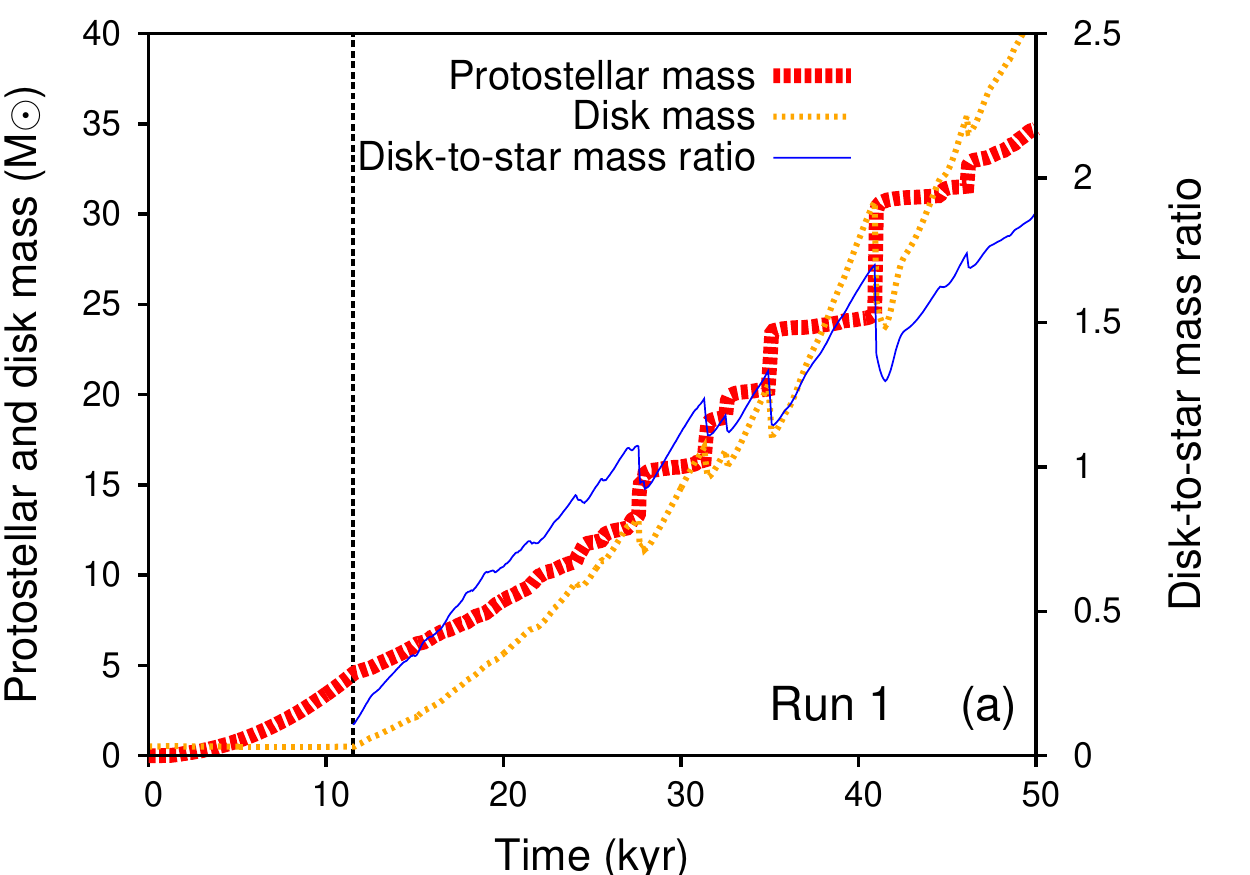}
	\end{minipage} 	\\
	\begin{minipage}[b]{ 0.48\textwidth}
		\includegraphics[width=1.0\textwidth]{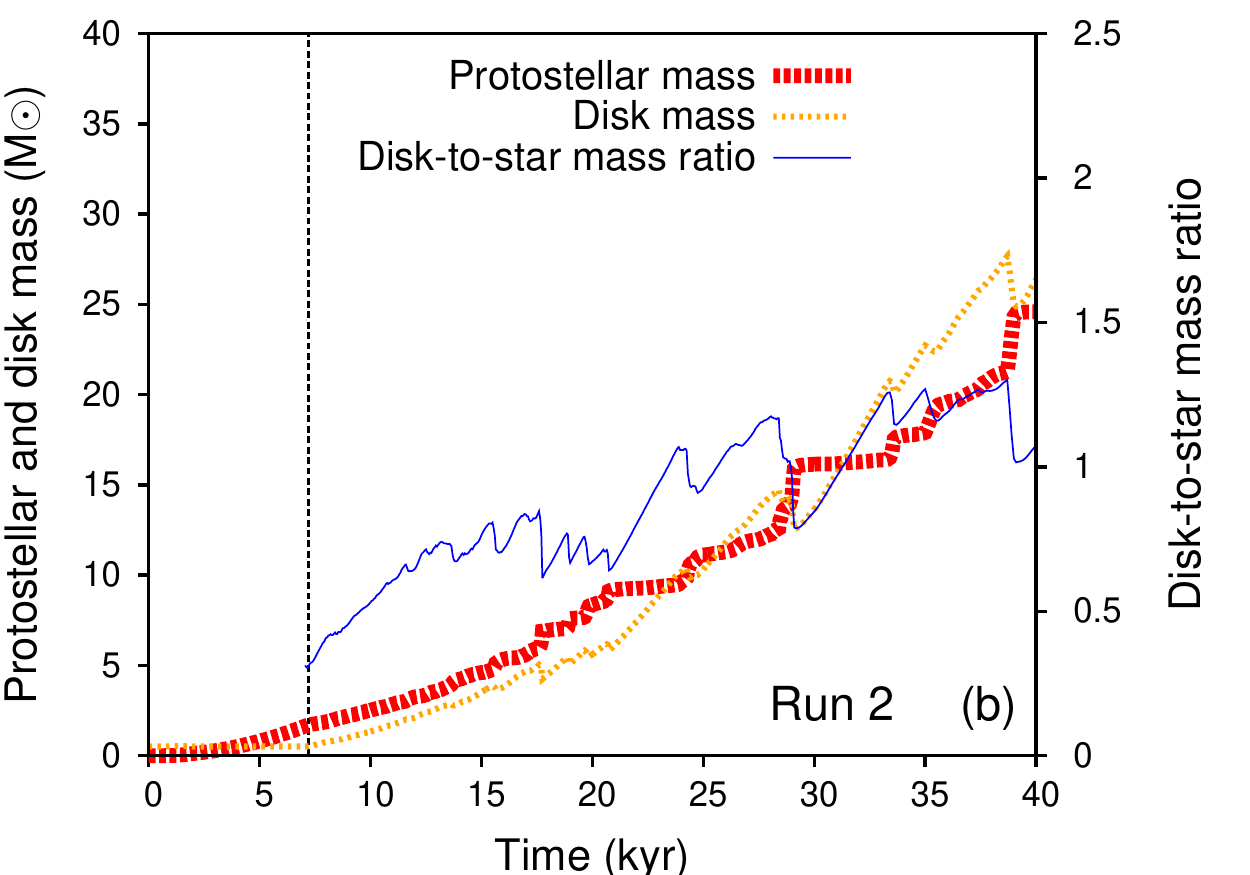}
	\end{minipage}		\\
	\begin{minipage}[b]{ 0.48\textwidth}
		\includegraphics[width=1.0\textwidth]{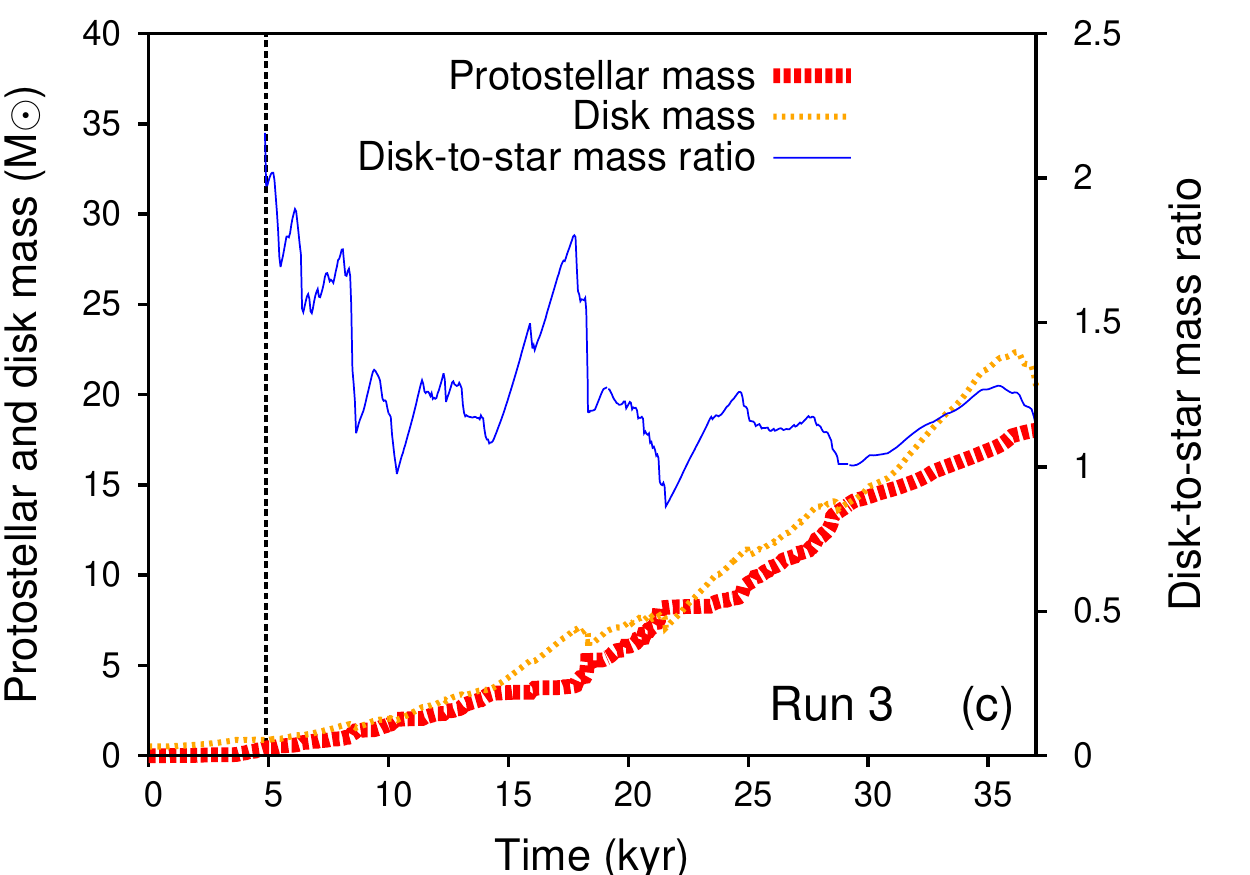}
	\end{minipage}	
	\caption{ 
		 \textcolor{black}{
		 Protostellar mass, disc \textcolor{black}{mass} and disc-to-star mass ratio 
		 evolving with time. Vertical lines mark the end of the free-fall 
		 collapse and the onset of the disc formation.
		 } 
		 }	
	\label{fig:disc_properties1}  
\end{figure}


\section{Accretion disc fragmentation}
\label{section:results_disc}

\textcolor{black}{In this section, we analyze} the properties of \textcolor{black}{the} 
collapsing pre-stellar cores. We follow the 
evolution of the density fields in the simulations and discuss their differences 
before \textcolor{black}{describing} the fragmentation of the accretion discs. For each model, we
look at the accretion rate onto the protostar and its \textcolor{black}{feedback} 
to the disc as accretion-driven outbursts.

\subsection{ Accretion disc fragmentation and protostar evolution }
\label{sect:disc}

In Figs.~\ref{fig:disc_evol1},~\ref{fig:disc_evol2} and ~\ref{fig:disc_evol3}, 
we plot the midplane density field (in $\rm g\,\rm cm^{-3}$) from our models 
Run~1,~2 and 3 at six different representative evolutionary times (in $\rm kyr$) 
elapsed since the beginning of the simulations. The figures show 
the inner region of $\approx$ few (k)AU around the protostar, while the 
computational domain is much more extended ($R_{\rm c} = 0.1 \rm pc$). 
Every \textcolor{black}{figure corresponds} to times \textcolor{black}{after the formation of the disc} at times $11.5$, $7.2$ 
and $3.7\, \rm kyr$ for our models Run 1, Run 2 and Run 3, respectively, \textcolor{black}{once 
pre-stellar} core material orbits the protostar with approximately 
Keplerian velocity. Despite of the fact that the overall evolution of the discs 
is globally similar, sensitive morphological differences appear. 
\textcolor{black}{The} accretion discs \textcolor{black}{develop} spiral arms under the effects of gravitational instability 
and the discs adopt the typical filamentary structure of self-gravitating systems 
under the effects of rotation~\citep{lodato_mc_30_2007}.

\textcolor{black}{
The discs are rather compact if $\beta_{\Omega}=0$ (Run 1) and more extended if 
the initial \textcolor{black}{angular velocity} distribution is steeper, because it favours the fast 
growth of large spiral arms, e.g. if $\beta_{\Omega}=-0.75$ (Run 3)}. All discs 
strongly fragment by developing inhomogeneous regions in their spiral arms, 
however, discrepancies arise with the distribution of those circumstellar 
clumps, as seen in Figs.~\ref{fig:disc_evol1}-\ref{fig:disc_evol3}. 
While orbiting the protostar in their host spiral arm, the clumps can migrate \textcolor{black}{to} larger radii or rapidly 
move inwards under the effects of gravitational torques, before \textcolor{black}{disappearing   
into} the sink cell. Note that the overdensity at $\approx 200\, \rm AU$ 
from the sink cell in the spiral arm of Fig.~\ref{fig:disc_evol3}d is subject to 
a rapid migration onto protostar that is described in detail 
in~\citet{meyer_mnras_464_2017}. This scenario of falling clumps repeats 
itself throughout the integration time of all models. Particularly, this 
mechanism is responsible for the variable accretion onto the protostar and 
it governs its mass evolution (see section~\ref{sect:bursts}).

Fig.~\ref{fig:disc_properties1} plots time evolution of the mass of the protostars 
(in $\rm M_{\odot}$, thick dotted red line). We measure the mass of the discs (in 
$\rm M_{\odot}$, thin dotted yellow line) and the disc-to-star mass ratio (thin 
\textcolor{black}{solid blue} line) in our simulations Run 1 (a), Run 2 (b), and Run 3 (c) by considering 
the material in a cylinder of radius $1500\, \rm AU$ and \textcolor{black}{height $500\, \rm AU$ 
centered \textcolor{black}{at} the origin and perpendicular to the disc, as suggested 
by~\citet{klassen_apj_823_2016}}. 
\textcolor{black}{Such a cylinder contains the discs throughout their evolution (see Fig.~\ref{fig:cylinder}).} 
The vertical dotted black line marks the time of the onset of the disc formation. The mass of 
our protostars monotonically increases as a function of time with a slope  
corresponding to accretion rate of circumstellar material. The periods of smooth 
mass increase, e.g. at times between $11.5\, \rm kyr$ and $26.5\, \rm kyr$ 
(Fig.~\ref{fig:disc_properties1}a) is produced by the baseline (but variable)  
accretion rate generated when non-fragmenting portions of spiral arms wrap and vanish 
into the sink cell (Fig.~\ref{fig:disc_evol1}a). This \textcolor{black}{dominates} the mass 
accretion at the \textcolor{black}{time of disc formation} and during the early disc fragmentation. 
At later times, the \textcolor{black}{stellar mass evolution} experiences sudden increases of up to a few 
solar masses, e.g. at times $27.6$, $35.1$ or $41.0\, \rm kyr$, which marks the accretion of clumps.

As our model with solid-body rotation \textcolor{black}{forms a disc that is smaller in size}, it produces more 
fragments, which in turn generate a larger number of step-like jumps in 
the protostellar mass history (Fig.~\ref{fig:disc_properties1}a) than the other runs 
(Fig.~\ref{fig:disc_properties1}b,c). The computational cost of our simulations 
increases as a function of the steepness in the initial velocity distribution, 
which is the only parameter \textcolor{black}{differentiating} Runs 1$-$3. Consequently, we can \textcolor{black}{have long}  
integration times for the model with solid-body \textcolor{black}{rotation} and form a $35\, \rm 
M_{\odot}$ protostar while the other stars reach only $25$ and $18\, \rm M_{\odot}$, 
respectively. The disc mass evolves accordingly, with sudden decreases when a clump 
leaves the disc to fall onto the protostar (see Section~\ref{sect:accretion}). 
The \textcolor{black}{disc-to-star mass} evolution 
reports the simultaneous increases of \textcolor{black}{the protostellar and disc mass}. 
\textcolor{black}{Importantly, the disc fragmentation does not} prevent accretion onto the protostar. 

\subsection{ Variable accretion onto the protostar }
\label{sect:accretion}

\begin{figure}
	\centering
	\begin{minipage}[b]{ 0.48\textwidth}
		\includegraphics[width=1.0\textwidth]{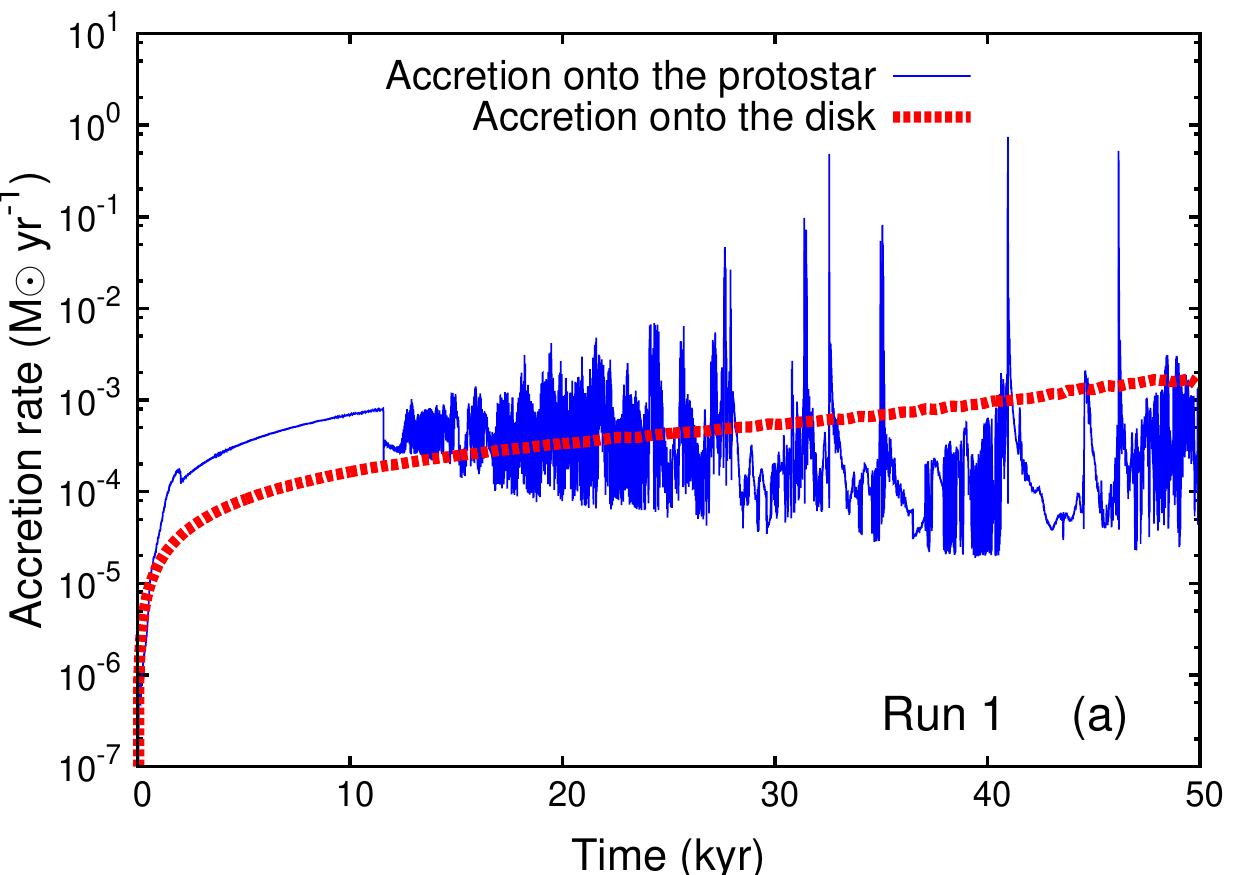}
	\end{minipage} 
	\begin{minipage}[b]{ 0.48\textwidth}
		\includegraphics[width=1.0\textwidth]{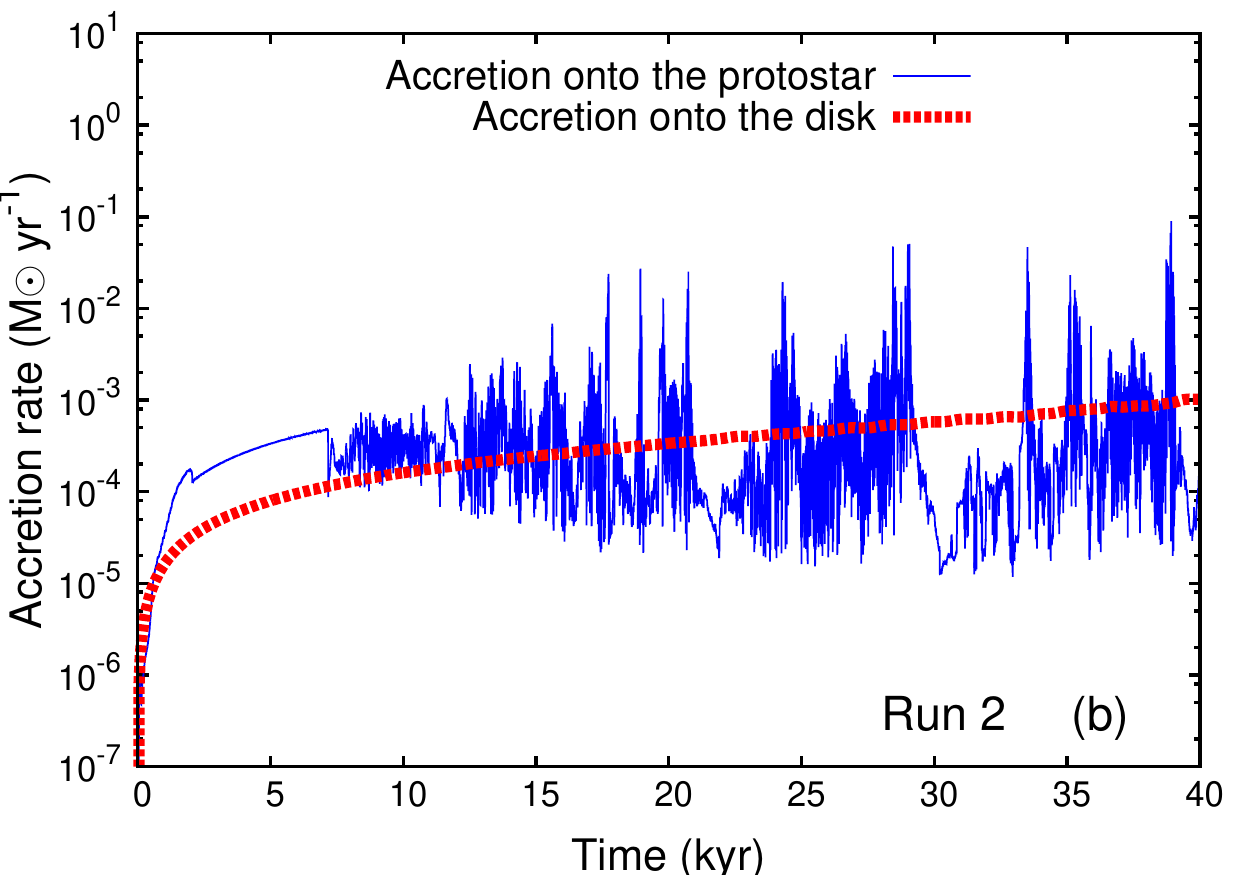}
	\end{minipage}		
	\begin{minipage}[b]{ 0.48\textwidth}
		\includegraphics[width=1.0\textwidth]{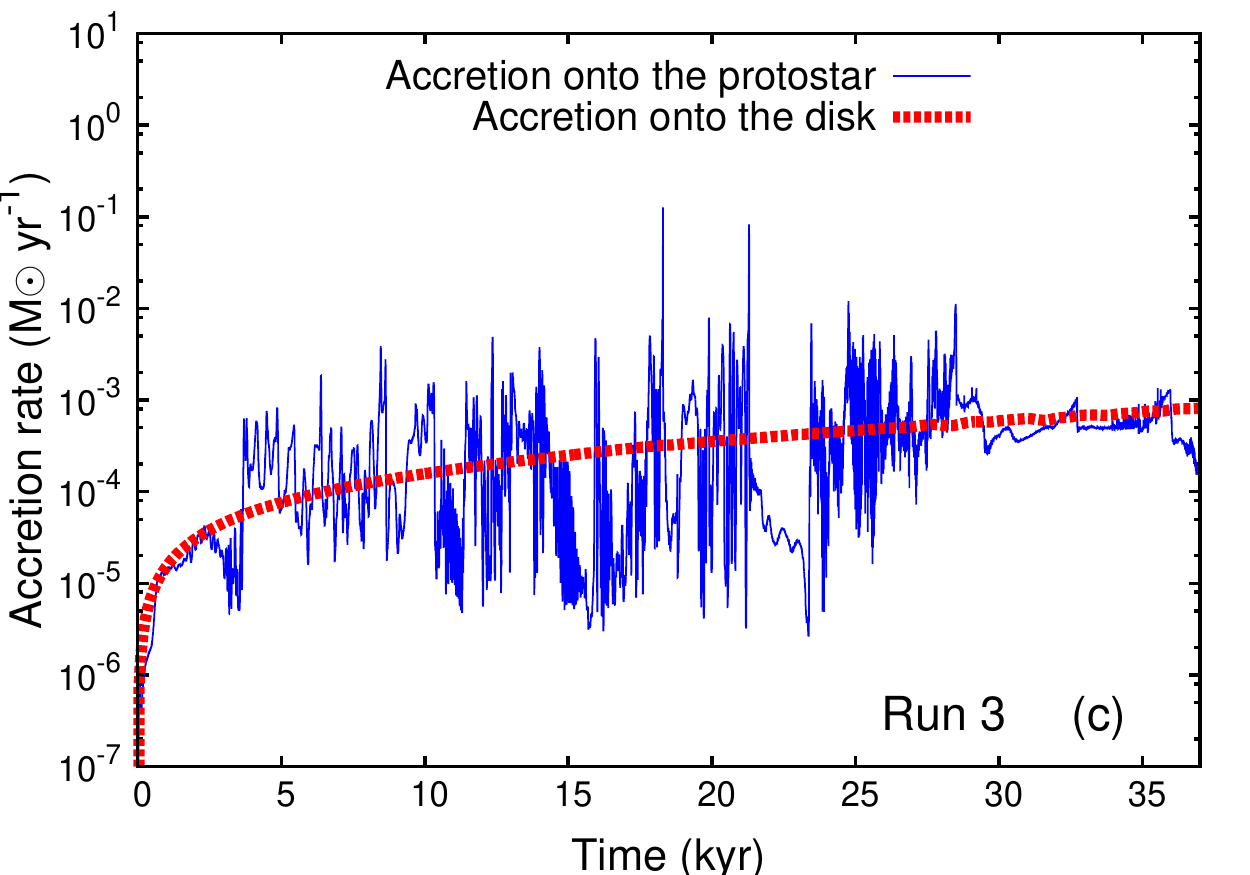}
	\end{minipage} 
	\caption{ 
		 Accretion rates onto the protostar-disc systems.  
		 }	
	\label{fig:disc_properties4}  
\end{figure}

Fig.~\ref{fig:disc_properties4} displays the accretion properties of the 
systems. The figure shows the accretion rate \textcolor{black}{(in $\rm M_{\odot}\, \rm yr^{-1}$,  
from the envelope onto the disc measured at 
$r=3000\, \rm AU$~\citep[cf.][]{meyer_mnras_464_2017}, thick  dotted red line)} and the 
accretion rate onto the protostar (in $\rm M_{\odot}\, \rm yr^{-1}$, thin solid 
blue line) in our Run 1 (a), Run 2 (b), and Run 3 (c). As a result of disc formation and its subsequent 
fragmentation, mass accretion onto the protostar is highly variable. The variety 
of lengths and size\textcolor{black}{s} of the \textcolor{black}{filaments and gaseous clumps} accreted by the protostar generates 
the variability of $\dot{M}$. This has already been found in several previous numerical 
studies, e.g. devoted to \textcolor{black}{low-mass star 
formation~\citep{vorobyov_mnras_381_2007,machida_apj_729_2011,vorobyov_apj_805_2015}, \textcolor{black}{and}
in the context of primordial star formation~\citep{voroboyov_aa_557_2013}}. When a dense circumstellar 
clump falls onto the protostar, violent accretion spikes happen thanks to the 
mechanism depicted in~\citet{meyer_mnras_464_2017}. All models have such 
remarkable accretion peaks, with increases of the accretion rate up to 
\textcolor{black}{about a few $10^{-1}\, \rm M_{\odot}\, \rm yr^{-1}$ over a time interval of 
$\approx\, 20\, \rm yr$}. 
The time interval separating the accretion spikes corresponds to a temporary 
damping of the oscillations of the accretion rate induced by the leftover of the 
clumps which are \textcolor{black}{gravitationally} swung away~\citep{meyer_mnras_464_2017}. 
The frequency of \textcolor{black}{the occurrence of accretion bursts} is about 
$M_{\rm cl}/\dot{M}_{\rm d}\approx 2$$-$$0.5\, \rm kyr$
with $M_{\rm cl}\approx0.5$$-$$\rm few\, \rm M_{\odot}$ the typical mass of 
a circumstellar clumps \textcolor{black}{and $\dot{M}_{\rm d}$ being the mass infall rate onto the disc}.

\subsection{ Protostellar luminosity and accretion-driven outbursts }
\label{sect:bursts}

\begin{figure}
	\centering
	\begin{minipage}[b]{ 0.45\textwidth}
		\includegraphics[width=1.0\textwidth]{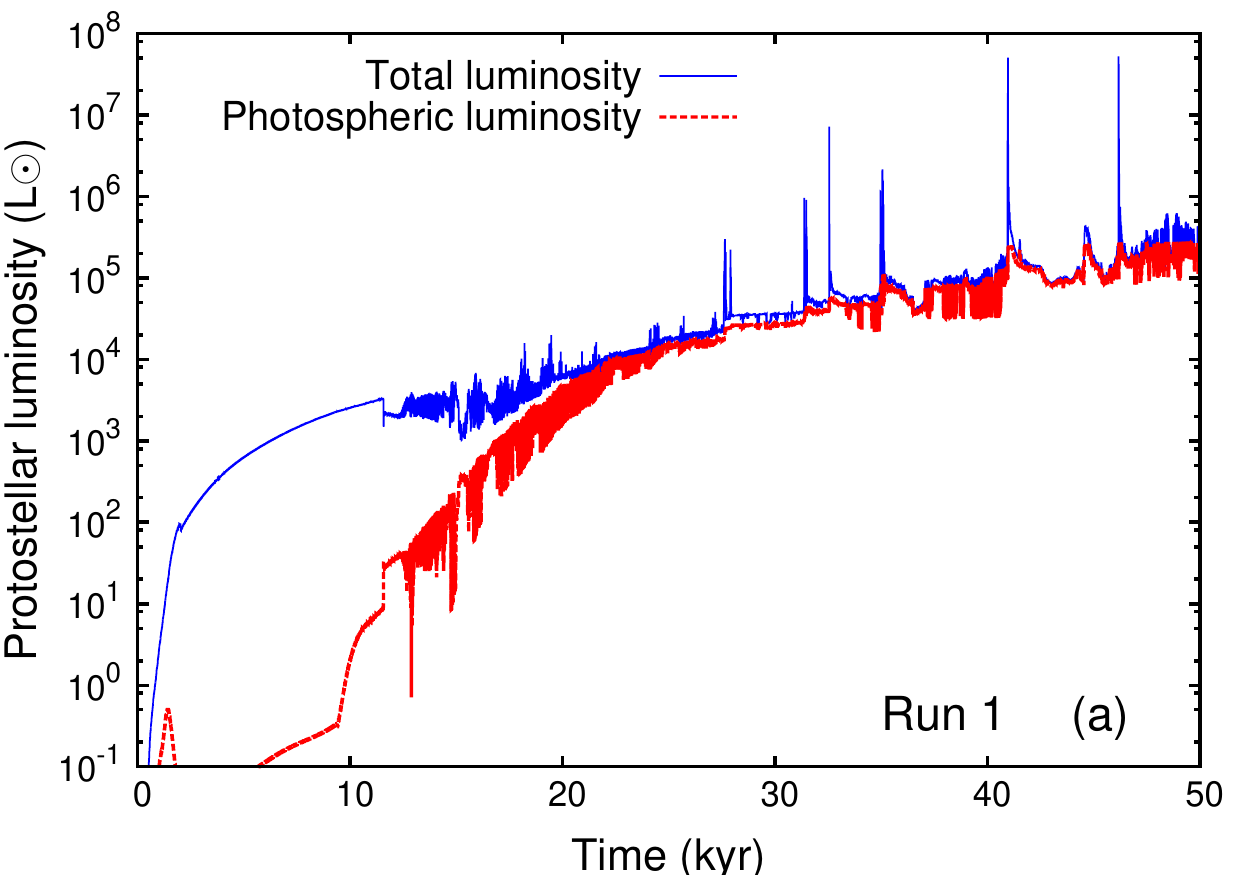}
	\end{minipage} 
	\begin{minipage}[b]{ 0.45\textwidth}
		\includegraphics[width=1.0\textwidth]{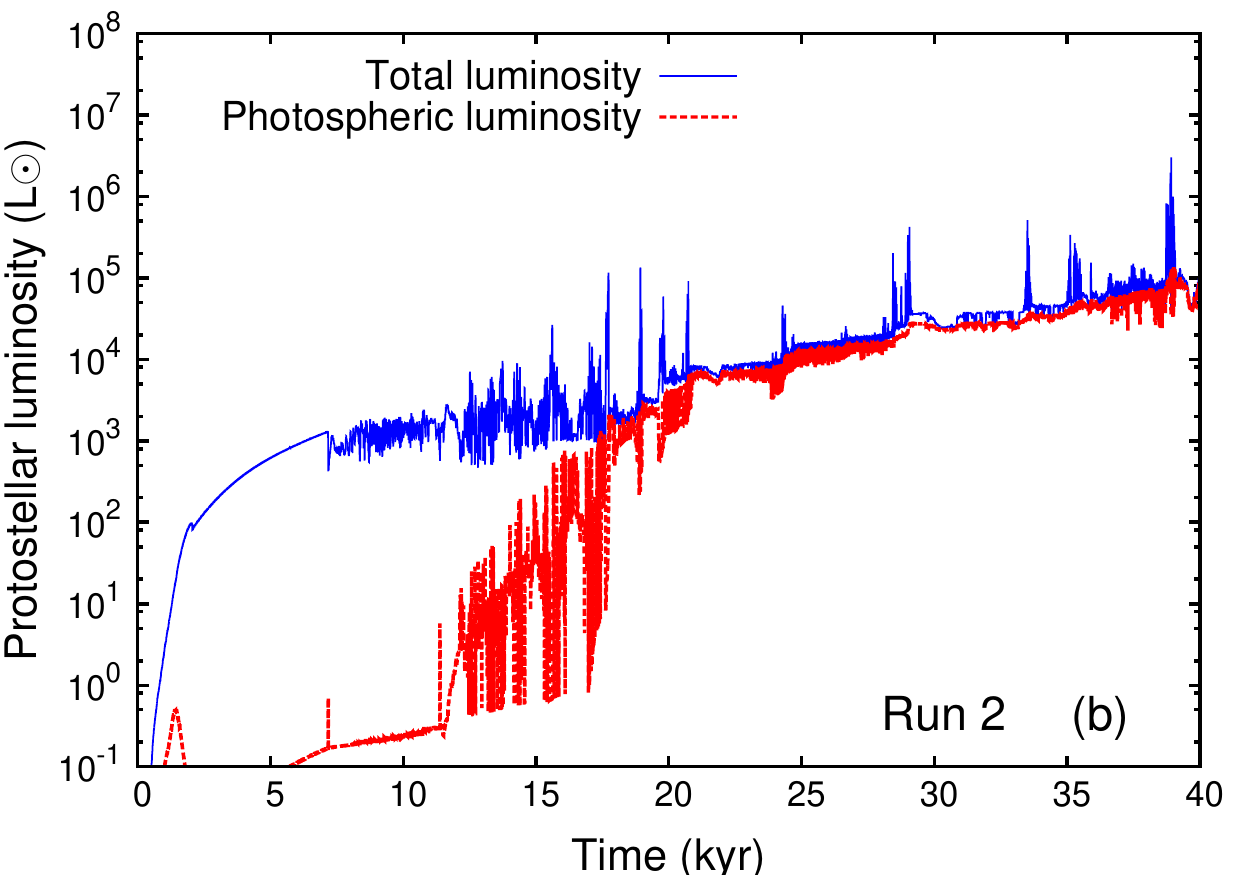}
	\end{minipage}		
	\begin{minipage}[b]{ 0.45\textwidth}
		\includegraphics[width=1.0\textwidth]{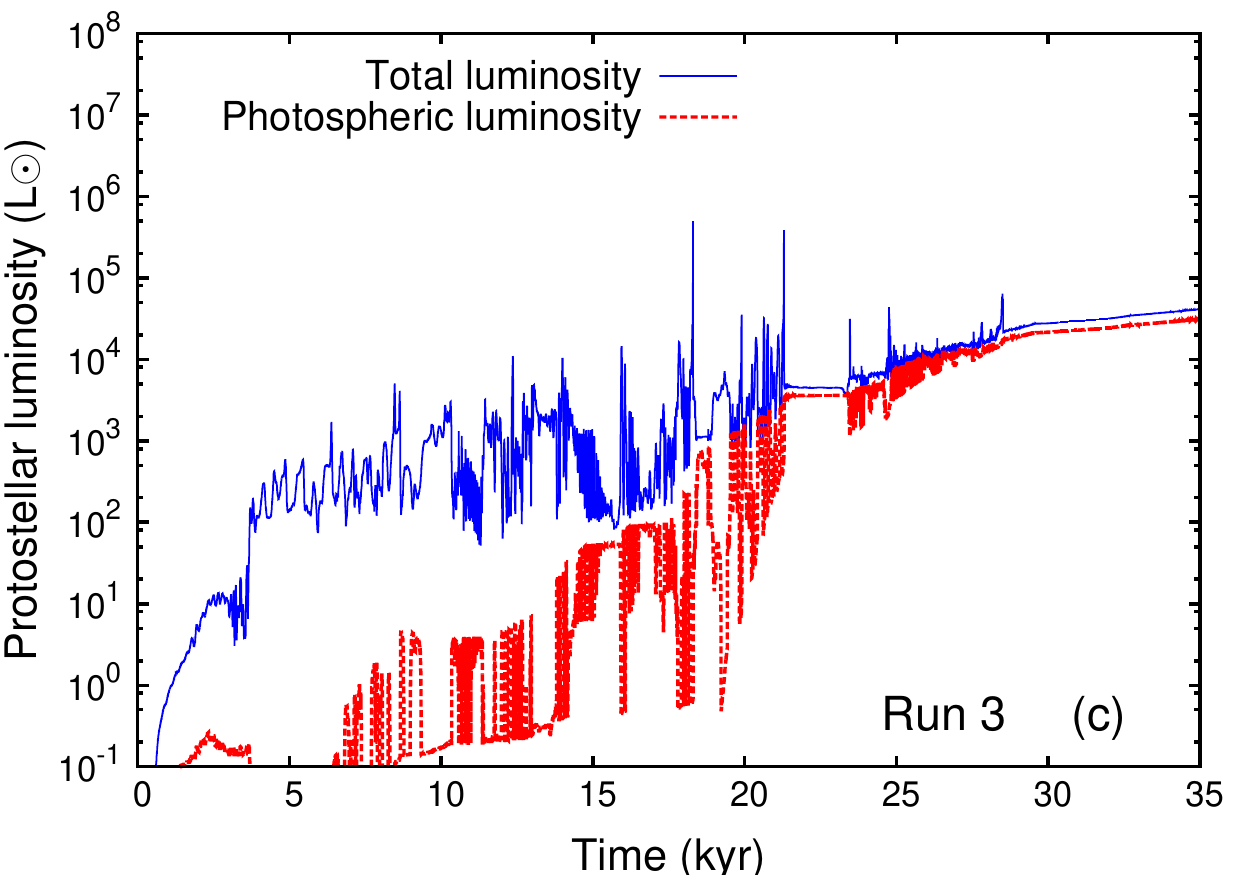}
	\end{minipage} 
	\caption{ 
	         \textcolor{black}{
		 Bolometric protostellar luminosities (blue solid lines) and photospheric emission 
		 of our protostars (dotted red lines). 
		 The lightcurves report the changes in the 
		 accretion rates of Fig.~\ref{fig:disc_properties4}.  
		 }
		 }	
	\label{fig:disc_properties5}  
\end{figure}

Fig.~\ref{fig:disc_properties5} \textcolor{black}{shows} the evolution of the photospheric 
luminosity $L_{\star}$ (dotted red line) and the total luminosity $L_{\rm 
tot}=L_{\star}+L_{\rm acc}$ (solid blue line) where \textcolor{black}{$L_{\rm acc}= 
GM_{\star}\dot{M}/2R_{\star}$} is the accretion luminosity of the protostar (in 
$L_{\odot}$) throughout our simulations Run 1 (a), Run 2 (b), and Run 3 (c).  To 
each strong increase of the accretion rate corresponds a rise of the bolometric 
luminosity of the protostar, that is clearly above the mean stellar luminosity. 
The intensity of those accretion-driven outbursts is also 
influenced by the protostellar radius, i.e. a bloated protostar decreases 
the intensity of \textcolor{black}{accretion-driven outburst 
since $L_{\rm acc} \propto 1/R_{\star}$~\citep{hosokawa_apj_691_2009}}. 
This explains why the relative intensity of the flares (governed by the accretion 
luminosity at times $\ge\, \rm 25$$-$$30\, \rm kyr$) does not correspond to the 
relative strength of the associated accretion peaks, see, e.g. the forest of 
peaks in Figs.~\ref{fig:disc_properties4}a and~\ref{fig:disc_properties5}a at 
times $\ge 30\, \rm kyr$, respectively. \textcolor{black}{The bloating of the young star has 
not finished, i.e. much} longer integration times of the 
system, ideally up to the zero-age main-sequence time, is necessary 
\textcolor{black}{in order to produce detailed statistics of accretion-driven} 
outbursts as a function of 
the initial conditions of our models. 

Interestingly, Run~3 does not have further accretion peaks after times $\ge 28\, 
\rm kyr$, at least for the integration times that we consider. This model 
generates our more extended disc \textcolor{black}{of radius $\approx 4\, \rm kAU$}, which develops structures resembling a second 
accretion disc bridged by a gaseous filament. \textcolor{black}{By second disc we mean 
that the \textcolor{black}{circumstellar clump concerned}, while rotating, migrates to radii larger than $\approx 2\, \rm kAU$ 
and begins accreting from the primary disc but also from the still collapsing pre-stellar core material, 
such that the clump forms \textcolor{black}{its} own accretion disc (not properly resolved by our logarithmic grid at this distance from the primary protostar), see 
also simulations of~\citet{offner_apj_725_2010,kratter_apj_708_2010}}. The midplane 
density field suggests that it may be due to the formation of a 
massive binary star of separation $\approx 2\, \rm kAU$ 
(Fig.~\ref{fig:disc_evol3}f), however, our logarithmically-expanding grid in the 
radial direction does not resolve the Jeans lengths at such \textcolor{black}{large} radii from 
the central protostar and, consequently, it does not allow 
us to conclude on this with more certitude. Our series of 
models show that the phenomenon of accretion-driven outbursts is a 
general feature of the parameter space in terms of initial \textcolor{black}{angular velocity}  
distribution. It stresses the very close \textcolor{black}{similarities} existing between the 
variability of forming massive stars and the other regimes of star formation, 
see the \textcolor{black}{extended literature about the formation of low-mass 
stars~\citep{vorobyov_apj_719_2010,vorobyov_apj_805_2015} and primordial 
stars~\citep{stacy_mnras_403_2010,greif_apj_737_2011,greif_mnras_424_2012,hosokawa_2015,sakurai_mnras_549_2016}}. 
Further work will provide us with more statistics on luminous accretion-driven 
outbursts to be compared with the FU-Orionis-like outburst 
of S255IR~\citep{burns_mnras_460_2016,caratti_nature_2016}.

\begin{figure*}
	\centering
	\begin{minipage}[b]{ 0.48\textwidth}
		\includegraphics[width=1.0\textwidth]{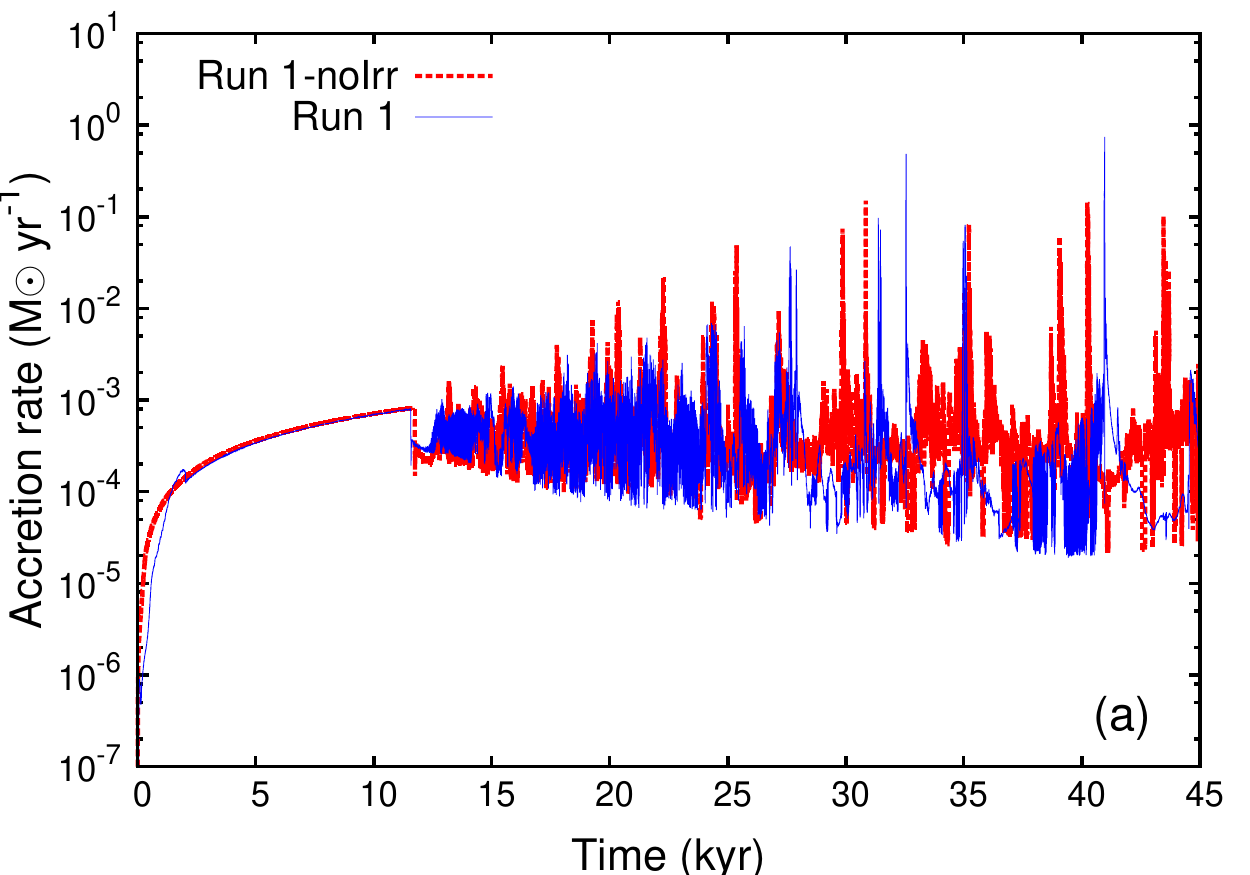}
	\end{minipage} 
	\begin{minipage}[b]{ 0.48\textwidth}
		\includegraphics[width=1.0\textwidth]{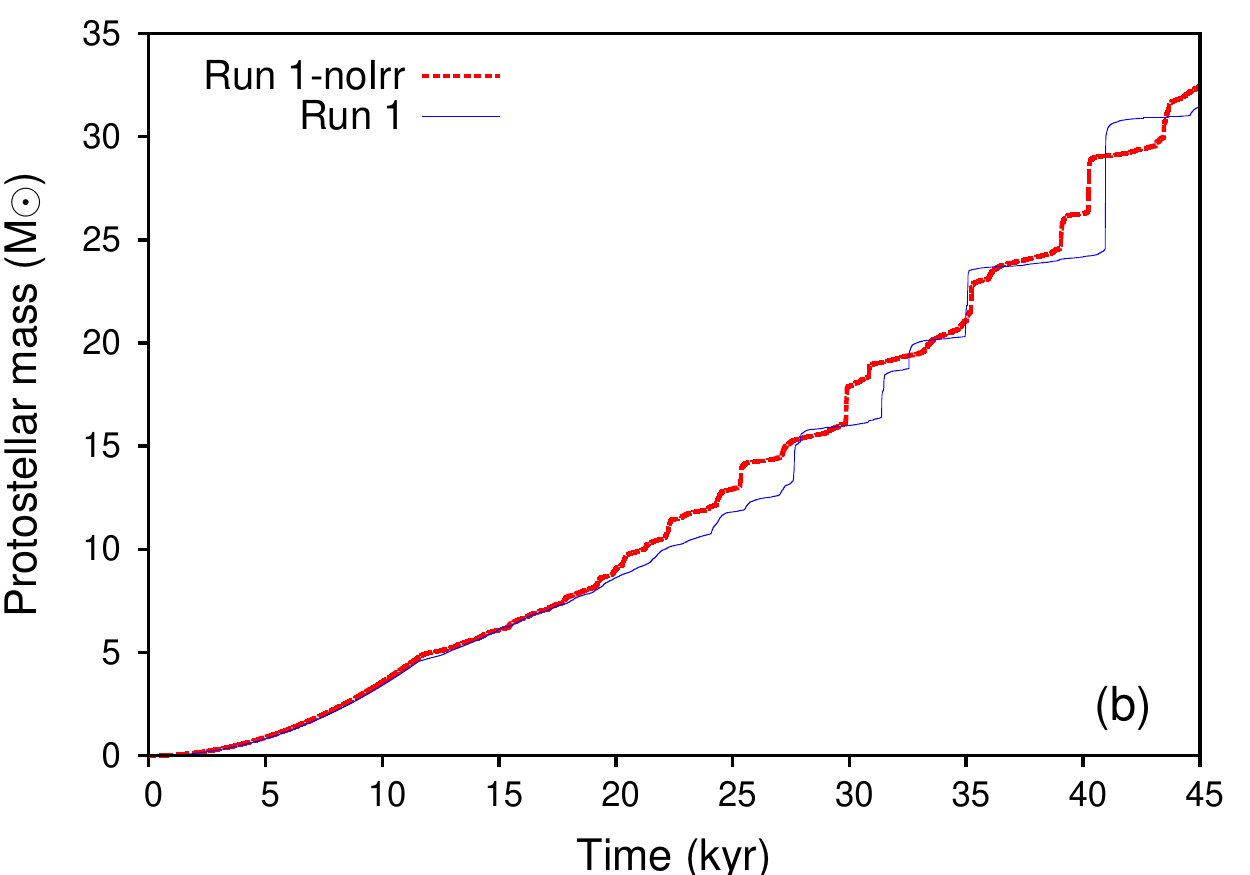}
	\end{minipage}		
	\caption{ 
		 \textcolor{black}{
		 Comparison between the accretion rate (a) and the mass evolution (b) of 
		 a model considered with (Run 1) and without irradiation (Run 1-noIrr).   
		 }  
		 }	
	\label{fig:irradation_qtt}  
\end{figure*}

\section{ Physical and numerical effects }
\label{section:numerics}

\textcolor{black}{
The role of the incident protostellar radiation and effects of the spatial resolution on the 
fragmentation of an accretion disc formed around a young high-mass star are examinated in 
this section. 
}

\subsection{ Effects of protostellar irradiation}
\label{sect:irradiation}

\textcolor{black}{
Fig.~\ref{fig:irradation_qtt} compares the accretion rate history (a) and the protostellar mass 
evolution (b) of two models with $\beta_{\Omega}=0$. Run~1 considers the protostellar radiation feeback 
in addition to the radiation transport in the accretion disc (thin blue solid line) while Run~1-noIrr ignores 
the protostellar irradiation (thick dotted red line). 
The mean accretion rate onto the central growing star is similar in both cases, with a 
mean value of about $10^{-4}-10^{-3}\, \rm M_{\odot}\, \rm yr^{-1}$ and lowest values of 
$\approx 5 \times 10^{-5}-5 \times 10^{-4}\, \rm M_{\odot}\, \rm yr^{-1}$. The absence/presence 
of stellar feedback does not prevent regular accretion spikes in the mass accretion rate from developing 
(Fig.~\ref{fig:irradation_qtt}a). 
In the non-irradiated case, the accretion bursts are more numerous, appear sooner and are of higher 
mean intensity than their irradiated counterparts, although no remarkable spikes reaching intensities of more 
than a few $10^{-1}\, \rm M_{\odot}\, \rm yr^{-1}$ happen, as in Run~1. 
}

\begin{figure}
	\centering
	\begin{minipage}[b]{ 0.36\textwidth}
		\includegraphics[width=1.0\textwidth]{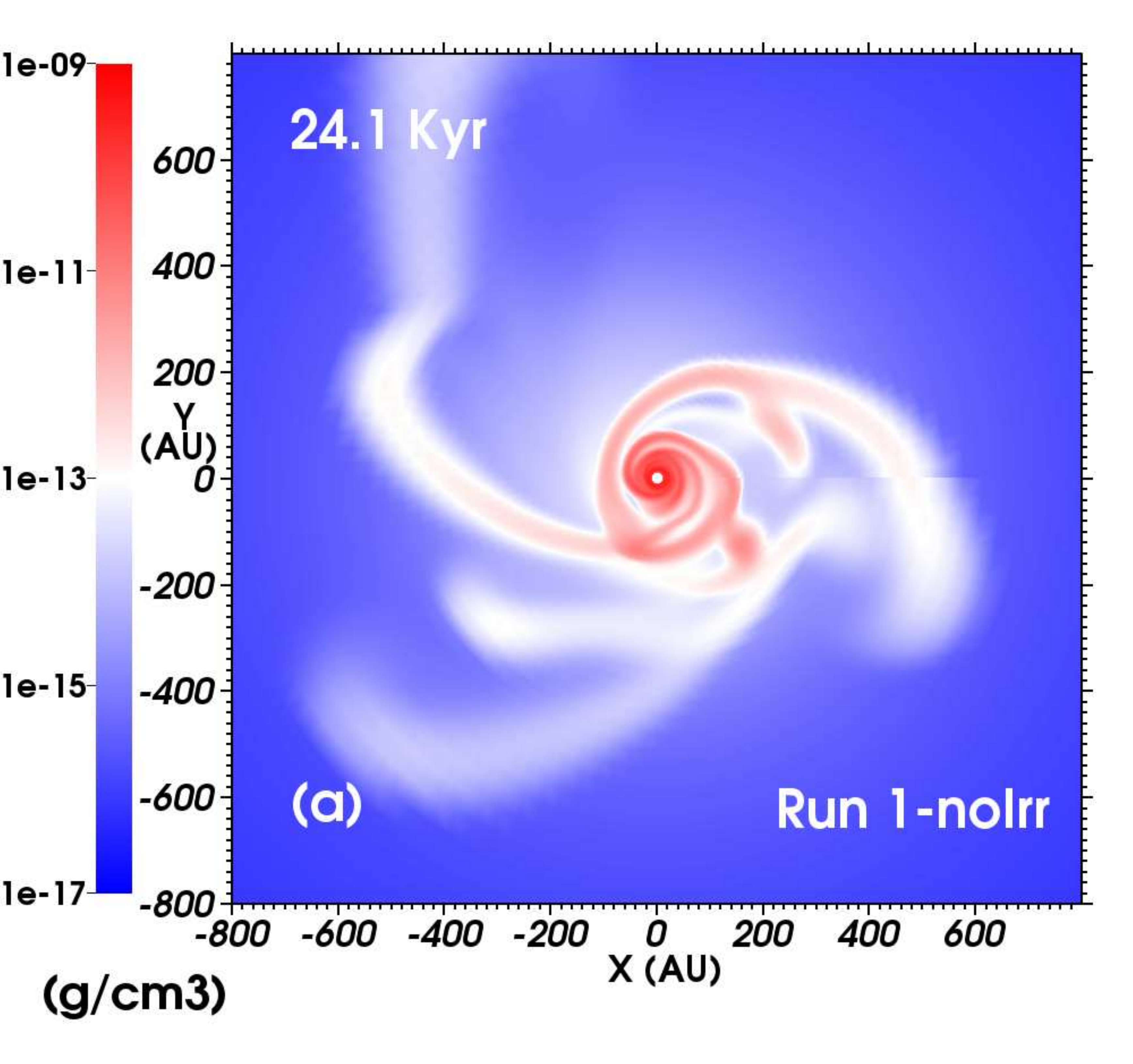}
	\end{minipage}	
	\begin{minipage}[b]{ 0.36\textwidth}
		\includegraphics[width=1.0\textwidth]{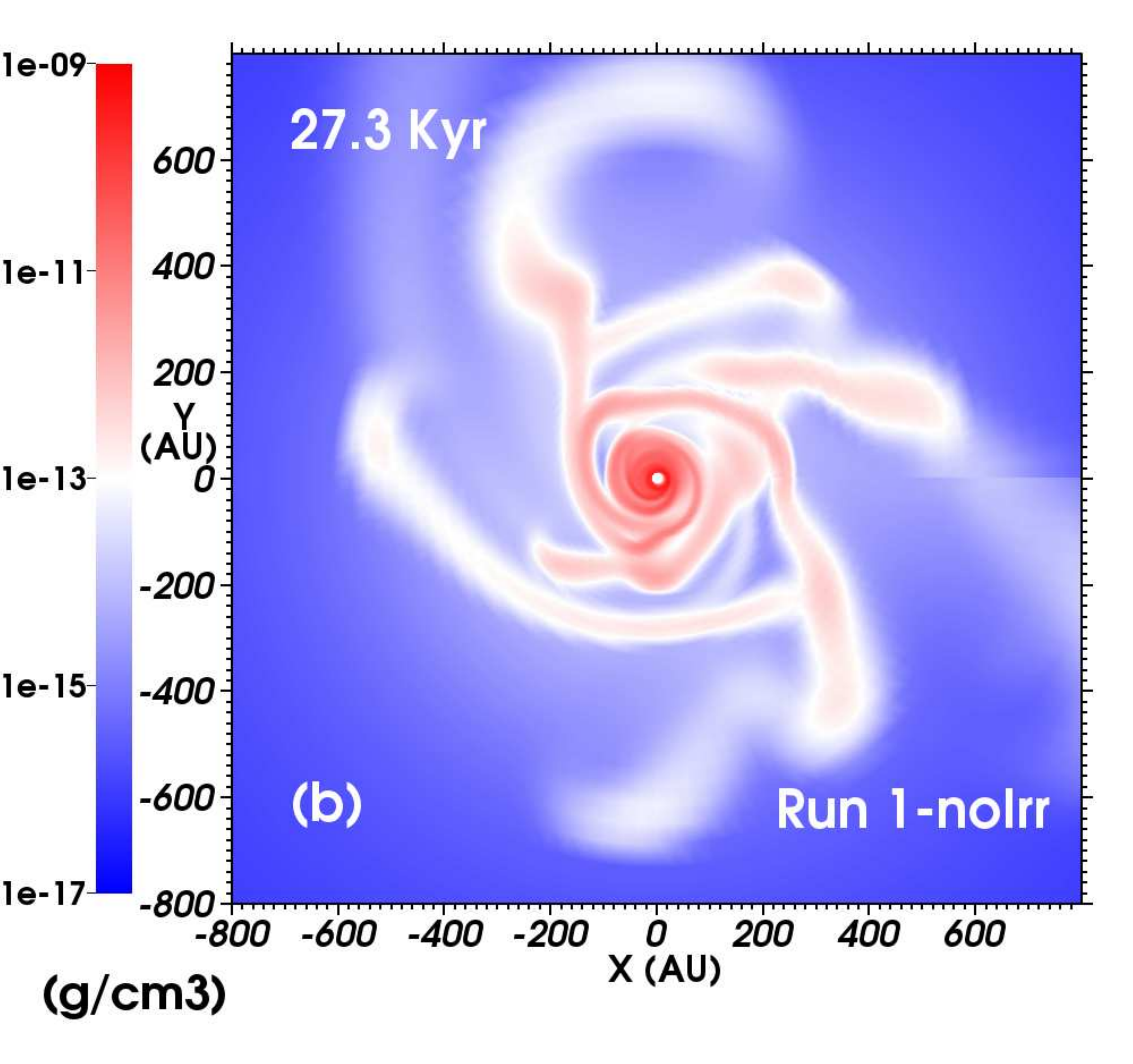}
	\end{minipage} 
	\caption{ 
		 \textcolor{black}{
		 Midplane density field of the model with initial solid-body rotation 
		 ($\beta_{\Omega}=0$) modelled without protostellar irradiation (Run 1-noIrr). 
		 }
		 }	
	\label{fig:irradation_fields}  
\end{figure}

\textcolor{black}{
This indicates an earlier but more violent fragmentation of the disc of Run 1-noIrr as compared to 
the irradiated disc of Run 1. Once fragmentation is triggered (at times larger than $30\, \rm kyr$), 
the variable accretion rate reaches values slightly higher than in the non-irradiated case. 
%
However, once the first accretion-driven outbursts take place, both stellar mass histories are very 
analogous for times $\le\, 18\, \rm kyr$ (Fig.~\ref{fig:irradation_qtt}b), \textcolor{black}{with} differences arising later. 
Indeed, despite of this similar qualitative stellar mass evolution ending with the formation of a star of mass 
$>30\, \rm M_{\odot}$ at times $\ge 45\, \rm kyr$, Run 1-noIrr globally reports \textcolor{black}{a higher number of} step-like increases of 
$\rm M_{\star}$. They correspond to the accretion of a larger number of dense circumstellar clumps. 
One can notice that the migrating gaseous clumps are of smaller individual mass in Run 1-noIrr, leading to less 
intense accretion spikes compare to those of Run~1 (Fig.~\ref{fig:irradation_qtt}a). 
This is consistent with the disc being of lower temperature, having smaller Jeans length and thus lower-mass clumps.
}

\textcolor{black}{
Fig.~\ref{fig:irradation_fields} shows two snapshots of the midplane density field of 
Run 1-noIrr, at times $24.1$ and $27.3\, \rm kyr$. The figures correspond to times before 
(a) and after (b) the initial fragmentation of the disc in Run 1 (Figs.~\ref{fig:disc_evol1}a,b). 
This time interval brackets the formation of the first fragment in the simulation of the 
accretion disc with irradiation (Run~1). 
When stellar feedback is considered, the disc is still stable at time a $24.1\, \rm kyr$ 
(Fig.~\ref{fig:disc_evol1}a) while in the case without incident protostellar radiation it 
has already multiple distinct spiral arms (Fig.~\ref{fig:irradation_fields}a) in which 
develop \textcolor{black}{overdense} regions (Fig.~\ref{fig:irradation_fields}b) that are responsible 
for the formation of (migrating) circumstellar clumps\textcolor{black}{,} producing the accretion-driven bursts 
in Fig.~\ref{fig:irradation_qtt}b. 
Earlier disc fragmentation happens since the close surroundings of the non-irradiated disc 
are colder and therefore fragment faster. Our comparison models illustrate the role of the direct protostellar 
heating in the stabilization of self-gravitating accretion discs around young high-mass stars. 
%
}

\begin{figure*}
	\centering
	\begin{minipage}[b]{ 0.32\textwidth}
		\includegraphics[width=1.0\textwidth]{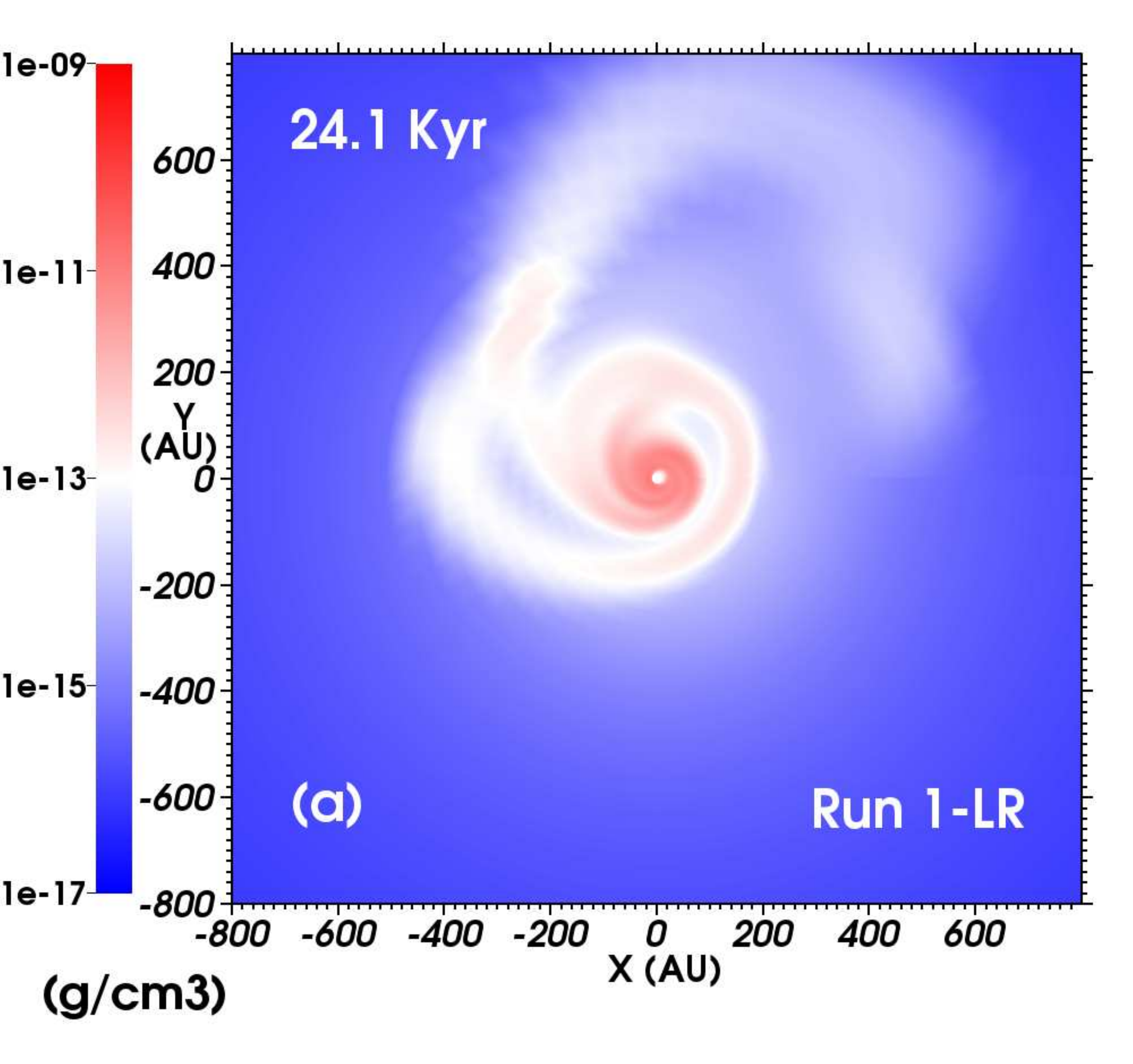}
	\end{minipage}	
	\begin{minipage}[b]{ 0.32\textwidth}
		\includegraphics[width=1.0\textwidth]{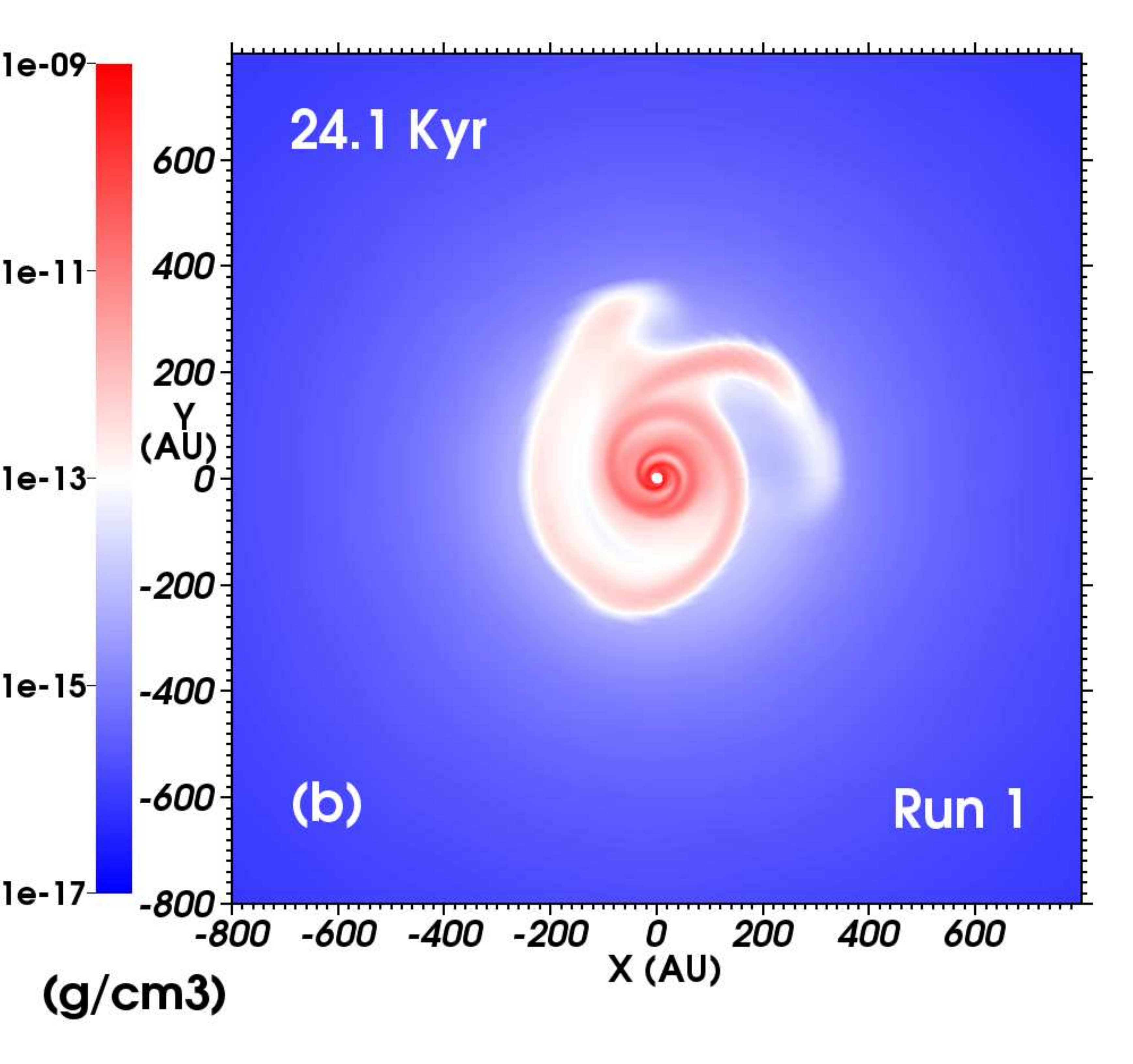}
	\end{minipage} 
	\begin{minipage}[b]{ 0.32\textwidth}
		\includegraphics[width=1.0\textwidth]{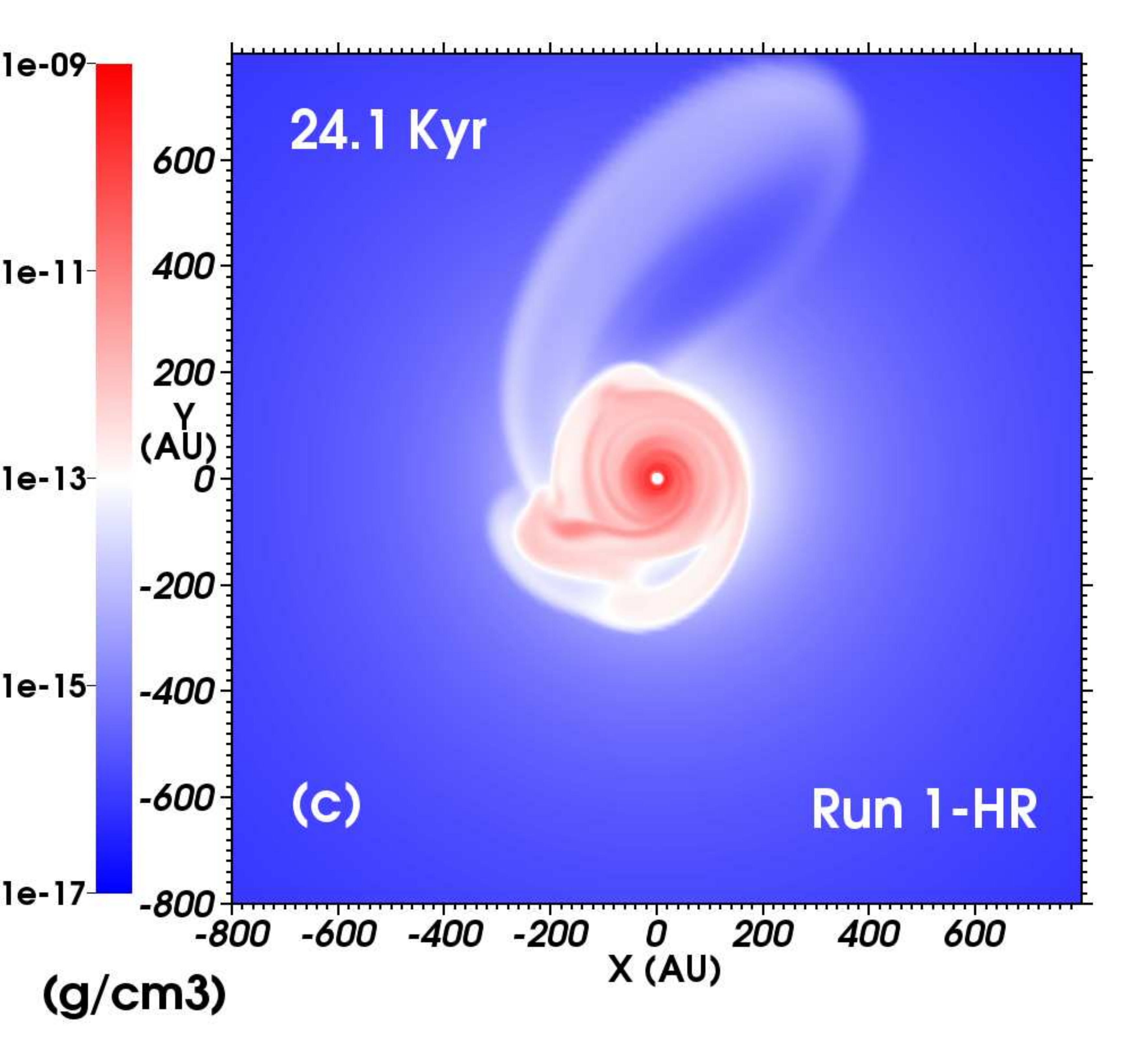}
	\end{minipage} \\ 	
	\begin{minipage}[b]{ 0.32\textwidth}
		\includegraphics[width=1.0\textwidth]{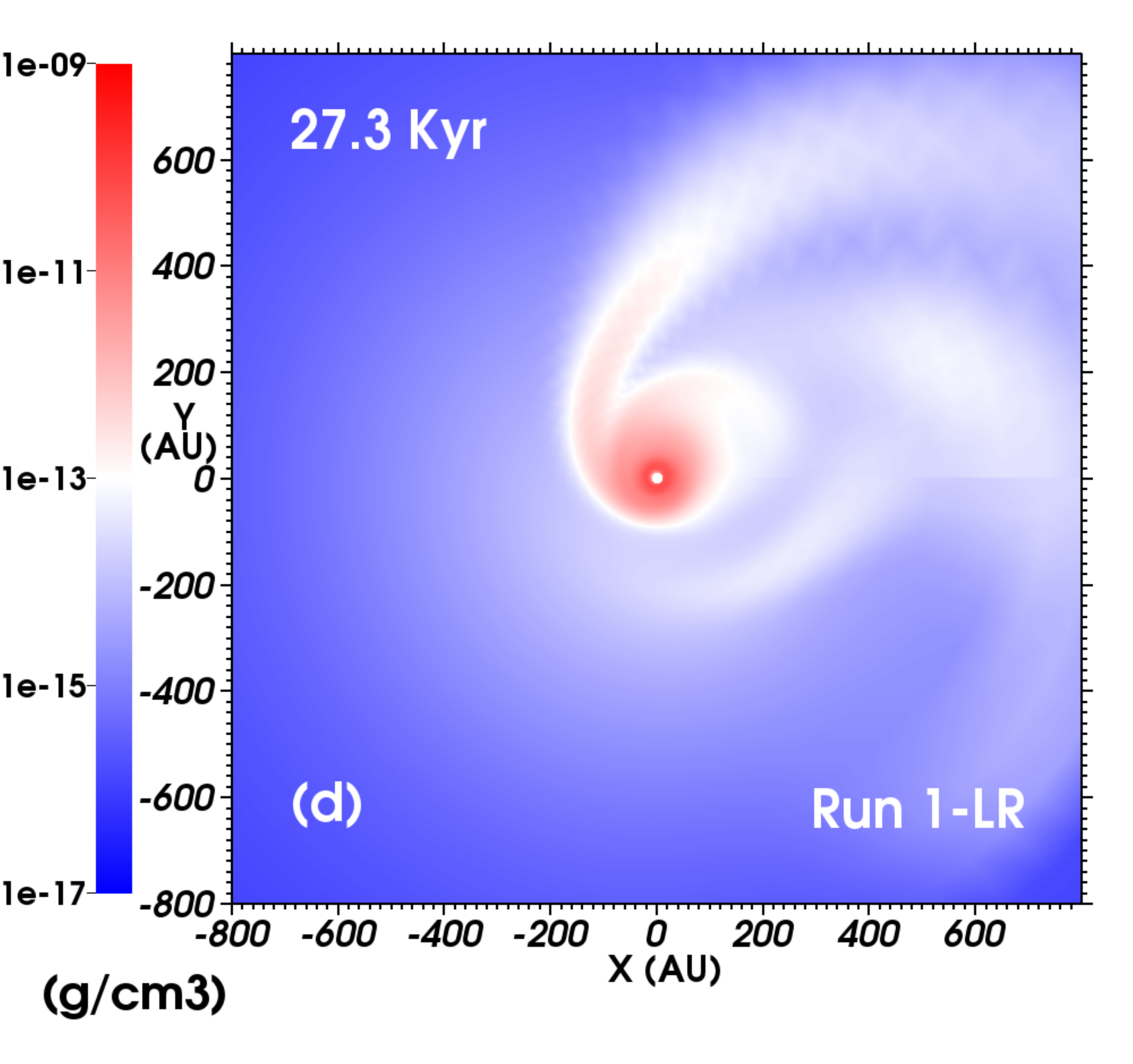}
	\end{minipage}	
	\begin{minipage}[b]{ 0.32\textwidth}
		\includegraphics[width=1.0\textwidth]{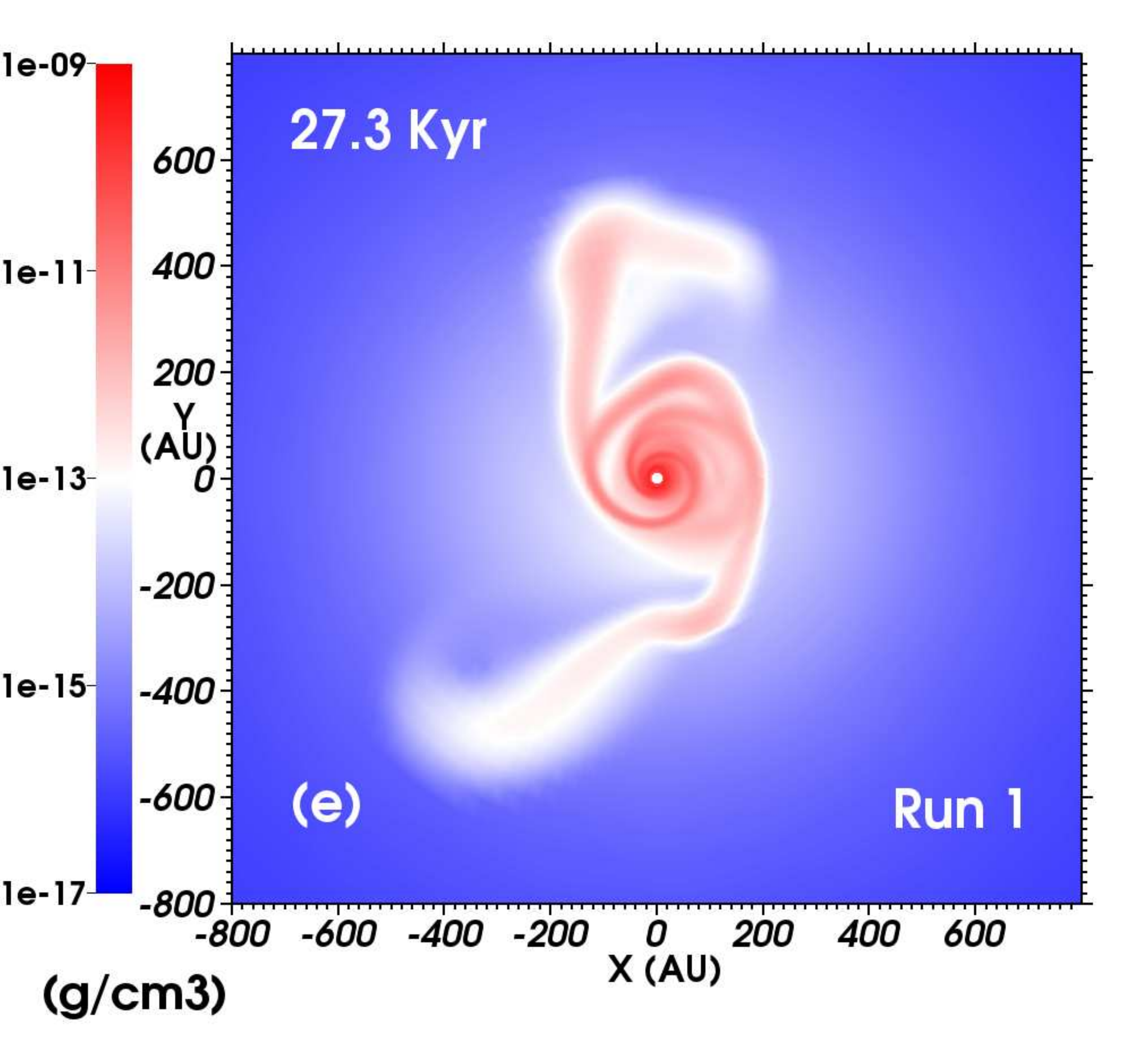}
	\end{minipage} 
	\begin{minipage}[b]{ 0.32\textwidth}
		\includegraphics[width=1.0\textwidth]{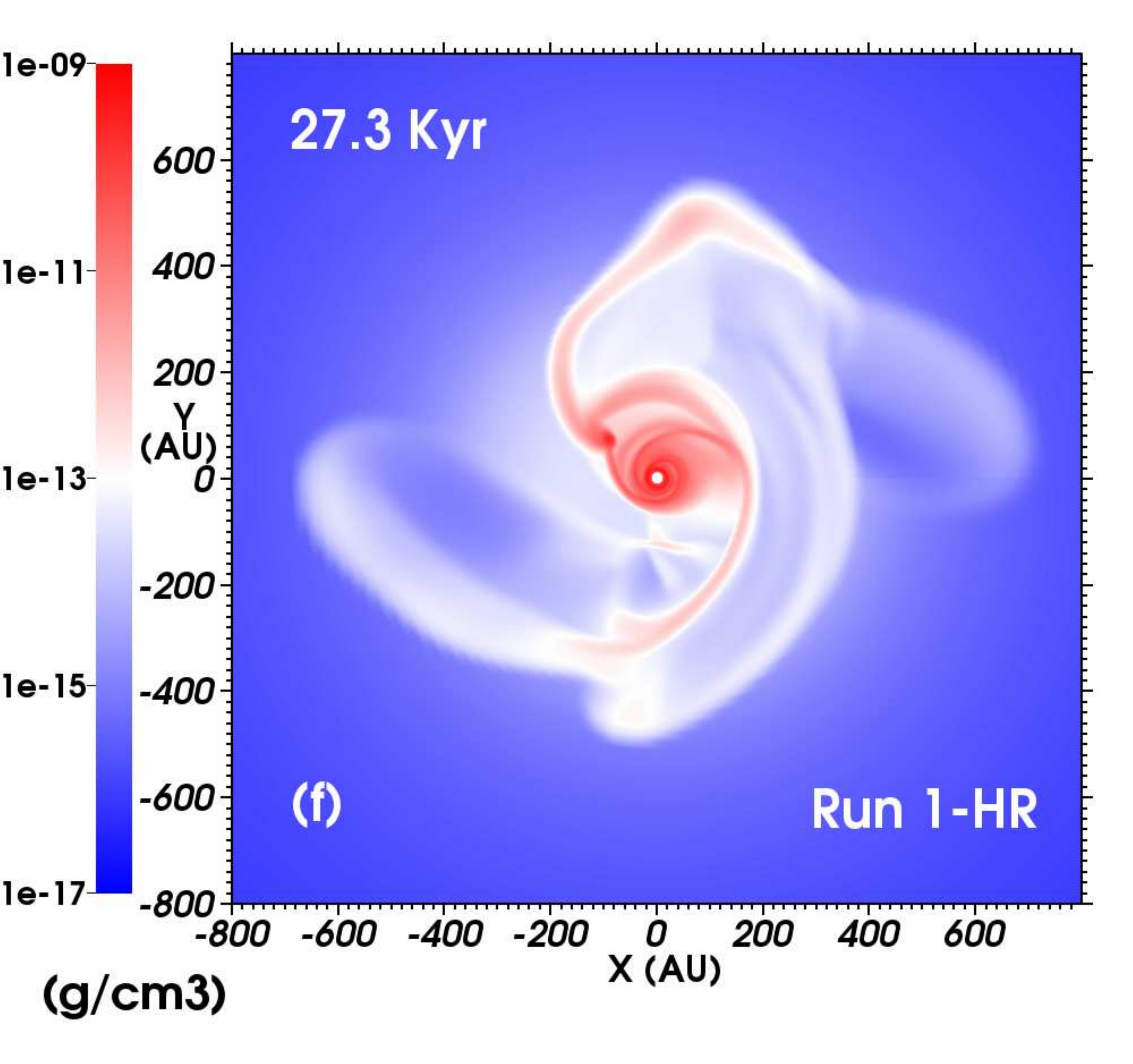}
	\end{minipage} \\ 
	\begin{minipage}[b]{ 0.32\textwidth}
		\includegraphics[width=1.0\textwidth]{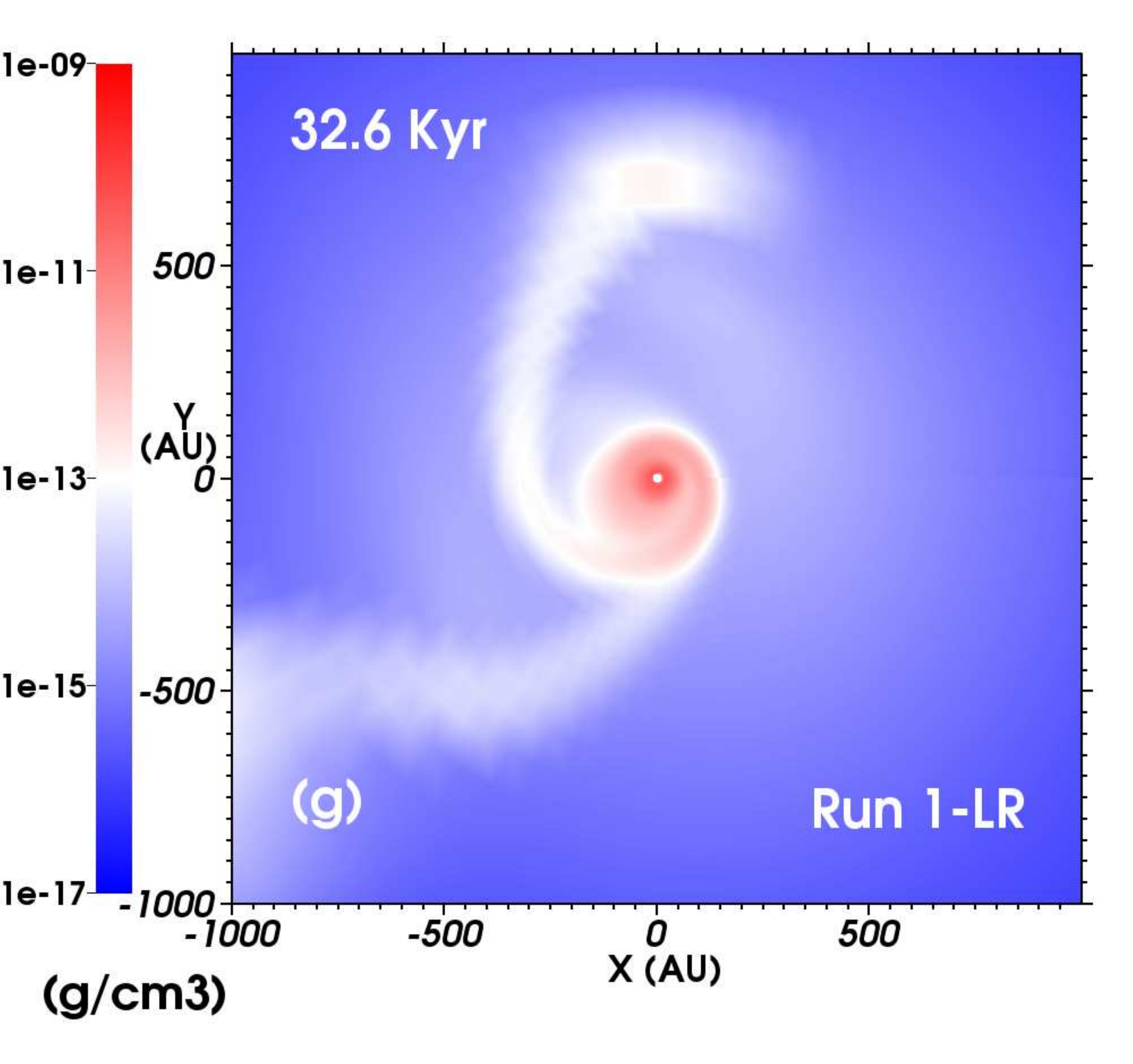}
	\end{minipage}	
	\begin{minipage}[b]{ 0.32\textwidth}
		\includegraphics[width=1.0\textwidth]{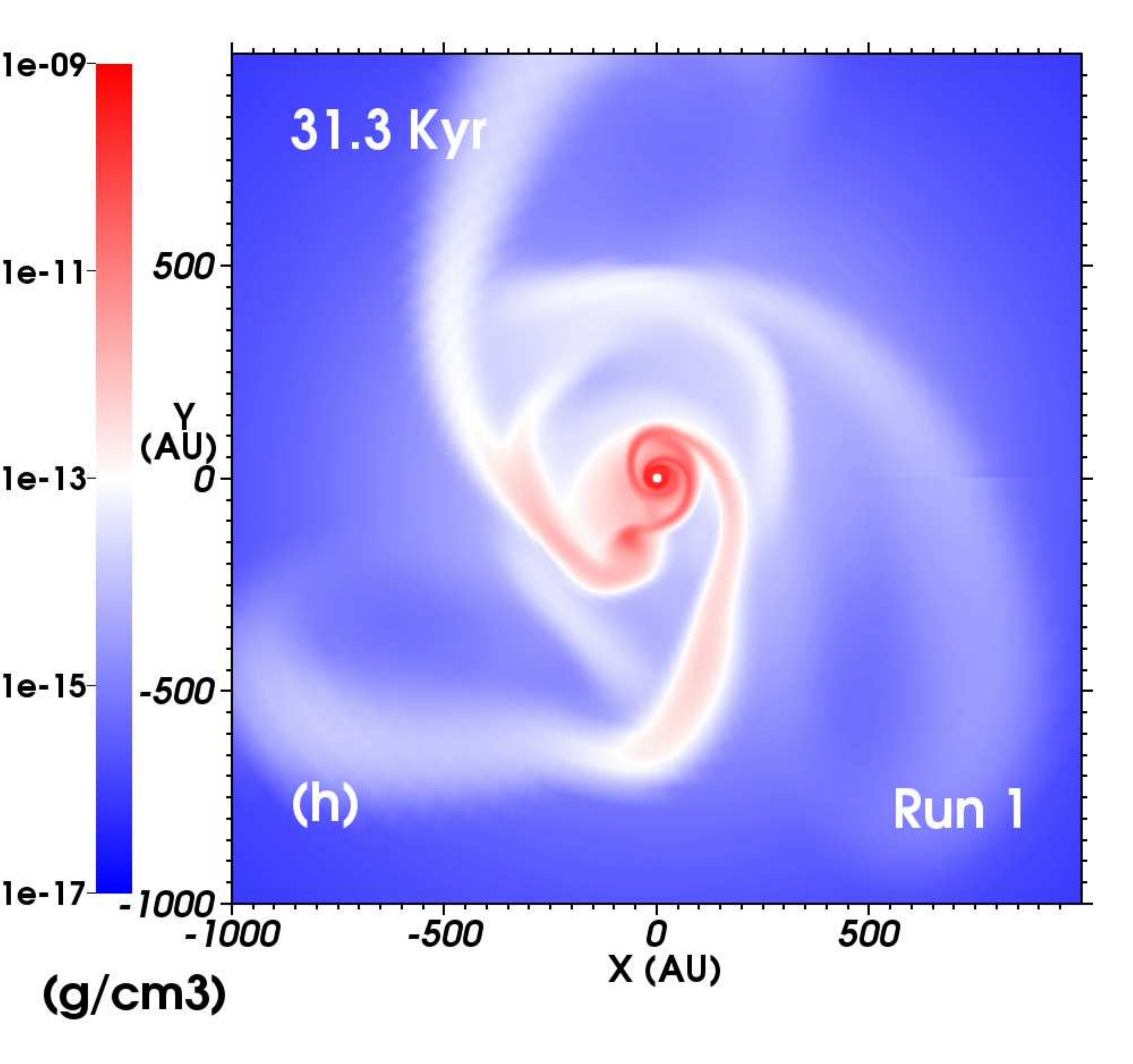}
	\end{minipage} 
	\begin{minipage}[b]{ 0.32\textwidth}
		\includegraphics[width=1.0\textwidth]{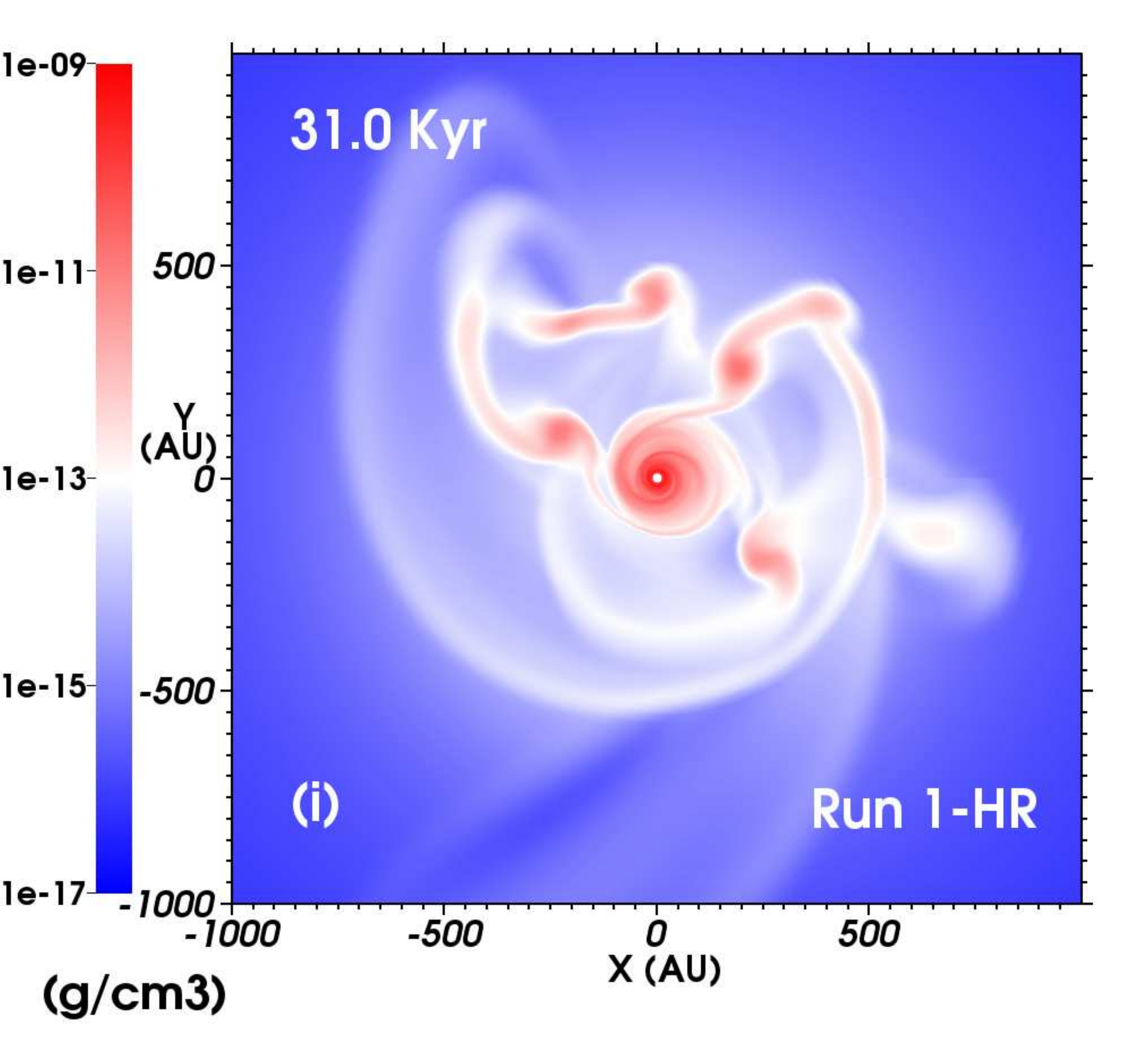}
	\end{minipage} 	
	\caption{ 
	        \textcolor{black}{
		 Density fields of models with initial solid-body rotation, considered with different 
		 spatial resolution (Run 1-LR, Run 1 and Run 1-HR).
		 }
		 }	
	\label{fig:resolution_fields}  
\end{figure*}

\subsection{ Effect of resolution} 
\label{sect:resolution}

\textcolor{black}{
Fig.~\ref{fig:resolution_fields} plots the midplane density field in Run~1-LR 
(left), Run~1 (middle) and Run~1-HR (right) at times $24.1$ (top panels), $27.3\, \rm kyr$ 
(cf. Run~1 in Fig.~\ref{fig:disc_evol1}a,b), \textcolor{black}{and at times} at times $>\, 30\, \rm kyr$ (bottom panels), 
when the fragmentation process is already triggered. 
In all three models, the disc at a time of $24.1\, \rm kyr$ has the shape of twisted spiral 
of distance about $350$-$400\, \rm AU$ from the protostar (Fig.~\ref{fig:resolution_fields}a-c). 
The innermost part of the disc is more resolved in Run~1 than in Run~1-LR and has 
numerous thin arms spiraling around the sink cell (Fig.~\ref{fig:resolution_fields}b), 
while Run~1-HR has just exhibited the first formation signs of denser substructure 
in its outer arm (Fig.~\ref{fig:resolution_fields}c).  
At time  $27.3\, \rm kyr$ one can see that extended spiral arms have already grown. 
The lowest resolution model (Run~1-LR) does not show traces of undergoing substructure 
formation (Fig.~\ref{fig:resolution_fields}d) while Run~1 has portions of spirals arms 
including the curved, denser sections (Fig.~\ref{fig:resolution_fields}e). 
Since the only difference between all models is the grid resolution, one can directly estimate  
the effect the spatial resolution has on disc fragmentation. The highest resolution model evolves similarly, 
except that the first fragment forms and migrates sooner (Fig.~\ref{fig:resolution_fields}f). 
At times $>\, 30\, \rm kyr$, the patterns show either spiral arms
(Fig.~\ref{fig:resolution_fields}g), or fragments in the case of the 
\textcolor{black}{most-resolved} model (Fig.~\ref{fig:resolution_fields}i). 
\textcolor{black}{
The bottom line of panels illustrates the effects of spatial resolution on the random development 
of the disc structures once the fragmentation process begins. 
}
Using the finest grid resolution to date, we point the prime importance for the treatment of the 
close stellar environment for reliability of numerical disc fragmentation studies. 
This was expected from comparison with disc fragmentation studies in the low-mass star 
formation regime where the irradiation is negligible, see e.g.~\citet{lichtenberg_aa_579_2015}, 
but also from models in the context of primordial star formation, in which stellar feedback 
is efficiently at work, see~\citet{greif_apj_737_2011,vorobyov_apj_768_2013,machida_mnras_448_2015} 
and the Fig.~3 of~\citet{hosokawa_2015} for a resolution study. 
}

\begin{figure}
	\centering 
	\begin{minipage}[b]{ 0.48\textwidth}
		\includegraphics[width=1.0\textwidth]{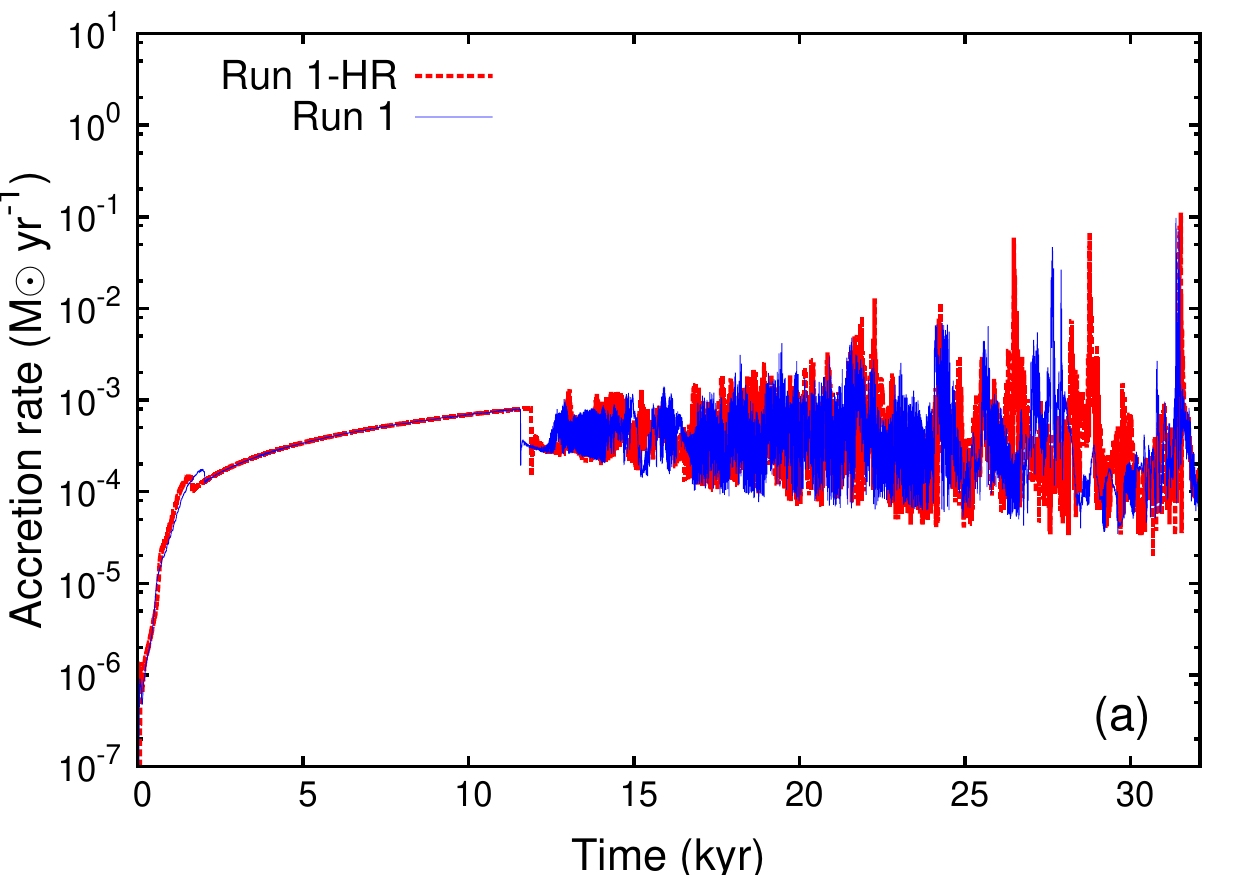}
	\end{minipage}	\\ 
	\begin{minipage}[b]{ 0.48\textwidth}
		\includegraphics[width=1.0\textwidth]{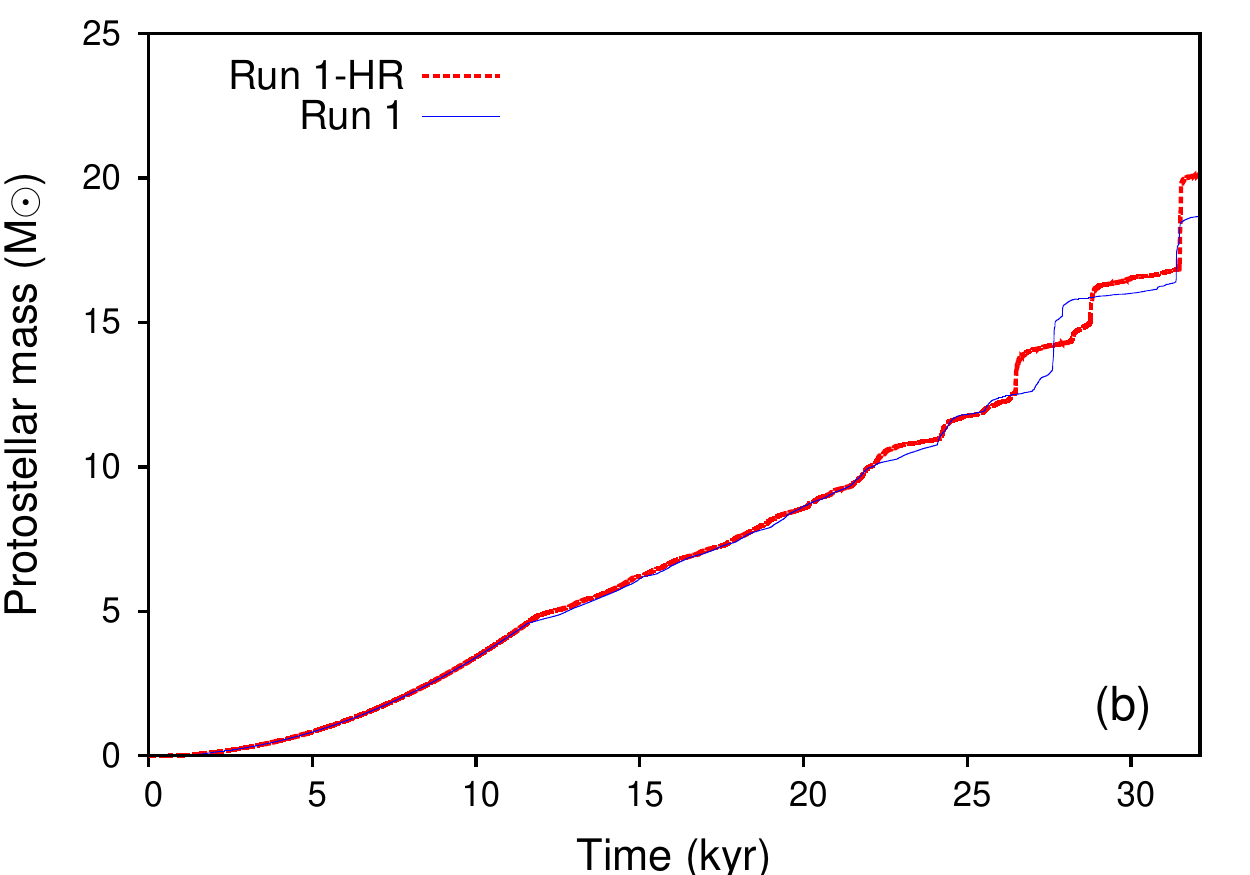}
	\end{minipage}	\\
	\begin{minipage}[b]{ 0.48\textwidth}
		\includegraphics[width=1.0\textwidth]{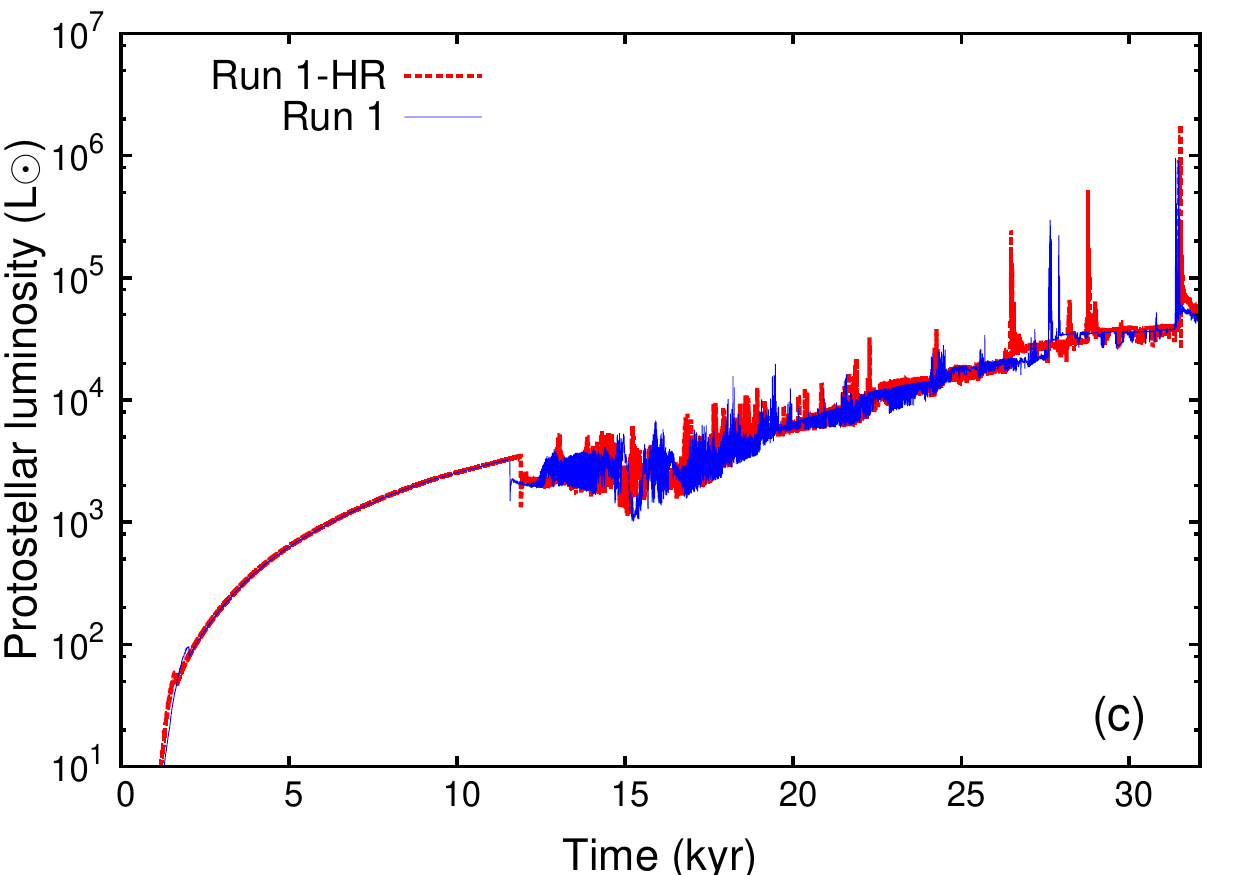}
	\end{minipage}	
	\caption{ 
		 \textcolor{black}{
		 Comparison between the accretion rate (a) and the mass evolution (b)
		 and the luminosity (c) of a model with several grid resolutions.   
		 }  
		 }	
	\label{fig:resolution_qtt}  
\end{figure}

\textcolor{black}{
Fig.~\ref{fig:resolution_qtt} is similar to Fig.~\ref{fig:irradation_qtt} for models 
Run~1 and Run~1-HR, which are our two most resolved models having an initial 
rigidly-rotating pre-stellar core ($\beta_{\Omega}=0$). 
%
%
One can notice the good agreement between their accretion rate historys (Fig.~\ref{fig:resolution_qtt}a). 
Especially, the first remarkable accretion spike happens 
similarly at the times of $\approx 22\, \rm kyr$ showing that the solution has converged up to 
the formation, fall and accretion of the first gaseous clumps. Further discs evolution logically differs 
in the sense that a higher resolution reveals the typical stochastic and fractal behaviour of 
\textcolor{black}{fragmenting} disc. 
The number of accretion-driven events slightly increases with the disc resolution because more 
clumps are formed, i.e. Run~1-HR has 3 separated spikes reaching $\ge 10^{-2}\, \rm M_{\odot}\, \rm yr^{-1}$, 
while Run~1 has only a twin one of the same intensity (Fig.~\ref{fig:resolution_qtt}c). 
The accreted mass per unit time is similar in both models and the mass evolution of the central 
protostar converges up to $\approx\, 22\, \rm kyr$, \textcolor{black}{the time} of the first clump migration. 
At times $<30\, \rm kyr$, the mean mass of a gaseous clump is smaller in Run~1-HR because more numerous but lighter fragments 
form in the disc, and large accretion events are replaced by the successive accretion of lighter fragments 
for a roughly equal total mass, see for example at times $\approx\, 26$-$29\, \rm kyr$ (Fig.~\ref{fig:resolution_qtt}b).  
%
}

\textcolor{black}{
We have performed a resolution study by increasing the resolution of our models up to the highest spatial 
resolution to date. The above presented elements show that our solutions are sufficient to extract the parameters 
required for our analyses and conclusions in terms of accretion variability, inner disk fragmentation and 
migrating gaseous clumps.   
}

\subsection{ Jeans length resolution} 
\label{sect:resolution_lambda}

\begin{figure}
	\centering
	\begin{minipage}[b]{ 0.44\textwidth}
		\includegraphics[width=1.0\textwidth]{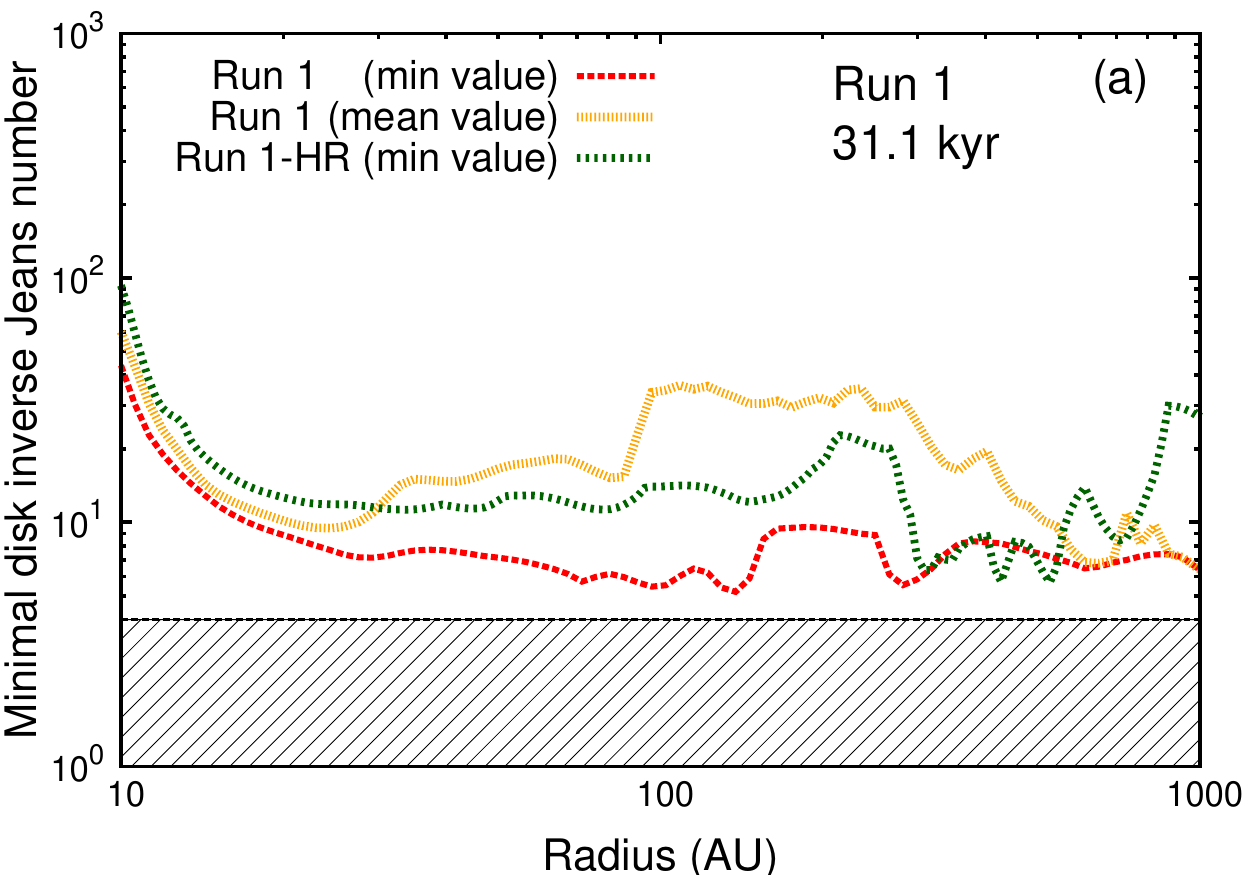}
	\end{minipage}	\\
	\begin{minipage}[b]{ 0.44\textwidth}
		\includegraphics[width=1.0\textwidth]{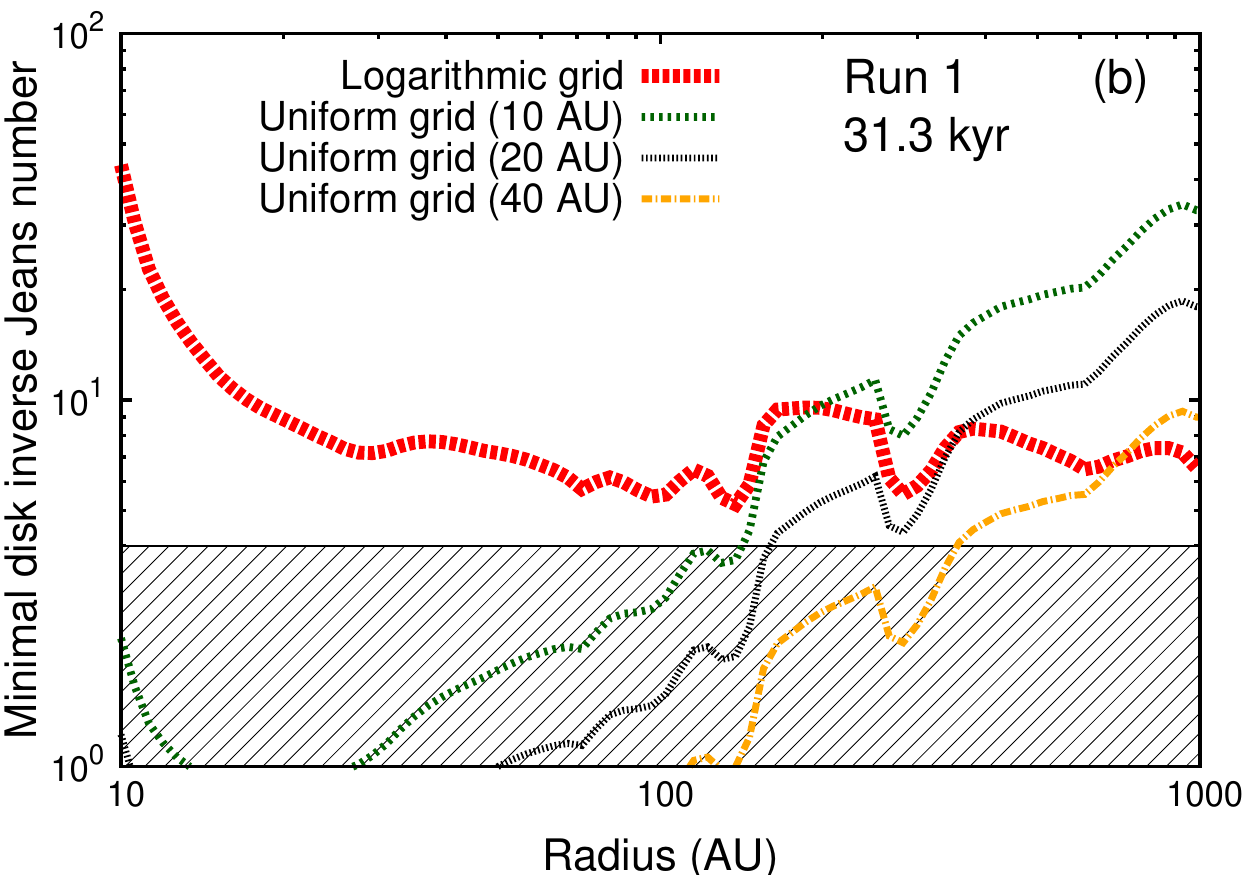}
	\end{minipage}	\\
	\begin{minipage}[b]{ 0.44\textwidth}
		\includegraphics[width=1.0\textwidth]{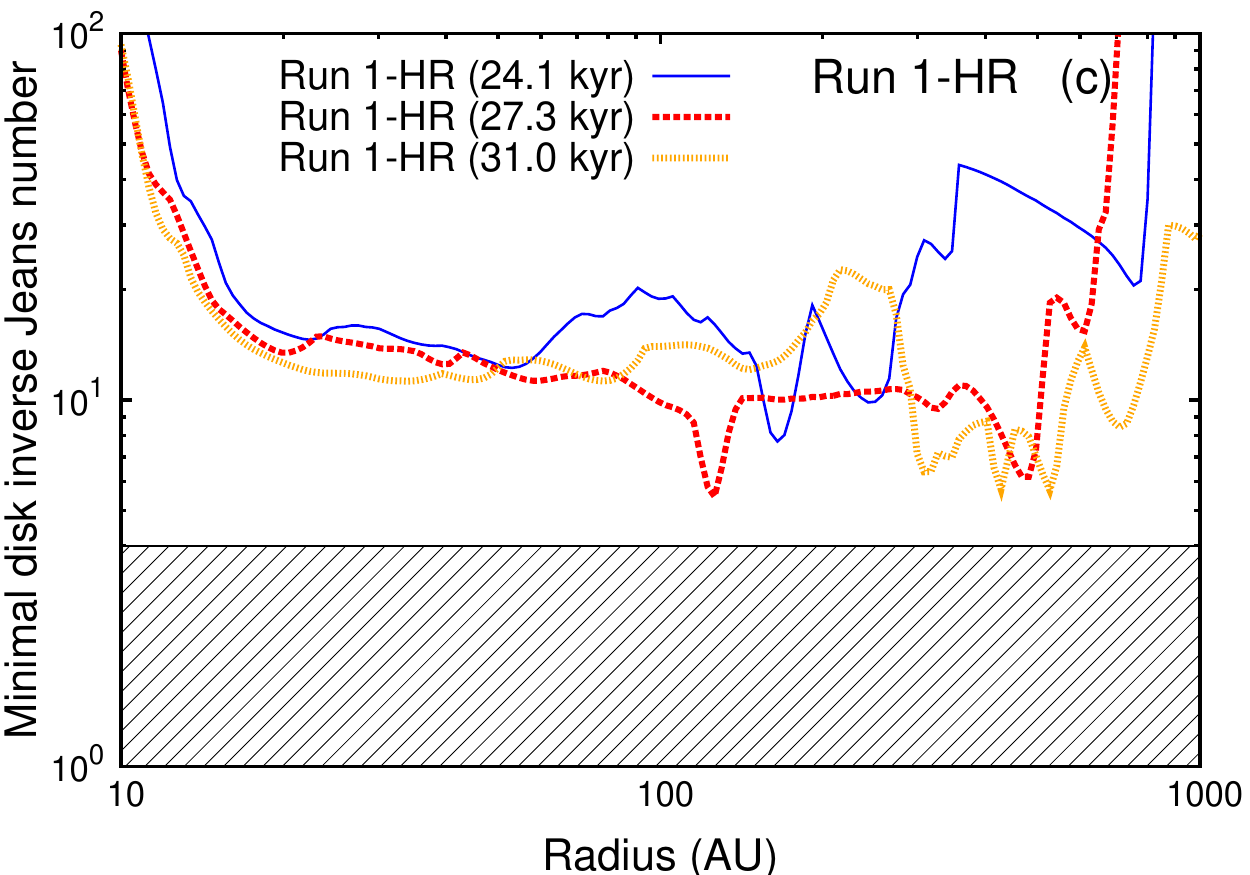}
	\end{minipage}		
	\caption{ 
		 \textcolor{black}{
		 Minimal inverse Jeans number in the disc midplane as a function of \textcolor{black}{radius}.  
		 It is plotted for the resolution study involving our initially solid-body rotating simulations (a), 
		 our disc in Run~1 as considered with different grids and resolutions (b)
		 and for our run~1-HR (c).  	
		 Truelove-unresolved regions are hatched. 
		 }
		 }	
	\label{fig:comparison_resolution}  
\end{figure}

\textcolor{black}{
This paragraph investigates our discs with respect of the Truelove criterion~\citep{truelove_apj_495_1998},  
which stipulates that the local Jeans length must be resolved by at least 4 grid cells. The so-called Jeans 
number is defined by, 
\begin{equation}
    n_{\rm J} = \frac { \Delta }{ \lambda_{\rm J} }, 
   \label{eq:jeans_number}    
\end{equation}
where, 
\begin{equation}
    \lambda_{\rm J} = \Big( \frac{ \pi c_{\rm s}^{2} }{ G \rho } \Big)^{1/2}, 
   \label{eq:jeans_lengths}    
\end{equation}
is the Jeans length and $\Delta$ is the grid cell size. Therefore the Truelove 
criterion reads,
\begin{equation}
    \frac{ 1 }{ n_{\rm J} } = \frac{ \lambda_{\rm J} }{ \Delta } \ge 4, 
   \label{eq:truelove}    
\end{equation}
and the clump-forming regions of the disc must satisfy this relation. 
Fig.~\ref{fig:comparison_resolution}a-c shows the minimal disc midplane inverse Jeans number 
$1/n_{\rm j}$, estimated from different simulation snapshots and plotted as a function of the 
radial direction for the inner $1000\, \rm AU$ which is the typical size of our discs. 
The hatched region in the figures indicates values $1/n_{\rm J}<4$ where the Truelove criterion 
is violated. 
%
%
Run~1 is fully resolved in the inner $1000\, \rm AU$ of the disc (red thick dotted line). 
The minimal inverse Jeans number exhibits variations corresponding to the presence of the 
dense spiral arms hosting circumstellar clumps (local minimum) and the underdense regions 
between the spiral arms (local maximum). Note that the mean value of the azimuthally averaged 
inverse Jeans number is much larger (thick solid orange line). 
High resolution simulation Run~1-HR is our best Truelove-resolved model (green thick dashed line). 
This criterion is a minimal requirement to consider simulations of self-gravitating 
discs are numerically reliable. It does not \textcolor{black}{consist of} a strict convergence rule, as \textcolor{black}{illustrated by} the 
density pattern differences of the Truelove-resolved models Run~1 and Run~1-HR.
}

\begin{figure}
	\centering
	\begin{minipage}[b]{ 0.4\textwidth}
		\includegraphics[width=1.0\textwidth]{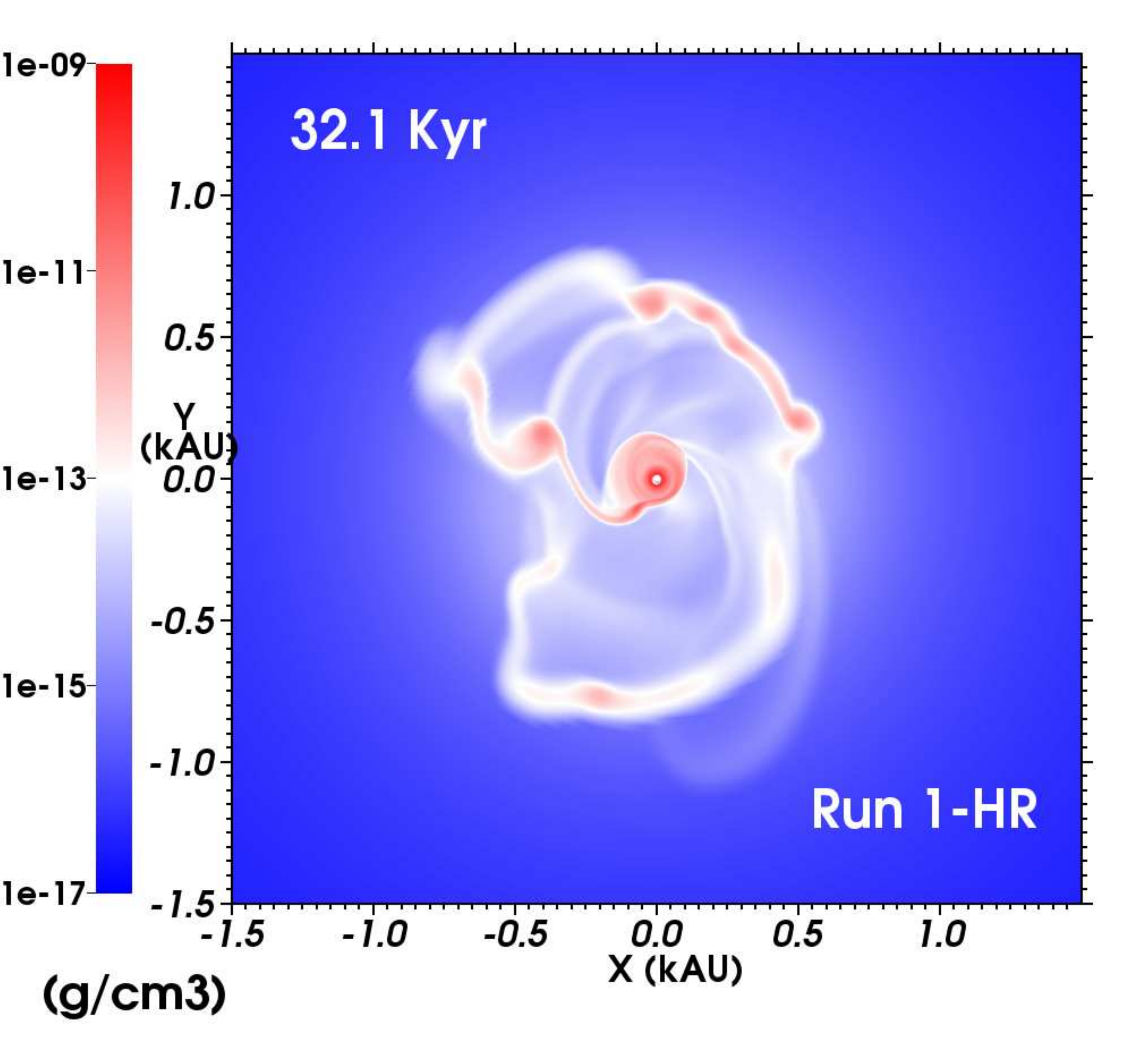}
	\end{minipage}	
	\caption{ 
		 \textcolor{black}{
		 Midplane density field of at the end of Run 1-HR. 
		 }
		 }	
	\label{fig:jeans_length}  
\end{figure}

\textcolor{black}{ 
Fig.~\ref{fig:comparison_resolution}b is similar to Fig.~\ref{fig:comparison_resolution}a but for Run~1, 
assuming different grid types and resolutions. The thick dotted red line assumes the logarithmic simulation 
grid while the other lines considers minimal inverse Jeans lengths calculated with our disc 
midplane density field together with uniform grids of resolution $\Delta=10$-$40\, \rm AU$, 
corresponding to the resolution typically reached by codes with Cartesian grids for this kind 
of simulations. 
The logarithmic grid is better than any considered uniform grid up to radii of about 
$150$-$200\, \rm AU$, which is similar to the inner disc size after simulation times 
of $\approx 23$-$30\, \rm kyr$, depending on the models. 
Particularly, the resolution $\Delta=40\, \rm AU$ is not able to reach our disc resolution  
up to radii of $\approx \, 600$-$800\, \rm AU$. This mean that substructures of the disc in Run~1 
can not be reproduced with such a grid size, as the onset of clump formation and migration happens 
in spiral arms at a typical distance of a few hundreds $\rm AU$ from the central 
protostar~\citep[see][]{meyer_mnras_464_2017}. This resolution-dependence of the numerical capture of 
accretion-driven outbursts phenomenon explains why it could not be captured by mesh refinement codes 
with a Cartesian grid since their maximum disc resolution of $10\, \rm AU$ is so far coarser in the inner disc than in 
our simulations~\citep[see e.g.][]{krumholz_apj_656_2007,klassen_apj_823_2016,rosen_mnras_463_2016}. 
%
%
}
\textcolor{black}{
Fig.~\ref{fig:comparison_resolution}c is similar to Fig.~\ref{fig:comparison_resolution}b but for Run~1-HR at 
different times of its evolution, corresponding to the snapshots in Fig.~\ref{fig:resolution_fields}\textcolor{black}{c,f and j}. 
Throughout the whole simulation time, this model is well Truelove-resolved ($1/n_{\rm j}\ge10$), even when 
clumps start flourishing at the outskirt of the disc ($1/n_{\rm j}\ge4$-$6$, see Fig.~\ref{fig:jeans_length}). 
%
%
The necessary resolution of the Jeans number in the clump cores is achieved partly because they evolve in the disc 
midplane, see Fig.~\ref{fig:comparison_resolution}a-c. In all our simulations, we generally find values of 
$\lambda_{\rm J}/\Delta \sim 6$-$8$, see details relative to the first migrating clump of Run~3 
in~\citet{meyer_mnras_464_2017}.  
}

\textcolor{black}{
The equatorial symmetry of our setup forces the clump to develop and evolve within the midplane. 
While this prevents clumps from potentially scattering to high disc latitudes.
Indeed, clumps are therefore resolved at all radii with the smallest grid cells permitted by the used 
cosine-like grid along the polar direction. 
The disc scale height in the clump-forming regions is $H\approx 10\, \rm AU$ and can increase with $r$ up to 
$H\approx 100\, \rm AU$ in the outer disc region. The choice of an expanding grid resolution 
$\Delta(r)$ allows us to resolve the disc scale height $H(r)\gg\Delta(r)$ and at the same time to 
reduce the number of cells and the computing time of the simulations. 
The Hill radii of the clumps, i.e. the surrounding regions that is gravitationally 
influenced by their mass, is $R_{\rm H}\approx 64-96\, \rm AU$ for a clump 
mass of $M_{\rm cl}\approx 1.0\, \rm M_{\odot}$ located at $r\approx 200-300\, \rm AU$ 
\textcolor{black}{from a star of} $M_{\star} \approx 10 \, \rm M_{\odot}$, whereas the grid resolution is of $\Delta \simeq\, \rm a\, \rm few\, \rm AU$ 
at that radius in our Runs~1-3 (and a factor of 2 smaller in Run~1-HR). 
The Hill radius goes as $R_{\rm H}\approx r (M_{\rm cl}/3M_{\star})^{1/3}$ and therefore increases 
at larger radii, which compensates the linear loss of resolution of the logarithmically expanding grid 
compared to the inner part of the disc. As the protostar grows, $M_{\star}$ increases and assuming 
$M_{\star} \approx 30 \, \rm M_{\odot}$ we find $R_{\rm H}\approx 45-67\, \rm AU$, which is still correctly 
resolved. As the clump \textcolor{black}{falls in, it loses} mass but the grid resolution increases; leading to similar 
conclusions as above. 
What is new in the present study is not the included microphysical processes, but the high 
spatial resolution allowed by the method in the innermost part of the accretion disc. 
We resolve the stellar surroundings better than any simulation before and explore the initial 
perturbations of the disc together with their effects on its subsequent overall instability. 
}


\section{How to characterize disc instability?}
\label{section:results_fragmentation}

\textcolor{black}{
Simulations with lower spatial resolution than used in our study often made use of 
sink particles to generate \textcolor{black}{nascent} stars in the accretion disc. Forming those sink 
particles underlines the assumption of disc fragmentation at this location, however, 
different simulation codes also apply different criteria, from density thresholds~\citep{krumholz_apj_656_2007} 
to the onset of local isotropic gravitational collapse~\citep{federrath_apj_713_2010}. 
Our different method, without sink particles, allows us to pronounce on the reliability of analytic 
criteria for disc fragmentation, by computing them for our resolved discs models and directly compare 
if \textcolor{black}{they result in disc fragmentation}. 
}
Therefore, we test \textcolor{black}{the} protostellar discs with respect to several criteria for the fragmentation 
of self-gravitating discs, i.e. the so-called Toomre, Gammie and Hill criteria. 
Particularly, we discuss whether those criteria are consistent with \textcolor{black}{our} results, 
and if they are necessary and/or sufficient to determine the unstable character 
of the discs, in the spirit of the analysis carried out in~\citet{klassen_apj_823_2016}.

\subsection{ Toomre criterion }
\label{sect:toomre}

The so-called Toomre parameter measures the unstable character of a self-gravitating 
disc by comparing the effects of the gravity~\citep{toomre_apj_138_1963} against 
the combined effects of the disc thermodynamics, i.e. the gas thermal pressure, together with 
the rotational shear induced by the Keplerian motion of the gas, 
providing a stabilizing force to the system. A region of a gaseous disc is Toomre-unstable 
(or $Q$-unstable) if the dimensionless quantity, 
\begin{equation}
    Q = \frac{ \kappa c_{\rm s} }{ \pi G \Sigma } \le Q_{\rm crit},
   \label{eq:toomre}    
\end{equation}
where $c_{\rm s}$ is the sound speed of the gas, $\Sigma$ the column mass 
density of the disc and $\kappa$ the \textcolor{black}{local epicyclic frequency. 
The aforementioned Toomre criterion 
was derived in the thin-disc limit for polytropic and axisymmteric discs, for which case 
$Q_{\rm crit}=1$.
In more realistic situations, the exact value of the critical Toomre parameter
$Q_{\rm crit}$ depends somewhat on the disc thickness, rotation curve and thermodynamics. 
Numerous numerical studies indicate that circumstellar discs become unstable to the growth of a spiral
structure if $Q$ becomes less than $Q_{\rm crit}=$1.5-2.0 \citep[e.g.][]{durissen_prpl_607_2007}. 
When the \textcolor{black}{local} $Q$-parameter becomes smaller than 1.0, spiral arms may fragment to form compact 
gaseous clumps.} This criterion is a necessary condition with respect to gravitational 
fragmentation, however, recent studies updated condition Eq.~(\ref{eq:toomre}) to $Q<0.6$, see 
the study of protoplanetary discs of~\citet{takahashi_mnras_458_2016}.

The middle column of panels of Fig.~\ref{fig:disc_overall_criteria} 
plots the $Q$-analysis of the disc in our Run 1 at several characteristic timescales of 
its evolution. The color scale indicates whether the disc is prone to fragment ($Q\le 1$, 
red), marginally unstable to fragmention ($Q\simeq 1$, white) or stable to fragmentation 
($Q> 1$, blue). 
Figs.~\ref{fig:disc_overall_criteria}e-h show that the Toomre criterion is 
satisfied by the denser regions of the spiral arms and by the circumstellar clumps 
in the accretion disc. The surface of the disc midplane which is prone to 
fragmentation increases as a function of time, as the circumstellar medium of 
the protostar loses axisymmetry (Figs.~\ref{fig:disc_overall_criteria}e,h). 
As noted in~\citet{klassen_apj_823_2016}, the disc's growing tendency to $Q$-instability 
comes along with an increase of the accretion variability. 
The \textcolor{black}{ejection of the leftover 
of the first accreted clump, producing the first accretion-driven 
outburst, provokes the formation of a dense filament propagating outwards} as described in fig.~1b-d 
of~\citet{meyer_mnras_464_2017}, which in its turn \textcolor{black}{favours} the generation of 
thin but extended spiral arms, enhancing the disc's instability and \textcolor{black}{susceptibility}  
to experience \textcolor{black}{further} fragmentation events. The more violent the disc 
fragmentation, the lower the value of the associated local $Q$-parameter.

\textcolor{black}{
At time $24.1\, \rm kyr$, the most Toomre-unstable clump located at a radius $\approx\, 100\, \rm AU$ 
from the protostar has $Q\approx0.25$-$0.4$ (Fig.~\ref{fig:disc_overall_criteria}d). This value is 
typical of the spiral arm in the discs which develop throughout our simulations. $Q$-values of 
$\approx2.6$-$3.2$ in the inner region of the disc ($\le\, \rm 100\, \rm AU$) indicate stability, 
except for some very thin (but still marginally Toomre-stable with $Q\approx1.7$) twisted spiral located 
at radii of $\approx\, 55\, \rm AU$. 
At a time of $27.3\, \rm kyr$, unstable spiral arms in the disc with $Q=0.8$$-$$0.86$ are noticeable, see for 
example the thin spiral arm at $\rm x=-200\, \rm AU$ and $\rm y=300\, \rm AU$ in Fig.~\ref{fig:disc_overall_criteria}e.  
The other regions of the disc which are fragmenting into blobs have characteristics such that $Q=0.8$$-$$1.2$. 
The snapshot in Fig.~\ref{fig:disc_overall_criteria}f draws similar conclusions with 
multiple $Q$-unstable clumps of $Q\le0.4$ linked to each other by spiral arms which 
central regions can also be locally unstable ($Q\le0.4$-$0.5$), however, most part of those 
arms are such that $Q=0.8$$-$$1.35$. 
Therefore, our results are consistent with the update of the Toomre criterion
reported in~\citet{takahashi_mnras_458_2016}, i.e. the only necessary condition for disc 
instability is $Q < 0.6$, as already confirmed in the context of the 
stable discs shown in~\citet{klassen_apj_823_2016}. 
}

\subsection{ Gammie criterion }
\label{sect:gammie}

\begin{figure*}
	\centering
	\begin{minipage}[b]{ 0.315\textwidth}
		\includegraphics[width=1.0\textwidth]{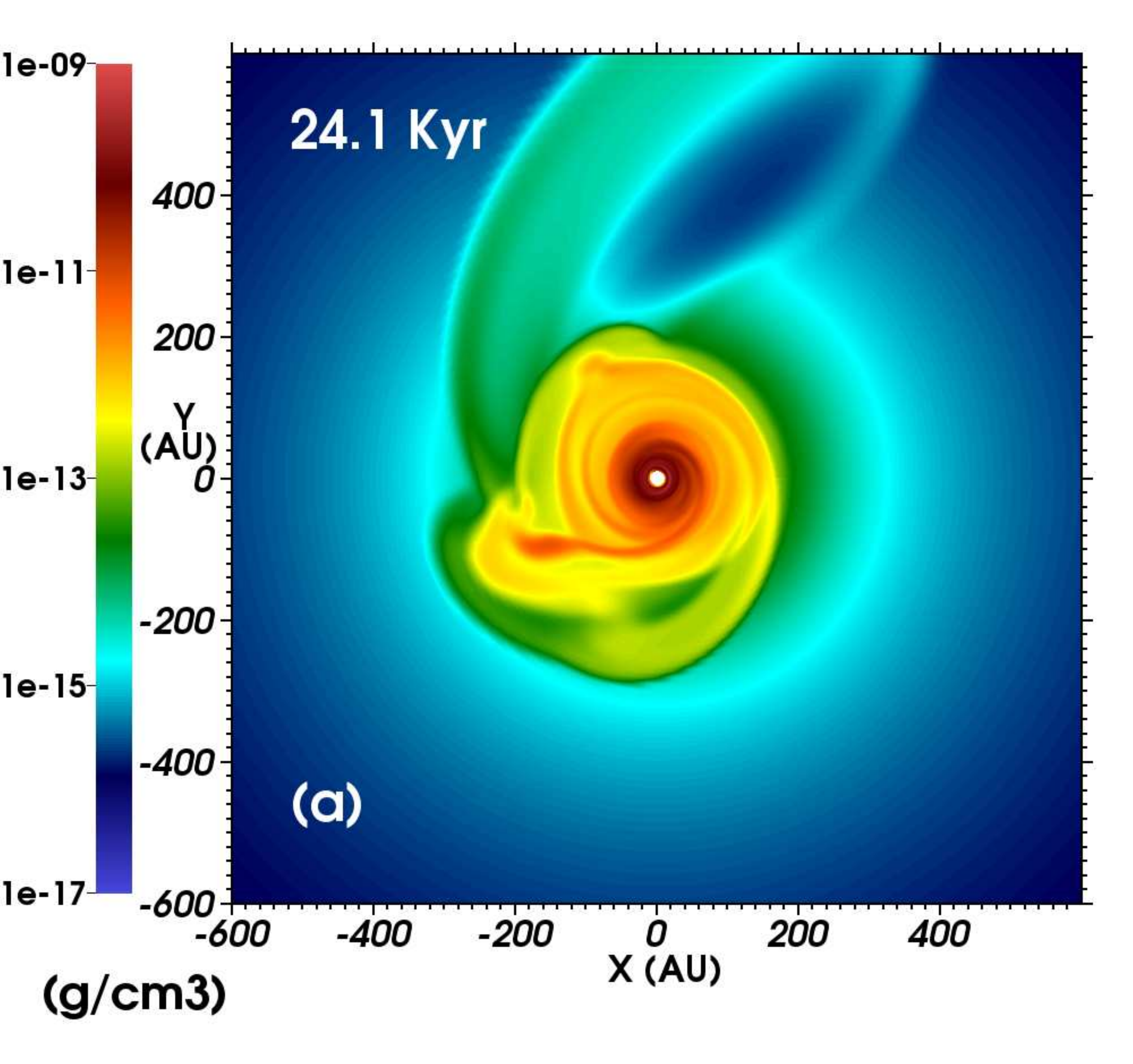}
	\end{minipage}	
	\begin{minipage}[b]{ 0.315\textwidth}
		\includegraphics[width=1.0\textwidth]{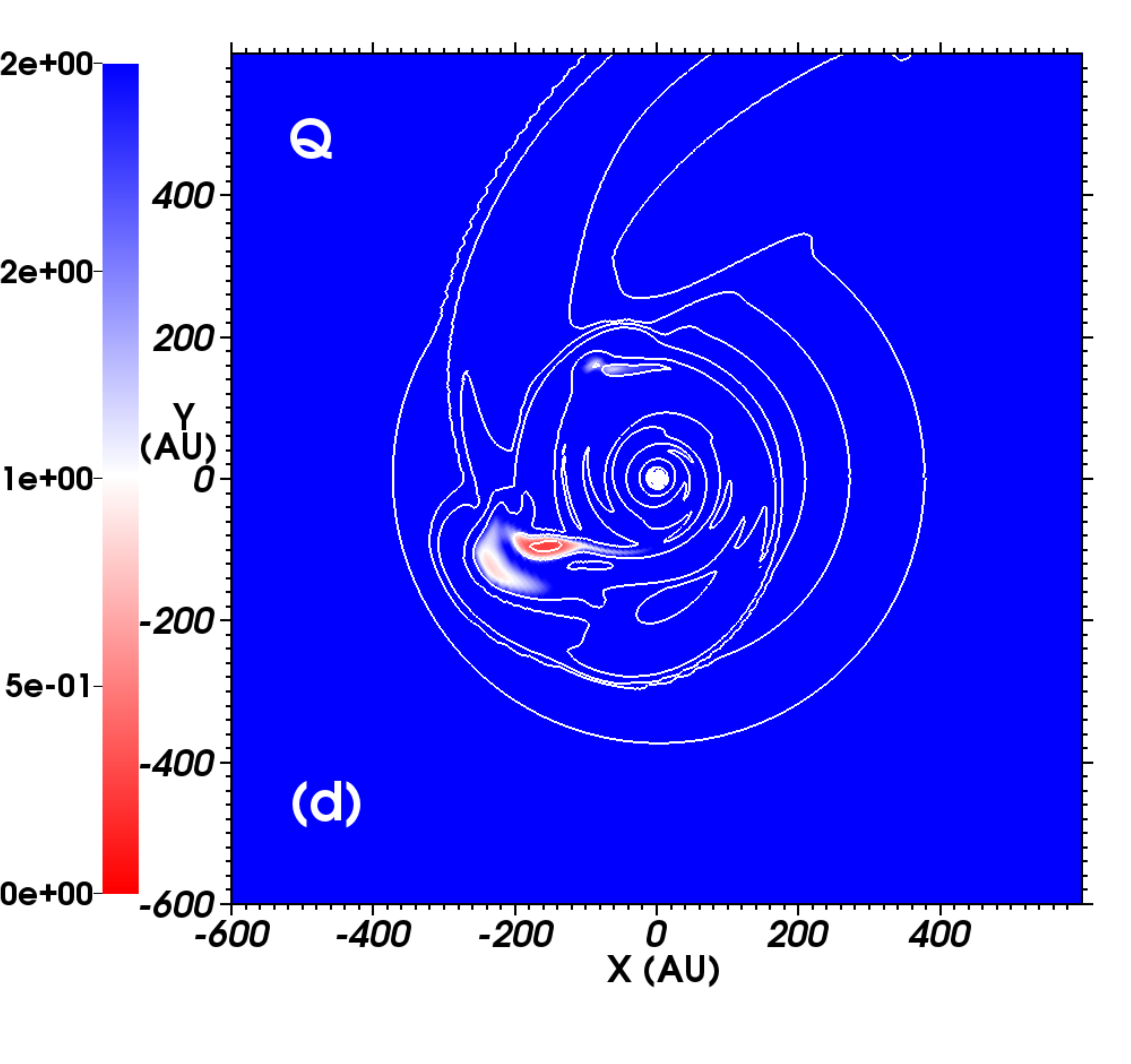}
	\end{minipage}	
	\begin{minipage}[b]{ 0.315\textwidth}
		\includegraphics[width=1.0\textwidth]{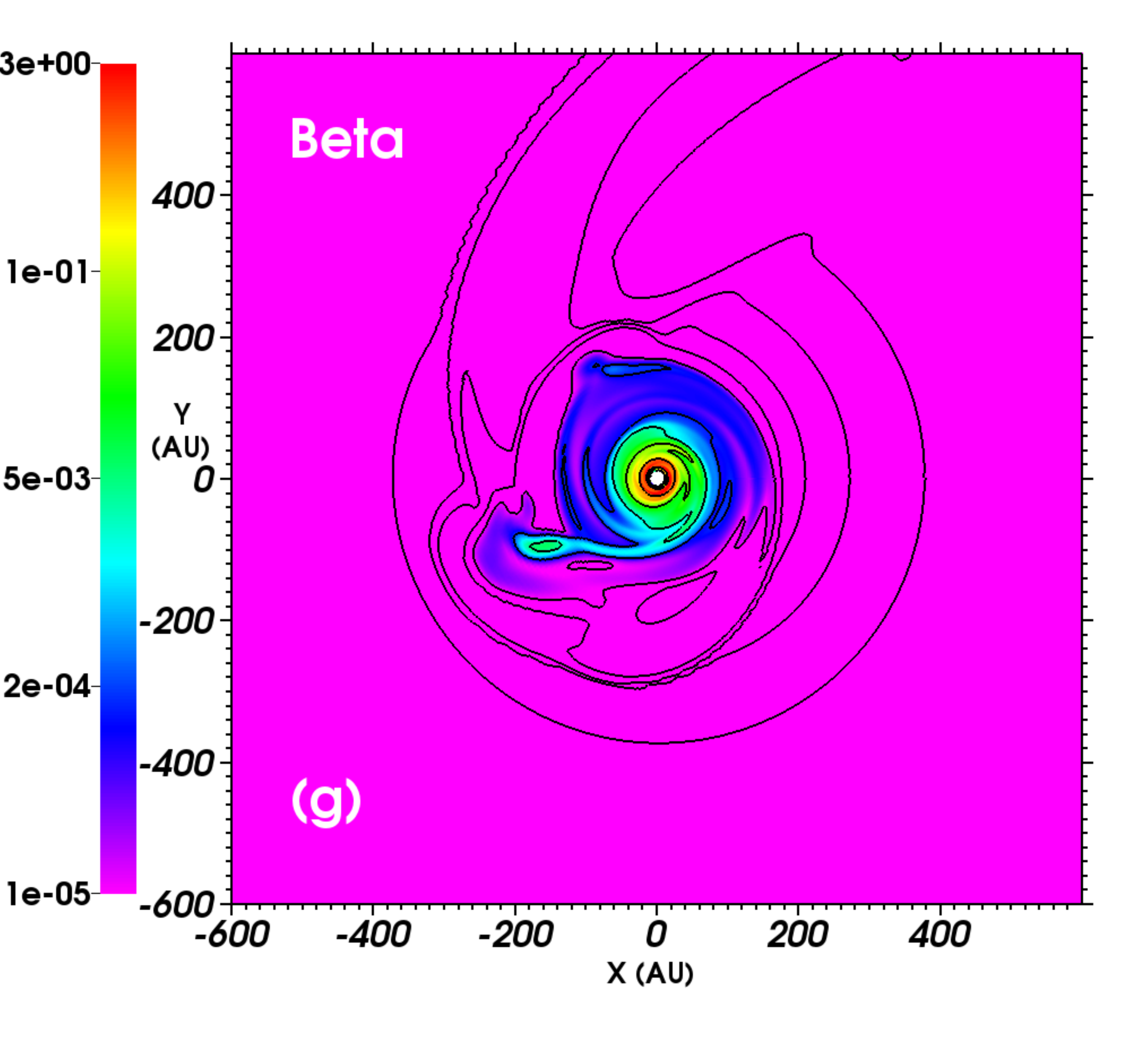}
	\end{minipage} \\
	\begin{minipage}[b]{ 0.315\textwidth}
		\includegraphics[width=1.0\textwidth]{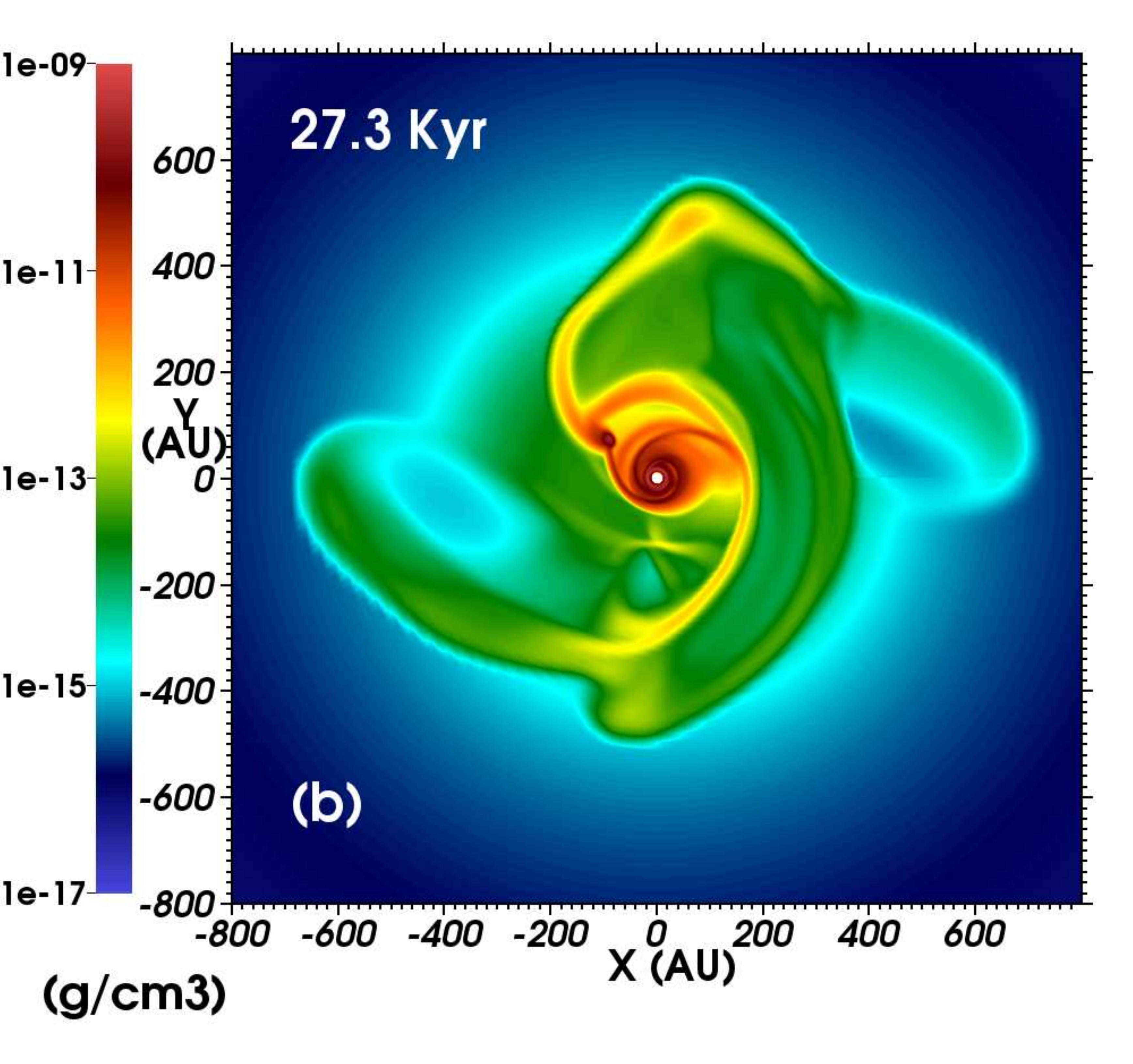}
	\end{minipage}
	\begin{minipage}[b]{ 0.315\textwidth}
		\includegraphics[width=1.0\textwidth]{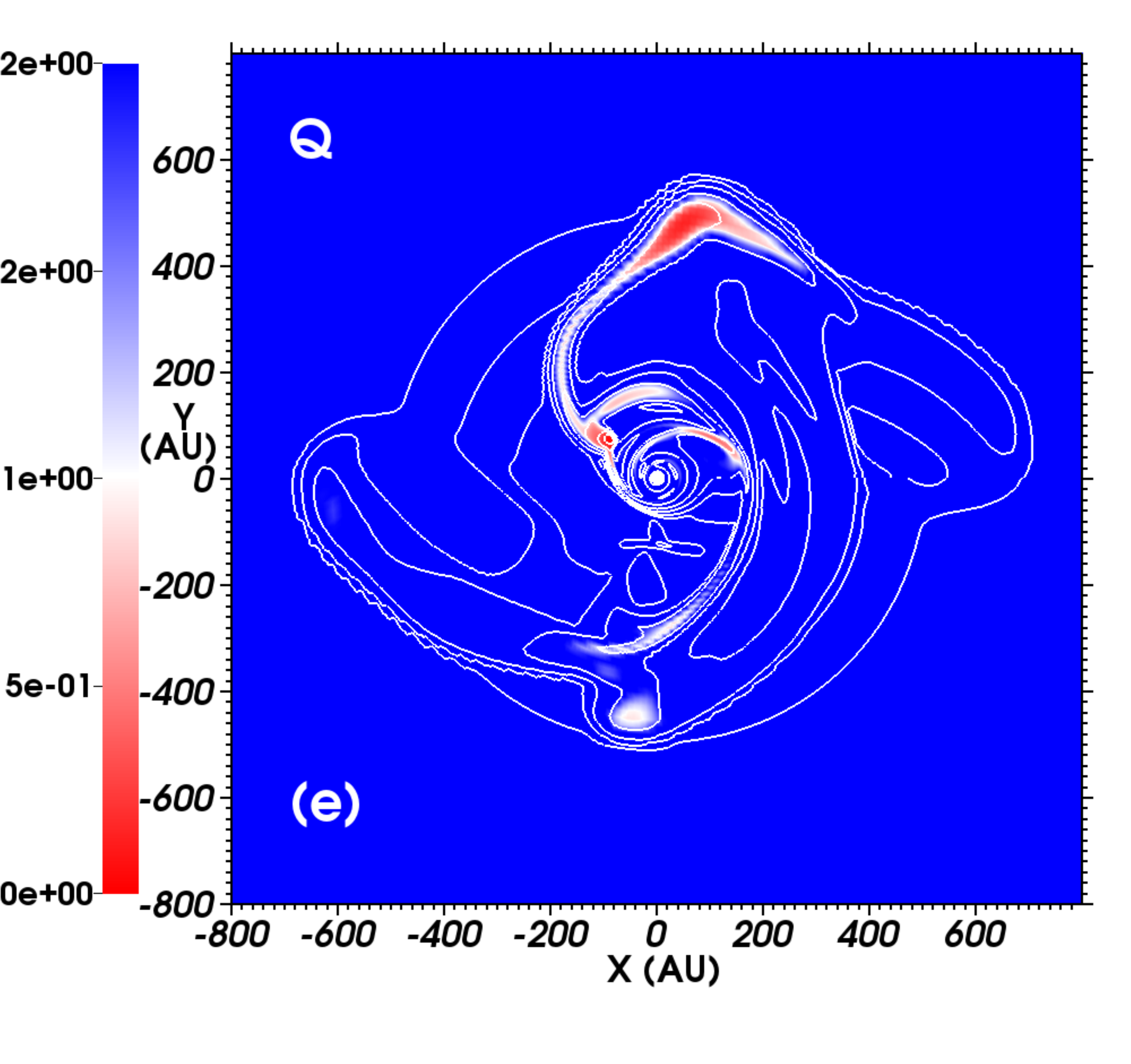}
	\end{minipage}
	\begin{minipage}[b]{ 0.315\textwidth}
		\includegraphics[width=1.0\textwidth]{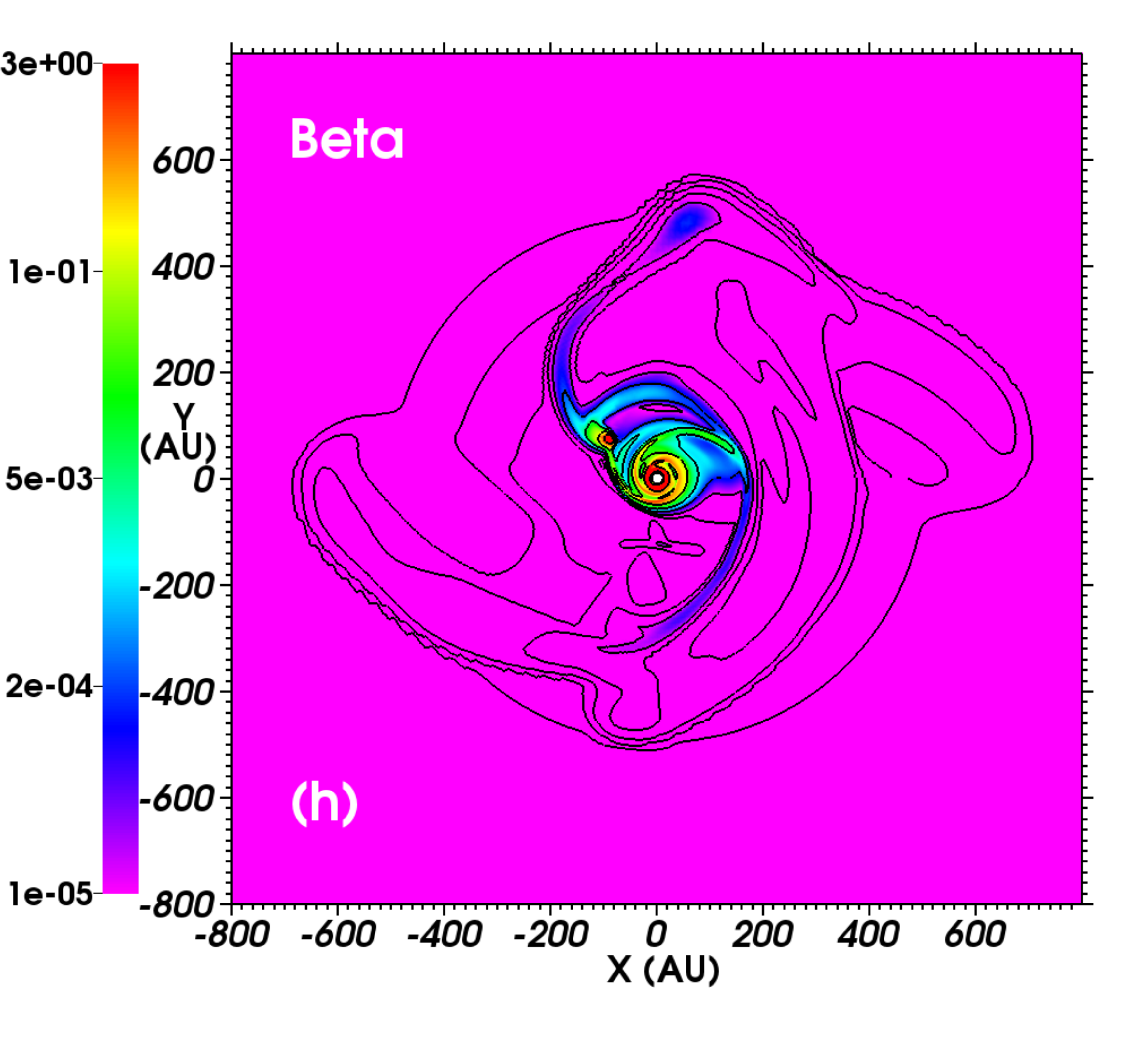}
	\end{minipage}	\\	
	\begin{minipage}[b]{ 0.315\textwidth}
		\includegraphics[width=1.0\textwidth]{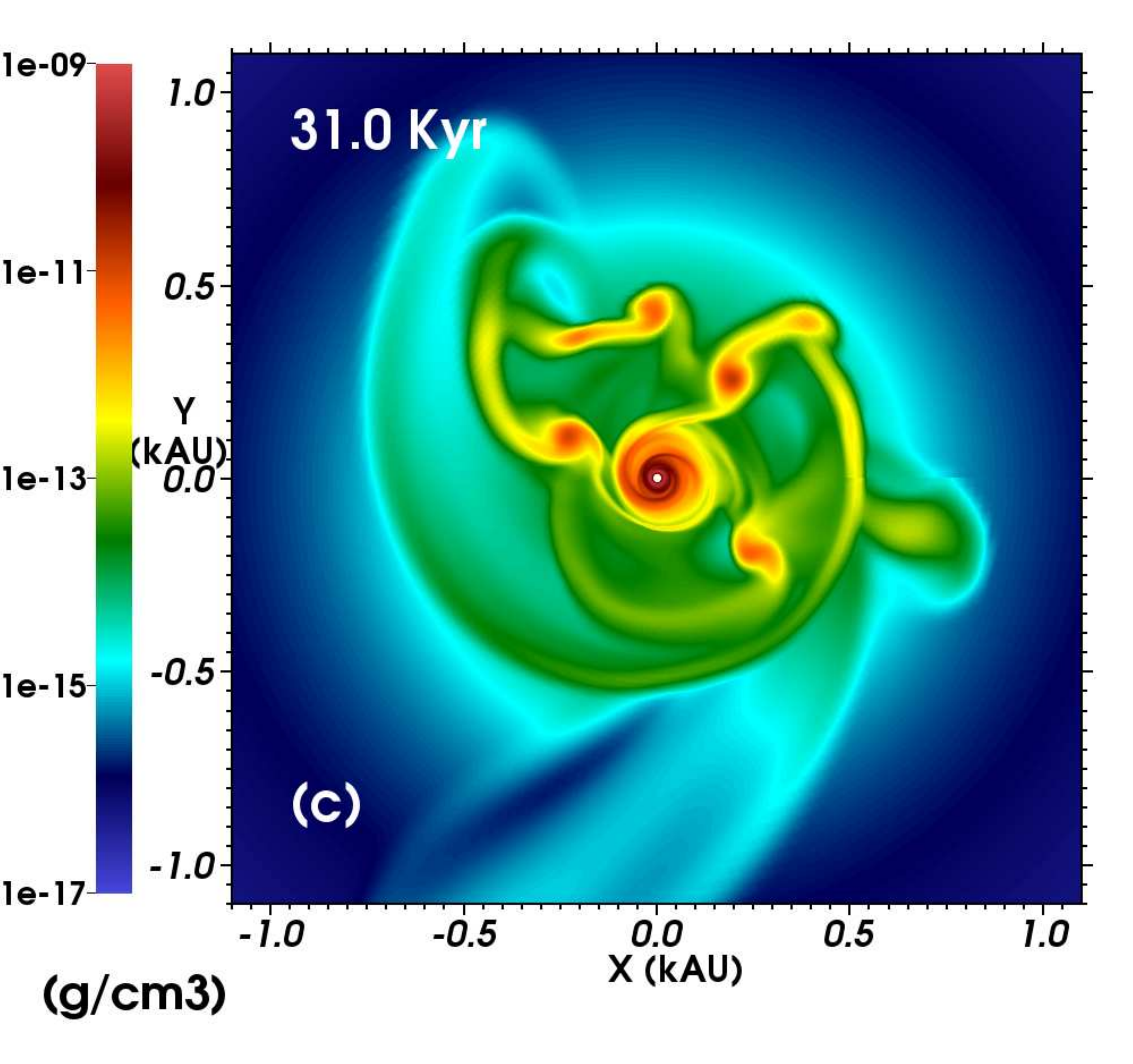}
	\end{minipage}	
	\begin{minipage}[b]{ 0.315\textwidth}
		\includegraphics[width=1.0\textwidth]{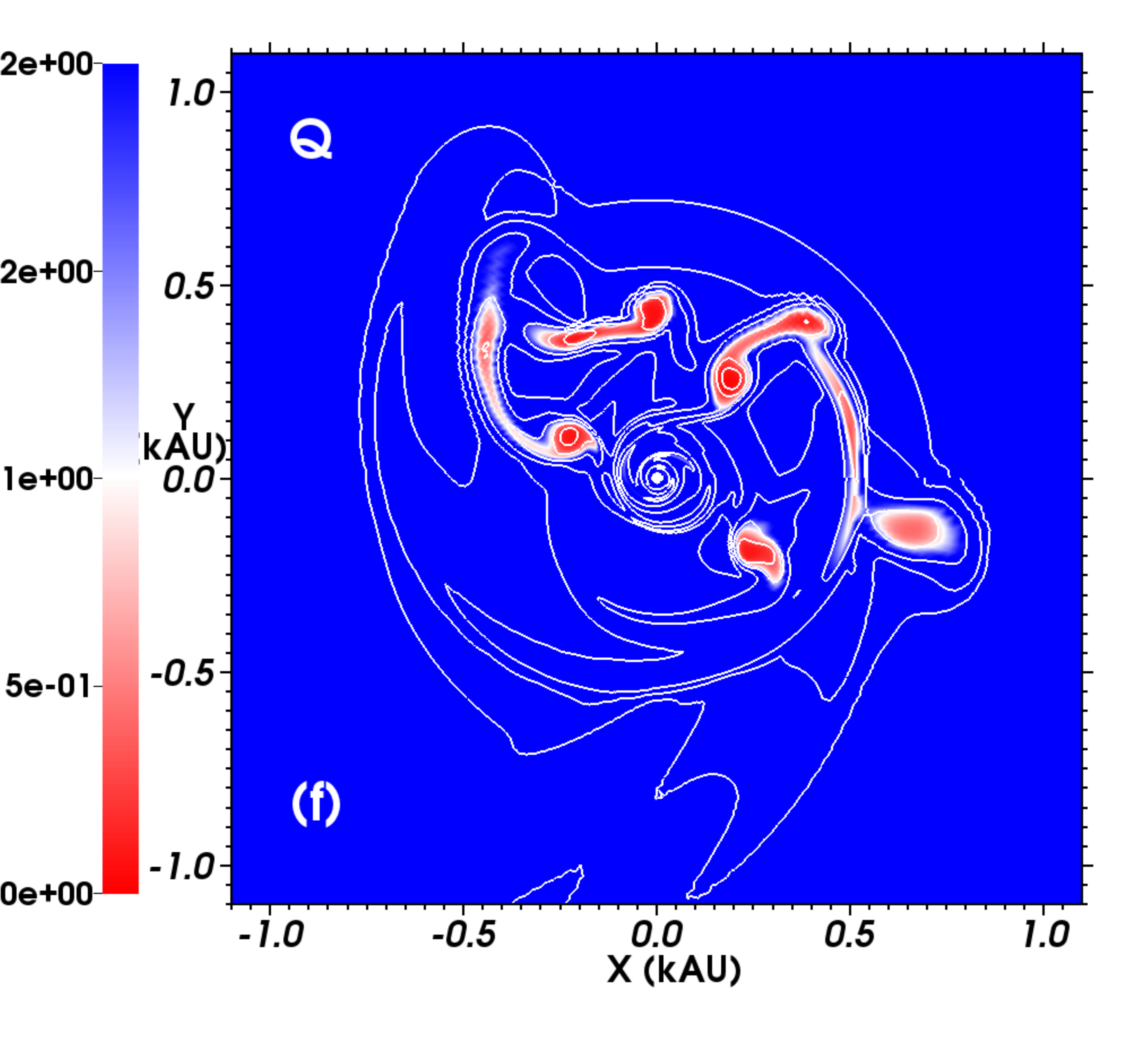}
	\end{minipage}	
	\begin{minipage}[b]{ 0.315\textwidth}
		\includegraphics[width=1.0\textwidth]{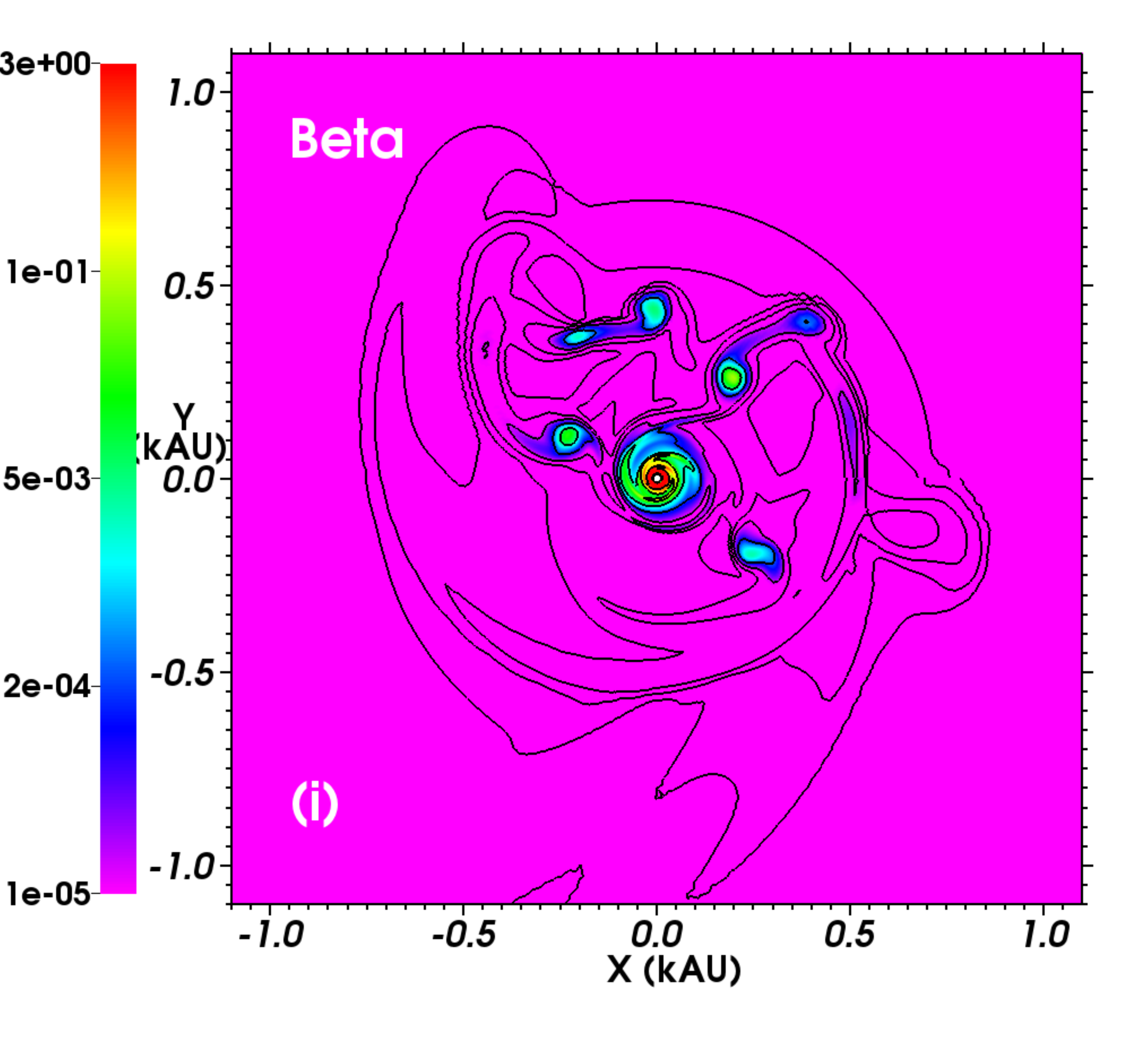}
	\end{minipage}  
	\caption{ 
		 \textcolor{black}{ 
		 Density fields of Run 1-HR (left, as in Fig.~\ref{fig:disc_evol1}a-c) together with 
		 corresponding Toomre $Q$-maps (middle) and 
		 Gammie $\beta$-maps (right).  
		 }
		 }	
	\label{fig:disc_overall_criteria}  
\end{figure*}

In the right column of panels of Fig.~\ref{fig:disc_overall_criteria} we show maps of 
the so-called Gammie criterion. It locally compares the effects of the gas cooling with 
the hydrodynamics, i.e. the heating of the gas pressure due to the shear due to the disc 
rotation~\citep{gammie_apj_462_1996}, and reads,  
\begin{equation}
     t_{\rm cool} \Omega \le \beta \approx 3-5,
   \label{eq:gammie}    
\end{equation}
\textcolor{black}{where the local cooling timescale $t_{\rm cool}$ depends on the disc thermodynamics. 
We compute it as,
\begin{equation}
     t_{\rm cool} = \frac{ E_{\rm int} }{ F }, 
   \label{eq:rafikov}    
\end{equation}                     
with $E_{\rm int}$ the internal energy in a disc column~\citep{klassen_apj_823_2016}
and $F$ the radiation flux leaving the disc surface. 
Our protostars are surrounded by discs that are not irradiated so that they adopt 
the vertical isothermal structure described in~\citet{rafikov_apj_704_2009}, and their midplane 
temperature still is governed by the pressure work of the spiral arms and the radiation flux 
leaving the disc can be estimated as $F=\sigma T^{4}/f(\tau)$ measured at the disc surface, where $\sigma$ is the Stefan-Boltzman constant and 
$f(\tau)=\tau+1/\tau$ is a function of the optical depth which interpolates between 
the optically thin ($\tau < 1$, outer part) and the optically thick ($\tau > 1$, 
inner part) regimes of the discs~\citep{rafikov_apj_662_2007,rafikov_apj_704_2009}.
The figure's color coding illustrates where the disc is $\beta$-stable ($t_{\rm cool} \Omega \ge 
3$$-$$5$, red) or $\beta$-unstable ($t_{\rm cool} \Omega < 3$$-$$5$, other colours).} This criterion 
has been numerically derived for self-gravitating discs around low-mass 
\textcolor{black}{stars} and it is subject to numerous discussions regarding the \textcolor{black}{critical}
value of $\beta$, which has been shown to be 
resolution-dependent~\citep{meru_mnras_411_2011,meru_mnras_427_2012} and 
equation-of-state-dependent~\citep{rice_mnras_364_2005, clarke_mnras_381_2007}. 
Therefore, the existence of an universal $\beta$--value remains 
unclear~\citep{lodato_mnras_413_2011, rice_mnras_420_2012, 
rice_mnras_438_2014}. 
The Gammie criterion can also be derived from the requirement that  the local cooling time 
$t_{\rm cool}$ be shorter than the fastest growth time
of gravitational instability $t_{\rm grav}=2\pi/(\Omega \sqrt{1-Q^2})$ \citep{shu_jbaa_102_1992},
so that the local density enhancements can get rid of excess heat generated during 
gravitational contraction. In this case,
\begin{equation}
t_{\rm cool} \Omega \le {2 \pi \over \sqrt{1-Q^2}}.
\end{equation}
This form of the Gammie criterion explains the recent results of \citet{boss_2017arXiv170104892B} arguing
that disc fragmentation can take place at higher values of $\beta$ given lower $Q$-values.

\textcolor{black}{
Figs.~\ref{fig:disc_overall_criteria}g-i show that our discs satisfy the Gammie criteriion
for instability, including the hot regions of spiral arms and gaseous clumps 
which are more $\beta$-stable than the interarm region, but nevertheless largely 
$\beta$-unstable. 
At the beginning of our simulations, the disc is mostly Gammie-unstable, testifying a rapid and efficient 
cooling of the gas, except in the very inner dense region of radius $\le\, 20$-$80\, \rm AU$, close 
to the protostar, see Fig.~\ref{fig:disc_overall_criteria}g,h and~\citet{klassen_apj_823_2016}. 
Throughout the simulations, the spiral arms have a typical value of $t_{\rm cool} \Omega = 0.01$$-$$0.005$  
(Figs.~\ref{fig:disc_overall_criteria}h,i) and even lower values ($t_{\rm cool} \Omega \ll 0.005$) 
in the interarm regions of the disc midplane, which means that the cooling timescale 
of the gas is shorter than the orbital period. 
Considering the both Toomre- and Gammie-unstable parts of the discs, we find that only the very innermost 
part of the disc does not satisfy the two criteria ($Q>1$ and $t_{\rm cool} \Omega \le $$3$-$5$). 
Our results therefore agree with the analysis of the stable accretion discs around massive stars 
of~\citet{klassen_apj_823_2016} who found a low $\beta$-value in their (clumpless) discs and 
concluded on (i) their capacity of undergoing future fragmentation and (ii) the non-sufficiency 
of the Gammie criterion to distinguish fragmenting from non-fragmenting regions of accretion 
discs around young massive stars. 
Note that our simulations are also in accordance with the predictions of Fig.~1 of~\citet{rafikov_apj_704_2009}, 
which concludes on the inevitable existence of a regime of disc fragmentation, located in regions at large 
distances from the central object in accretion discs with high accretion rates like ours, 
with $\dot{M} \ge 10^{-4}$-$10^{-3}\, \rm M_{\odot}\, \rm yr^{-1}$ and $\Omega \simeq 10^{-11}$-$10^{-10}\, \rm s^{-1}$. 
Section~\ref{sect:comparison} further discusses the usefulness of a Gammie criterion for discs around young massive stars. 
}

\subsection{ Hill criterion }
\label{sect:hill}

The so-called Hill criterion compares the effects of the 
self-gravity of a spiral arm with respect to the shear produced by the stellar 
tidal forces~\citep{roger_mnras_423_2012}. It has been derived in the context \textcolor{black}{planet 
embryos in protostellar discs} by examining its local sphere of influence, i.e. the 
capacity of segment of spiral arm to further accrete planetesimals and it has been 
applied with success in the context of the circumstellar medium of massive protostars~\citep{klassen_apj_823_2016}. 
A segment of a spiral arm of a given thickness $l$ is Hill-unstable if,
\begin{equation}
  \frac{l}{2 R_{\rm Hill}} < 1, 
\end{equation}
where,
\begin{equation}
    R_{\rm Hill} \approx \sqrt[\leftroot{-1}\uproot{2}\scriptstyle 3]{ \frac{G \Sigma l^{2} }{ 3 \Omega^{2} } }  
   \label{eq:hill}    
\end{equation}
is the Hill radius. The circumstellar gas that is not within twice the Hill radius of 
the fragment has its evolution governed by the gravity of the central protostar~\citep{roger_mnras_423_2012}. 
Note that \textcolor{black}{their} work considered an isolated disc that was not further accreting from its 
pre-stellar core. This criterion turned to be more general than the Toomre and 
Gammie criteria because it was consistent with the apparent stability of $Q$- and $\beta$- 
unstable discs~\citep{klassen_apj_823_2016}. 
We calculate the Hill number of a selected number of cross-sections of spiral arms in our Run~1-HR. 
\textcolor{black}{
The blobby arm of Fig.~\ref{fig:disc_overall_criteria}a gives $l=40\, \rm AU$, 
$\Sigma \approx 4.19\times 10^{3}\, \rm g\, \rm cm^{-2}$ and $\Omega=3.0\times 10^{-10}\, \rm s^{-1}$, respectively, 
which makes $R_{\rm Hill} \approx 48\, \rm AU$ and $l/2R_{\rm Hill} \approx 0.41 <1$. 
We repeat this analysis with the northern arm of Fig.~\ref{fig:disc_overall_criteria}b, 
the southern arm of Fig.~\ref{fig:disc_overall_criteria}b, the western arm of 
Fig.~\ref{fig:disc_overall_criteria}c and the eastern arm of Fig.~\ref{fig:disc_overall_criteria}c. 
With the values of $l=29, 14, 42, 39\, \rm AU$, 
$\Sigma \approx 3.47\times 10^{2}, 2.09\times 10^{2}, 1.57\times 10^{2}, 3.5\times 10^{2}\, \rm g\, \rm cm^{-2}$ 
and $\Omega=2.3\times 10^{-10}, 2.7\times 10^{-10}, 8.05\times 10^{-11}, 5.01\times 10^{-11}\, \rm s^{-1}$, respectively, 
and therefore $R_{\rm Hill} \approx 20.1, 9.43, 54.8, 49.6\, \rm AU$ 
and $l/2R_{\rm Hill} \approx 0.70, 0.74, 0.38$ and $0.39 <1$, respectively. 
}
Those numbers are in accordance with the unstable appearance of our discs and identical 
analysis at different times give similar results. 
We conclude that the Hill criterion is a reliable tool to predict 
the stability of self-gravitating discs \textcolor{black}{around young massive stars}.


\section{ Spectroscopic massive binaries formation by disc fragmentation }
\label{section:closebin}

\textcolor{black}{
Massive binaries are fundamental in the understanding of high-mass stellar evolution~\citep{sana_sci_337_2012}. 
The case of a massive proto-binary made of two young high-mass stars has been predicted~\citep{Krumholz_sci_2009} 
and observed~\citep{kraus_apj_835_2017}. However, the formation scenario of massive binaries made of a high-mass 
component and a close low-mass component is so-far unexplained, although it is suspected to be closely correlated to 
the hierarchical multiple systems of protostars formed by disc fragmentation~\citep{krumholz_apj_656_2007,rosen_mnras_463_2016}. 
This study explores such hypothesis in the context of the observed spectroscopic massive binaries~\citep{2013A&A...550A..27M}, a particular case 
of close massive binaries made of an O star with a short-period, low-mass companion which we explain by showing that 
disc fragmentation is a viable road for the formation of proto-O star with a short-period, low-mass companion. 
This section details that our simulations are consistent with the disc fragmentation channel of close/spectroscopic 
massive proto-binaries formation that evolve towards close/spectroscopic massive binaires once the most massive component 
enters the main-sequence phase of its evolution. Finally, we connect it with the phenomenon of accretion-driven outbursts 
in the high-mass regime of star formation. 
}

\subsection{Accretion-driven outbursts from young massive stars}
\label{sect:bursts}

\textcolor{black}{
Using the same numerical method as in this study, we demonstrated in~\citet{meyer_mnras_464_2017} 
that disc fragmentation around young high-mass star is followed by the formation of circumstellar 
clumps prone to migrate and fall onto the central protostar. 
}
We previously interpreted those 
observable as a tracer of disc fragmentation and postulated that the recent 
burst from the young high-mass star S255IR-SMA1~\citep{caratti_nature_2016} 
could be caused by the accretion of a circumstellar clump onto the protostellar 
surface. However, the fate of the clumps once they crossed the inner $10\, \rm AU$ 
was not calculated within our method and our conclusions \textcolor{black}{are} interpreted 
(i) by analogy with other studies on FU-Orionis(-like) ourbursts from young stars
and (ii) calculating the centrifugal radius of one of those circumstellar clumps. 
\textcolor{black}{We have drawn the conclusion} that accretion-driven outbursts \textcolor{black}{are} 
a general feature of massive star formation, already well-known in low-mass star formation as the 
FU-Orionis phenomenon~\citep{smith_mnras_424_2012,vorobyov_apj_805_2015} and in 
primordial star formation~\citep{hosokawa_2015}.

\textcolor{black}{
This study extends the work of~\citet{meyer_mnras_464_2017} to different angular velocity 
distributions of initial collapsing pre-stellar cores. Our set of simulations includes the case of a 
\textcolor{black}{core in solid-body rotation} that has recently been associated, with the help of several fragmentation criteria, 
as possessing all \textcolor{black}{characteristics required for disc fragmentation, despite the clear lack} of fragments \textcolor{black}{captured} in 
the hydrodynamical model~\citep[see][\textcolor{black}{and discussion on sink-particle algorithms in Section~\ref{sect:sink}}]{klassen_apj_823_2016}. 
Differences between simulations of~\citet{klassen_apj_823_2016} and our models mainly lies in that we avoid 
using sub-grid models (sink particles, see Section~\ref{sect:sink}) and a higher spatial resolution than 
previously allowed by mesh-refinement codes in the close environment of the young high-mass protostar. 
Our models show fragmenting accretion discs \textcolor{black}{whose} evolution results in a pattern of clumps and filaments. 
This allows us to stress \textcolor{black}{similarities} between our results and simulations devoted 
to \textcolor{black}{low-mass and primordial} star formation.  
The remaining open questions consist in knowing more in detail the number of circumstellar clumps formed in 
massive discs and the exact subset of them migrating down to the central protostar. Other explanations \textcolor{black}{of the} 
formation of circumstellar clumps are plausible, such as the gravitational attraction of gaseous clumps formed 
within the pre-stellar core or by fragmentation of a neighbouring accretion disc, e.g. during the formation 
of two massive binaries, and subsequently ejected away by the gravitational sling effect~\citep{vorobyov_aa_590_2016}. 
In this paper, we continue the interpretation of the clumps' fate. They can lose their envelope and produce 
accretion-driven events once the material reaches the stellar surface, however, if the clump continues to 
collapse and does not fall directly to the protostar, it will end up as a close, low-mass companion. 
An example of a clump having such properties is detailed below. 
}

\subsection{Forming close/spectroscopic massive binaries by clump migration}
\label{sect:bin}

\begin{figure}
	\centering
	\begin{minipage}[b]{ 0.45\textwidth}
		\includegraphics[width=1.0\textwidth]{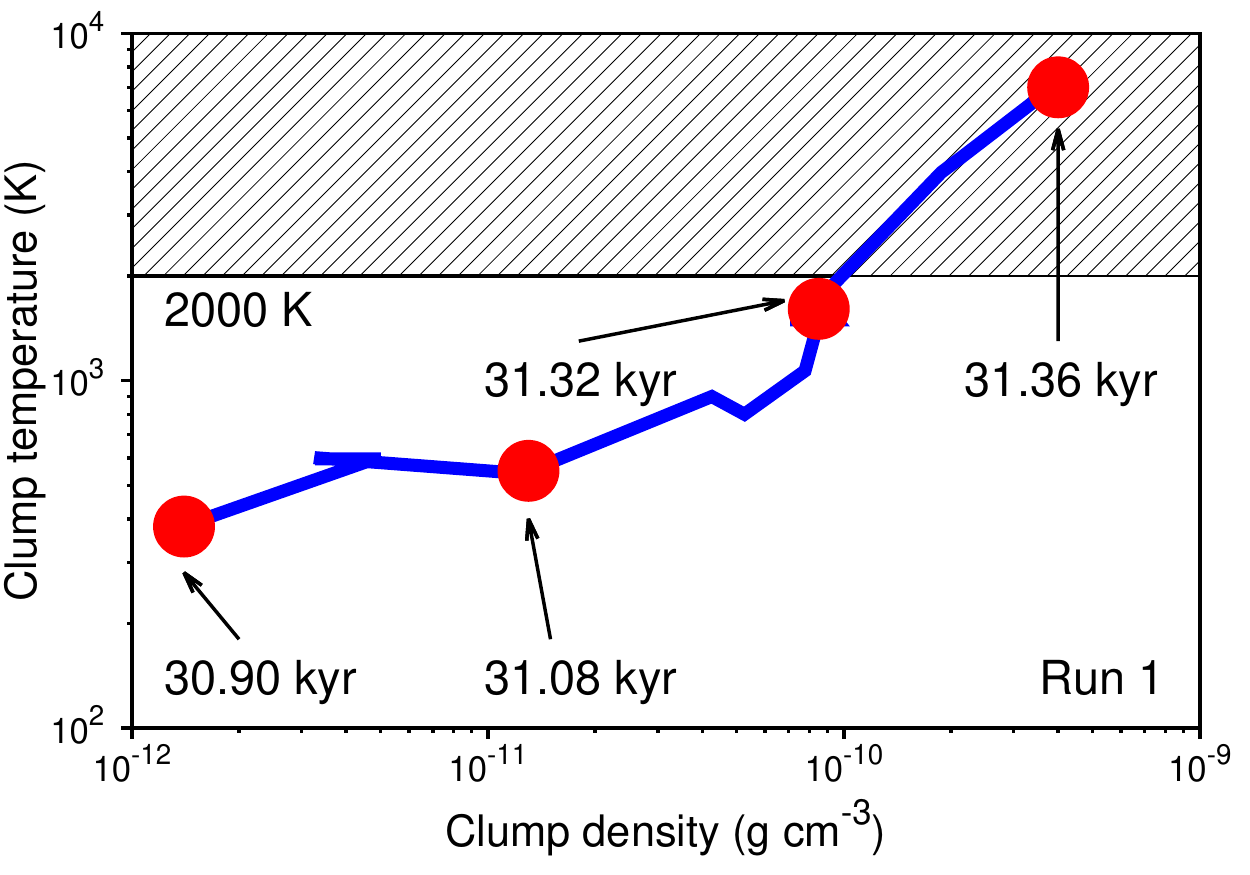}
	\end{minipage}		
	\caption{ 
		 Temperature-density evolution of the circumstellar clump respesponsible 
		 for the first ourburst of our Run~1, showing the characteristics of a 
		 gaseous blob on the path to low-mass star formation. The hatched region 
		 fall under \textcolor{black}{dissociation}, not included in our simulations.
		 }	
	\label{fig:blob2}  
\end{figure}

Fig.~\ref{fig:blob2} \textcolor{black}{shows} the evolution of the densest grid cell of the clump 
responsible for the burst happening in Run~1 at $31.4\, \rm kyr$. 
We follow its evolution between the times \textcolor{black}{$30.9$} and $31.4\, \rm kyr$. 
At $31.32\, \rm kyr$, the core of the clump reaches a density of 
$\rho \approx 8.6\, \times 10^{-11}\, \rm g\, \rm cm^{-3}$ and 
$T \approx 1724\, \rm K$. About 24 years later, those numbers are 
$\rho \approx 4.13\, \times 10^{-10}\, \rm g\, \rm cm^{-3}$ and a temperature 
$T \approx 4638\, \gg 2000\, \rm K$ because no dissociation 
is included into our equation of state. \textcolor{black}{This is sufficient to consider 
this clump as to be on \textcolor{black}{its} path to star formation, indicating that our simulations 
are consistent with the formation scenario of a massive binary by disc fragmentation.   
The central protostar is therefore in an embryonic 
\textcolor{black}{binary system of a current separation of 10 AU and below} when the hot disc 
fragment disappear into the sink cell, which is much smaller than found in previous 
calculations~\citep[a few $100\, \rm AU$$-$$1000$, see][]{krumholz_apj_656_2007}.
}

Furthermore, given the clump's properties and assuming angular momentum 
conservation, we estimate at a time $31.36\, \rm kyr$ its centrifugal radius, 
i.e. the Keplerian orbit at which it falls \textcolor{black}{\citep[see analysis in][]{meyer_mnras_464_2017}}. The clump of 
$M_{\rm c} \approx 1.2\, \rm M_{\odot}$ is located at a radius 
of $r=38\, \rm AU$ from the protostar, has an azimuthal velocity of 
$v_{\phi}\approx 1.8\times 10^{6}\, \rm cm\, \rm s^{-1}$ and the protostellar 
mass is $M_{\star} \approx 16.36\, \rm M_{\odot}$ with a disc mass in the region 
$\le 38\, \rm AU$ of $M_{\rm d} \approx 3.16\, \rm M_{\odot}$. \textcolor{black}{We 
find that, in the midplane, $R_{\rm c} = (rv_{\phi})^{2}/(G(M_{\star}+M_{\rm d})) \approx 27.01\, \rm AU$, 
and 20 years later at a time $31.38\, \rm kyr$, we find 
$R_{\rm c} \approx 4.64\, \rm AU \approx 996\, \rm R_{\odot} \gg\, R_{\star} \approx 100-10\, \rm R_{\odot}$.} 
\textcolor{black}{
Under our assumptions, it shows that this clump does not directly land onto the protostar as in~\citet{meyer_mnras_464_2017}.  
The value of $R_{\rm c}$ calculated just before the clump enters the sink cell corresponds to a 
Keplerian period of $P\approx 842\, \rm d$ which is of the order to the period range observed 
in some massive binaries~\citep{2014ApJS..213...34K}. 
As discussed in~\citet{meyer_mnras_464_2017}, this value of $R_{\rm c}$ is an upper limit because 
of the clump's angular momentum loss during its inward migration. Therefore, the corresponding 
system will be tighter and result in a so-called short-period binary 
($P \sim 10-100\rm d$). \textcolor{black}{This naturally evolves towards a massive spectroscopic binary 
($P \sim 1-10\rm d$)~\citep{2014ApJS..213...34K}, when the central protostar leaves the pre-main-sequence 
phase and becomes an O-type star. }
%
}

\subsection{ Accretion-driven outbursts from high-mass star-forming regions as tracer of close/spectroscopic massive binary formation}
\label{sect:tracer}

Since envelope and core of the clumps may separate during the 
\textcolor{black}{infall due to tidal stripping}, both phenomenon -- 
accretion-driven outbursts and close binary formation -- may happen together, 
and the bursts would be, in this case, a tracer of close/spectroscopic binary formation. 
A follow-up study will statistically investigate the properties of the accretion 
spikes in \textcolor{black}{our} simulations, as 
in~\citet{vorobyov_apj_805_2015}. \textcolor{black}{Although,} our study demonstrates its consistency 
within our assumptions, we nevertheless can not give the definitive proof that 
close binaries can form by the disc fragmentation in the surroundings of the 
protostar, since the migration of circumstellar clumps can only be followed up 
to $r_{\rm in}=10\, \rm AU\gg \rm R_{\star}$. 
Discriminating the fate of the 
clumps between close binaries and accretion-driven outbursts is difficult since 
it requires multi-scale simulations resolving at the same time the disc 
fragmentation at \textcolor{black}{$\sim 100\, \rm AU$} and the fall of clumps 
down to $\sim~1000\, \rm R_{\odot}$ for a close/spectroscopic binary, and/or eventually to 
$\sim 100\, \rm R_{\odot}$ for an accretion burst. Such simulations are far 
beyond the scope of this work.

\section{Discussion}
\label{section:discussion}

This section is devoted to a general \textcolor{black}{discussion} of our results. We review 
the limitation of our setup, especially in terms of microphysical processes that we may 
include into future studies in order to tend towards a more realistic description 
of the surroundings of massive protostars. \textcolor{black}{We briefly compare our results with 
previous studies, compare them with results obtained with sink particles methods and we 
discuss what could influence the fragmentation of our accretion discs.} 
\textcolor{black}{Finally, we predict what our results would look like when observed.}

\subsection{Limitation of the model}
\label{sect:microphys}

Despite of reaching sub-AU resolution within the inner region of the accretion discs, our simulations 
still suffer from a lack of spatial resolution. 
Increasing the grid resolution would allow us to \textcolor{black}{calculate in more detail}, 
e.g. the internal physics of the migrating circumstellar clumps. 
%
%
This problem equally affects codes with static grids or with mesh-refinement techniques such as 
{\sc flash}~\citep{benerjee_apj_660_2007}, {\sc orion}~\citep{krumholz_apj_656_2007} or 
{\sc ramses}~\citep{commercon_aa_510_2010}.  
\textcolor{black}{
Although chosen in accordance with current numerical studies on disc fragmentation in star formation, 
several other assumptions of our method could be subject to further improvements. In addition to the 
treatment of the radiation itself, it mainly concerns: 
}

\begin{enumerate}

\item \textcolor{black}{ 
The utilised prescription to estimate the accretion luminosity. 
In~\citet{meyer_mnras_464_2017}, the accretion luminosity is assumed to be a fraction $f=1$ of the gravitational 
potential energy of the material crossing the inner boundary. This fraction is finally converted into stellar 
radiation instead of being absorbed and mixed to the stellar interiors. 
Motivated by observations,~\citet{offner_703_apj_2009} constrained this value to $f=3/4$, that,   
in our case, will make our accretion spikes less intense. This will change our estimate of the increase in 
accretion luminosity during the bursts and make our results even more consistent with the observations 
of~\citet{caratti_nature_2016}, strengthing their interpretation as a manisfestation of the burst mode of massive star formation. Complex 
star-disc interaction simulations calculating in detail the accretion and penetration of disc material 
into the interior of the protostar in the vein of~\citet{1996ApJ...461..933K} are required to determine $f$ more accurately. 
}

\item \textcolor{black}{ 
The initial shape of the pre-stellar core and/or the choice of perturbation seeds for the gravitational 
instability. We initialize the simulation with spherically-symmetric pre-stellar cores, however, observations 
\textcolor{black}{have} revealed their non-isotropic, filamentary shapes. Particularly, the study on massive star formation 
of~\citet{banerjee_mnras_373_2006} uses initial conditions taken as a turbulent core structure extracted from 
a previously-calculated cluster-forming clump. 
The considered symmetry breaking of the gravitational collapse in our case is numerical~\citep[cf.][]{hosokawa_2015}, 
while many studies initially sparse the density field of the pre-stellar core with a particular seed 
mode~\citep[m=2, see][]{benerjee_apj_660_2007} that is azimuthally imposed on the top of the density profile. 
}

\item \textcolor{black}{
The stellar motion. The gravitational influence of the disc onto the star is neglected in our setup in which the 
star is static. The effect of the disc inertia onto the protostar becomes important when discs develop heavy and/or 
extended spiral arms displacing the barycenter of the star/disc system at radii larger that the size of the sink cell, 
which in its turn influences the stellar motion. 
This is intrinsically taken into account in simulations utilizing moving sink particles as representations of forming stars 
and it will be included in future studies with the method described in~\citet{hosokawa_2015} and~\citet{regaly_aa_601_2017}. 
This is a well-known inconvenience of simulations of accretion discs with static central object. Also note that 
numerical simulations with grid-refinement \textcolor{black}{methods} do not strictly conserve angular momentum and suffer from grid-alignment effects. 
However, since the models of~\citet{klassen_apj_823_2016} do not fragment to the same degree as ours, we conclude that 
the preponderant issue for disc fragmentation is spatial resolution. 
Including the stellar motion should make the \textcolor{black}{discs have net angular} momentum that is lower than in the case 
without stellar motion. The disc will be of smaller radius and them less prone to fragmentation. 
However, this effect can easily be counterbalanced by a mild increase of the initial rotation rate 
in the parental cloud, as shown in~\citet{regaly_aa_601_2017}. 
}

\item \textcolor{black}{
Simulation time.  
The numerical cost of simulations of massive stars like ours is huge and currently obliges investigations to be restricted 
to the early formation phase. However, since long-term evolution simulation of low-mass star formation~\citep{matsumoto_apj_839_2017} 
highlighted that, under some circumstances, discs modify their shape and orientation, one can wonder whether totally new disc 
morphologies will arise around more evolved young high-mass stars.
}

\item \textcolor{black}{
Missing physical mechanisms.
Apart from covering the parameter space which is still far from being fully explored in 
terms of initial pre-stellar core mass $M_{\rm c}$, size $R_{\rm c}$ and ratio of kinetic per gravitational energy 
$\beta$, neglected microphysical processes may also change our results. 
Particularly, the ionization feedback of the protostar filling the bipolar outflows which develops perpendicular 
to the discs and giving birth to bipolar \hii regions that can be observed~\citep{campbell_apj_282_1984}, 
may experience intermittency~\citep{2017arXiv170607444T} as in primordial star formation~\citep[see][]{hosokawa_2015}. 
Thus, the inclusion of the magnetization of the pre-stellar core is also necessary to tend towards a global picture 
of star formation because it influences at the same time the jet morphology~\citep{pudritz_prpl_277P_2007,frank_prpl_451_2014} 
and the fragmentation of accretion discs~\citep{hennebelle_apj_830_2016,fontani_aa_593_2016}, see Section~\ref{sect:fragmentation}.  
}

\item \textcolor{black}{
The midplane-symmetric nature of our simulations. This can be improved by considering the full three-dimensional 
evolution of the protostellar disc. 
We force the disc to develop symmetrically with respect to the plane normal to the rotation axis, see Fig.~\ref{fig:cylinder}. 
Full "4$\pi$" models would allow us to take into account the vertical bending of the disc visible, for example in Fig.~13 
of~\citet{krumholz_apj_656_2007}. 
Note also, that, if our simulations with static stars do not allow us to appreciate the effects of the disc wobbling, 
models with Cartesian grids \textcolor{black}{cannot} qualitatively pronounce on the absence of stellar motion on the disc dynamics. 
For all those reasons, the assumptions of midplane-symmetry of our simulations are 
reasonable and our method does not overestimate fragmentation from this point of view.
What causes disc bending modes and how that can affect disc fragmentation deserves a separate focused study. 
%
%
}

%
%

\item 
\textcolor{black}{
The size of the sink cell. 
Indeed, our method neglects the fate of the clumps once they cross the innermost boundary of 
$10\, \rm AU$~\citep[see also][]{meyer_mnras_464_2017}, which has so far not 
been modeled up to the boundary layer of the disc. 
The presence in the close environment of protostars of more \textcolor{black}{(smaller)} clumps 
in chaotic and mutual interaction would change the geometry of the 
accretion flow onto the protostar, that are influenced by complex radiation processes, 
such as the line-ablation mechanisms acting in the surroundings of O/Be 
stars~\citep{kee_mnras_458_2016}. 
However, studies devoted to low-mass star formation demonstrated that while about 
half of the clumps are gravitationally supported and remain in the disc, the falling 
clumps may be ejected from the disc or further fragment within the innermost disc 
regions~\citep{greif_mnras_424_2012,meru_mnras_454_2015,vorobyov_aa_590_2016}. This may be enhanced 
by $\rm H_2$ cooling and collisional emission, not treated in our simulations. 
}

\end{enumerate}

However, such improvements would only be possible at the 
cost of even more computationally-intensive simulations 
\textcolor{black}{requiring higher spatial resolution and huge memory resources. 
Particularly, this will give us access} 
to the probable chaotic behaviour of the circumstellar clumps in the inner 
region of the disc~\citep[see discussion in][]{meyer_mnras_464_2017}.

\subsection{General discussion}
\label{sect:comparison}

The study of low-mass star formation \textcolor{black}{has led to a large literature on its theory}, but the 
three-dimensional modelling of the birth of high-mass stars is still a relatively 
recent field of research. A few works only tackled the problem and their 
differences principally lie in the way to treat the stellar feedback and/or in the 
choice of pre-stellar core initial configuration. Most simulations 
assume a density distribution $\rho(r) \propto r^{-3/2}$ but models such as 
in~\citet{kuiper_apj_732_2011} and~\citet{kuiper_apj_763_2013} \textcolor{black}{additionally} consider 
$\rho(r) \propto r^{-2}$. The other fundamental parameter of the pre-stellar core 
is its initial velocity field. While some studies considered non-rotating 
pre-stellar \textcolor{black}{environments} in which turbulence was initially 
driven~\citep{krumholz_apj_656_2007}, non-solid body rotating pre-stellar cores 
have begun to be investigated in~\citet{meyer_mnras_464_2017} with 
$\Omega(R)\propto R^{-0.75}$ and our work continues this with $\Omega(R)\propto 
R^{-0.35}$. The initial mass of pre-stellar cores is typically taken to $M_{\rm 
c} \approx 100\, \rm M_{\odot}$ and the ratio of rotational-to-gravitational 
energy to $\beta\approx$ few $\%$ as in our study, but models have been calculated 
with $M_{\rm c}=1000\, \rm M_{\odot}$~\citep{peters_apj_711_2010} and $\beta \ge 10 
\%$~\citep{seifried_mnras_417_2011,klassen_apj_823_2016}. A more detailed review 
of the initial conditions of the pre-stellar cores in terms of $M_{\rm c}$ and 
$\beta$-ratio can be \textcolor{black}{found} in~\citet{meyer_mnras_464_2017}.

\textcolor{black}{The manner of treatment of radiation feedback differs greatly from study to study}. 
Pre-calculated cooling curves are less accurate \textcolor{black}{but allow the coverage of 
large parameter spaces}~\citep{benerjee_apj_660_2007,seifried_mnras_417_2011}, e.g. including the 
magnetization of turbulent pre-stellar core\textcolor{black}{s}~\citep{seifried_mnras_432_2013},
\textcolor{black}{whereas radiation transport}
algorithms taking  into account the physics of 
dust are able to self-consistently model the dust 
sublimation front~\citep{kuiper_apj_722_2010}. \textcolor{black}{Our method is a modification of the {\sc pluto} 
code~\citep{mignone_apj_170_2007,migmone_apjs_198_2012} described 
in~\citet{kuiper_aa_511_2010,kuiper_apj_732_2011}}. Recently, a similar scheme 
has been implemented into the {\sc flash} and {\sc orion} codes, using a 
mesh-refinement grid~\citep{klassen_apj_797_2014,rosen_jcp_330_2017}. 
\textcolor{black}{In that sense, our study has the advantage to have a highly-resolved grid 
in the inner parts of the disc together with a radiative transfer solver more accurate than a 
simple cooling law, but it has the disadvantage not to properly resolve clumps potentially forming at 
larger radii. Although the frequency-dependent ray-tracing method has been successfully benchmarked with respect 
to our utilised grey solver~\citep{kuiper_aa_555_2013}, its requirement in terms of numerical resources was too high. 
However, radiation transport methods making use of a frequency-dependent ray-tracing solver have been developed and 
used in the context of three-dimensional massive star formation simulations~\citep{kuiper_apj_732_2011,rosen_mnras_463_2016}.  
Particularly, the effects of the interplay between such radiation transport methods and grid resolution 
on the modelling of (Rayleigh-Taylor unstable) bipolar outflows from massive stars is largely 
discussed in~\citet{kuiper_aa_537_2012}, in~\citet{rosen_mnras_463_2016} and in~\citet{harries_2017}. 
}

\textcolor{black}{
Another similarity between our simulations and recent studies is their consistency 
with respect to analytical criteria predicting disc fragmentation. 
Numerical solutions of the surroundings of high-mass protostars is typically compared to 
the Toomre, Gammie and Hill criteria. As for other kind of self-gravitating discs, the 
Toomre criterion characterise well fragmenting 
\textcolor{black}{regions~\citep[$Q \le 0.6$, see][]{takahashi_mnras_458_2016,klassen_apj_823_2016}} whereas the Gammie 
analysis alone is in most regions fulfilled and therefore not sufficient to discriminate between 
fragmenting and non-fragmenting scenario, \textcolor{black}{e.g. at the early stage of the disc formation}. 
Additionally, accordance between numerical simulations and the Hill analytic prediction were found. It 
applies to our \textcolor{black}{unstable} models as it did in the disc simulations of~\citet{klassen_apj_823_2016} that 
are stable to fragmentation, but unstable to development of spiral modes. Consequently, this criterion is the most reliable 
tool so far to characterize the stability of spiral arms in \textcolor{black}{discs} from young stars. 
}

Fragmentation as a response of accretion discs against gravitational instability 
has been widely investigated analytically and numerically in different astrophysical contexts. Particularly, the 
study of~\citet{kratter_mnras_373_2006} considers the \textcolor{black}{antagonistic} effects of both 
angular momentum transport, leading discs to fragmentation and viscous heating tending  
to stabilize them. Under those assumptions which "underestimate the prevalence 
of disc fragmentation", they found that discs \textcolor{black}{around} high-mass stars ($\ge 10\, 
\rm M_{\odot}$) are prone to fragmentation and that this probability increases with 
$M_{\star}$. Our Run~1 with solid-body rotation is consistent with such 
prediction since the disc fragments at a time $\approx 30\, \rm kyr$ when the 
protostar has reached $\approx 17\, \rm M_{\odot}$, see 
Fig.~\ref{fig:disc_evol1}c and Fig.~\ref{fig:disc_properties1}a). However, 
our disc models initially deviating from rigid rotation fragment before 
the protostar reaches such mass, e.g. at a 
time $\approx 30\, \rm kyr$ when the central protostar is only $\approx 7\, \rm 
M_{\odot}$ (Run~2, see Fig.~\ref{fig:disc_evol2}b and 
Fig.~\ref{fig:disc_properties1}b). 
Finally, we recall that the episodic feedback of secondary objects have been \textcolor{black}{shown} to 
play a role in further enhancing disc fragmentation shortly after accretion 
driven-outbursts~\citep{mercer_2016arXiv161008248M}. This process may apply to 
our discs \textcolor{black}{around} massive protostars and further enhance their 
fragmentation \textcolor{black}{(see Section~\ref{sect:sink})}.

\subsection{ Self-consistent versus sink particle simulations }
\label{sect:sink}


\textcolor{black}{
The sink particle approach consists in coupling a discrete N-body-like method to a continuous resolution 
of the hydrodynamics equations in a grid-based code. Therefore, it is a violation of the self-consistent treatment 
of the different physical processes at work in the computational domain. 
Such approach is nevertheless unavoidable, since low-mass stars such as brown dwarfs or 
solar-type stars appear as point-mass objects, once they undergo second collapse from first 
hydrostatic cores down to stellar densities. 
All the subtlety lies in (i) applying sufficiently strong sink-creation criteria in order to 
avoid fictious particle formation leading to overestimating the number of stars and (ii) 
introducing them as late as possible during the secondary gravitational collapse, ideally 
when the grid size is of the order of the radius of the new-born star, not to artificially 
influence the local protostellar disc dynamics. Indeed, 
introducing sinks modifies the gas dynamics, itself a function of the resolution of the 
simulation at the moment of the particle creation. 
}

\textcolor{black}{
Criteria allowing codes to generate sinks greatly vary between numerical schemes and simulations. 
A too sophisticated but very realistic particle-creation algorithm will rarefy secondary star formation, 
leading to the conclusion of disc stability, despite of a $Q$- and $\beta$-unstable disc~\citep{klassen_apj_823_2016}, 
whereas a simplified one will produce over-numerous and unphysical star formation. 
Indeed,~\citet{rosen_mnras_463_2016} notes that ''multiplicity results are sensitive to the 
physics included, radiative transfer treatment used, and sink creation and merging criteria employed". 
Moreover, choosing a (set of) criterio(n-a) for sink particles creation is equivalent to a preference 
for particular star formation laws, still largely under debate. 
Particularly,~\citet{machida_mnras_438_2014} showed with non-turbulent, (non-)ideal magnetohydrodynamics 
simulations of rotating Bonnor$-$Ebert spheres of initial mass ranging from $1.0$ to $100\, \rm M_{\odot}$, 
\textcolor{black}{how sensitive disc solutions around low-and high-mass stars can be} 
as a function of both the accretion radius of sink particles and the density threshold constituting 
their principal creation condition~\citep[see also discussion in][]{padoan_apj_730_2011}. 
They concluded on the necessity of sink accretion radius $\le1.0\, \rm AU$ and a sink 
particule density threshold $\ge 10^{13}\, \rm cm^{-3} \simeq 10^{-11}\, \rm g\, \rm cm^{-3}$. 
The properties of the \textcolor{black}{companion} core in Fig.~\ref{fig:blob2} is consistent with this assertion, however, 
one can clearly see that our clump becomes a star when its core is denser than 
$10^{-10}\, \rm g\, \rm cm^{-3}$, revealing (at least in this case) the insufficiency of the criterion 
of~\citet{machida_mnras_438_2014} since our gaseous clump is not yet hot enough to be considered as a 
secondary star. Perhaps a more reliable criterion for sink particle creation could be the temperature 
of the clump's core?
Also pointed by~\citet{hennebelle_aa_528_2011}, the "use of sinks may alter significantly the 
evolution of the calculations in particular in the presence of magnetic field", which is an 
additional drawback of numerical models like in~\citet{myers_apj_766_2013,myers_mnras_439_2014}. 
}

\textcolor{black}{
Our method gets rid of those caveats since the local collapse is self-consistently captured in 
the density and temperature fields. 
%
We justify the use of a spherical coordinate system by our scientific goal, which is to understand the 
formation of spectroscopic companions to massive OB stars, by exploring star formation and migration in 
the inner part of the disc. The main caveat of our disc dynamics is therefore the lack of 
spatial resolution in their outer region. However, as discussed in Section~\ref{section:numerics}, 
this is acceptable since the migrating clumps develop in the inner and intermediate region of the protostellar disc. 
Our alternative method to simulations with Cartesian grids and sink particles self-consistently produce discs 
fragmenting into clumps hot enough to verify the condition of secondary star formation, which is a confirmation 
of the discovery of multiplicity in the massive protostellar context in~\citet{krumholz_apj_656_2007}. 
With both methods, the number of companions forming in the disc decreases with the consideration of 
the pre-stellar core's magnetization~\citep{machida_mnras_438_2014,commercon_apj_742_2011}, see also 
Section~\ref{sect:fragmentation}. 
%
%
No formation of twin massive binaries~\citep{Krumholz_sci_2009} is observed directly in our models, although Run~3 
shows the formation of a massive disc-like clump (Fig.~\ref{fig:disc_evol3}).   
%
}

\textcolor{black}{
In the case of models with "smart" particles possessing their own feedback, high numerical 
resolution is needed to resolve the accretion discs around the formed secondary 
stars, in which the complex flow induces accretion rate variability and \textcolor{black}{makes} the feedback of the 
secondary stars episodic by the mechanisms largely described in~\citet{vorobyov_apj_719_2010} 
and~\citet{Bae_apj_795_2014}. This will subsequently affect the results by further favouring violent 
fragmentation of the discs~\citep[see][]{mercer_2016arXiv161008248M}. 
%
%
Even more realistic sink particles should take into account jet-launching 
mechanisms perturbing the disc dynamics by creating disc cavities or outflows, and/or, 
in the case of, e.g. intermediate-mass secondary star formation~\citep{2017arXiv170604657R}. 
Located beyond the dust sublimation radius, their \hii region will intermittently release 
ionized gas into the radiation-shielded part of the central protostar's disc and enrich it with 
momentum and energy, perturbing its internal thermodynamics. 
}

%
%

\subsection{What can we expect on disc fragmentation by improving on the limitations of the models?}
\label{sect:fragmentation}

\textcolor{black}{
With the help of a grid more resolved in the stellar close environment than ever 
before,~\citet{meyer_mnras_464_2017} revealed the fragmenting character of circumstellar discs.
}
\textcolor{black}{
Since our models showed the clump-forming nature of discs around massive stars, one may ask, whether 
additional physical effects, not resolved in our simulations may arise and change the global picture of our solutions? 
As an example, increasing the spatial resolution in low-mass star formation simulations enhanced the 
possibility of circumstellar clumps to be ejected out of their host accretion disc~\citep{vorobyov_aa_590_2016}.  
High-resolution studies in the context of low-mass and primordial star formation have also shown that further 
fragmentation may take place in the very innermost disc regions modelled by our static sink cell. Those 
phenomenons are provoked by migrating falling fragments or by $\rm H_{2}$ cooling and/or by collisionally 
induced emission, see~\citet{stacy_mnras_403_2010,greif_mnras_424_2012} but also~\citet{meru_mnras_454_2015}, 
however, they did not suppress physical conclusions that were previously drawn in their respective context. 
Therefore, the same may apply in the massive protostellar context. 
}

\textcolor{black}{
Higher resolution simulations may highlight other detailed features in the fragmentation dynamics
such as very inner accretion disc fragmentation occurs during massive star formation. This will therefore 
result in a pattern of very small gaseous clumps that can merge together, orbit on to the central protostar 
or be dynamically ejected away by gravitational sling, revealing amongst other clump-clump interactions as 
those discussed in detail in~\citet{zhu_apj_746_2012}.
Although this would change the geometry of the accretion and affect the detailed variability of the protostellar luminosity, 
the trajectory heavy clumps such ours that migrate fast, hosting the birth of \textcolor{black}{a secondary star}, will not be greatly perturbed. They will either fall directly 
onto the stellar surface (accretion burst), orbit around it (close binary formation) or lose their envelope 
while reaching a Keplerian orbit around the central protostar (close binary formation accompanied by accretion-driven event). 
One should also keep in mind that such criticisms are also valid regardless the used coordinate system, 
the grid refinement criteria and/or the utilised sink particles creation algorithms. Cartesian simulations 
of massive star formation like in, e.g.~\citet{krumholz_apj_656_2007} or~\citet{klassen_apj_823_2016}, but allowing more 
grid refinement levels and/or comparing different sink-particle algorithms are therefore highly desirable. 
}

{

\textcolor{black}{ 
The above discussion refers to the aggravation of fragmentation once it has been triggered in the core, 
however, some physical processes can directly influence the fragmentation by modifying the disc temperature. 
Two other processes can indeed, under some circumstances, directly control whether fragmentation itself \textcolor{black}{occurs}, 
namely the presence of magnetic fields and/or the initial turbulence of the pre-stellar core. 
Often neglected for the sake of simplicity, the magnetization of the gravitationally collapsing structures can 
modify the disc dynamics. This has been reported at different length scales, spanning from giant molecular clouds 
to low-mass star-producing cores~\citep{commercon_aa_510_2010,tomida_apj_714_2010}. 
The effects of the pre-stellar core magnetic properties on delaying, promoting, or preventing disc fragmentation, 
together with a description of the variety of the forming different low-mass binaries systems are discussed 
in~\citet{tomida_apj_801_2015}. 
Particularly, the interplay between the magnetic breaking of the infalling gas and the radiative 
feedback of the protostar can inhibit the initial fragmentation processes of collapsing pre-stellar 
cores. Note that this applies to the regime of a strong magnetic field strength~\citep{commercon_apj_742_2011} 
under which our results do not fall as we neglect magnetization.
}

\textcolor{black}{ 
These magnetized models in the study of~\citet{commercon_apj_742_2011} propose a channel to generate isolated O stars. 
However, it also underlines that two secondary fragments "associated with a relatively high Jeans mass reservoir" form, 
and consequently one could expect "this early fragmented system [to] give rise to a close massive binary system". 
Such conclusions were beforehand only drawn by Lagrangian particles simulations of~\citet{bonnell_mnras_362_2005} 
and by the hydrodynamical disc fragmentation model of~\citet{krumholz_sci_323_2009}. Although using two totally 
different methods, they both demonstrate the possibility of producing twin massive binaries by pre-stellar core 
gravitational collapse followed by disc fragmentation, under the assumption of a neglected magnetization of the gas. 
%
%
Finally, note that since star-forming regions are dense, the non-ideal character of magneto-hydrodynamical processes 
such as Ohmic dissipation, Hall effect and ambipolar diffusion can, as an additional effect, influence the gas dynamics. 
Their significance on the physics of interstellar filaments~\citep{ntormousi_aa_589_2016} and on low-mass star 
formation~\citep{masson_apjs_201_2012,tomida_apj_801_2015,masson_aa_587_2016,marchand_aa_592_2016} implies that 
they, no doubt, should impact high-mass star formation as well. 
}

\textcolor{black}{
Turbulence is another key ingredient in the theoretical determination of star formation 
rates~\citep{hennebelle_743_apj_2011}. In the case of the gravitational collapse of a 
single, isolated pre-stellar cloud, turbulence modifies the gravitational collapse 
itself but also directly affects its magnetization. 
Particularly, it has been suggested that the so-called magnetic breaking catastrophe is a natural 
consequence of the over-idealized initial conditions of low-mass star formation simulations, 
and that native pre-stellar turbulence naturally leads to Keplerian disc formation~\citep{seifried_mnras_423_2012},  
although the work of~\citet{wurster_mnras_457_2016} brings a non-ideal but non-turbulent solution to this problem. 
Additionally, turbulence strongly influences the pre-stellar core magnetic field coherence~\citep{seifried_mnras_446_2015}. 
The study of~\citet{rosen_mnras_463_2016} includes simulations with stellar radiation and turbulent initial conditions 
which evolve towards complete disorganization of the disc-bubble system in the protostellar surroundings. As Keplerian 
discs~\citep{johnston_apj_813_2015} and clear disc-outflow structures~\citep{zinchenko_apj_810_2015} have been observed 
around young high-mass stars, those models indicate that other feedback processes stabilizing the pressure-driven 
radiative bubble such as outflow launching mechanisms~\citep{kuiper_apj_800_2015,kuiper_apj_832_2016} must be at 
work efficiently during the formation of massive stars. 
\textcolor{black}{Since} sub-AU-resolved, three-dimensional gravitation-radiation-magnetohydrodynamic, turbulent simulations of high-mass 
star formation have not been conducted yet, the question is therefore to know by how much fragmentation in the context 
of our pre-stellar cores would be affected if considered with both (non-ideal) magnetization and turbulence. 
Those questions will addressed in a subsequent study. 
}
%
%

\subsection{Prediction of observed emission}
\label{sect:observations}

Although observational evidences \textcolor{black}{for} bipolar \hii regions \textcolor{black}{~\citep{francohernandez_apj_604_2004}} and accretion 
flows~\citep{keto_apj_637_2006} around young high-mass stars became more 
numerous over the past years, the direct imaging of their accretion discs was 
until recently under debate~\citep{beuther_aa_543_2012}, mostly because of 
the high opacity of the parent environment in which they 
form \textcolor{black}{and large distances to these objects}~\citep{zinnecker_araa_45_2007}. Recent observations from the 
{\it ALMA} interferometer suggested the presence of a Keplerian accretion disc 
in the surroundings of the early O-type star AFGL 
4176~\citep{johnston_apj_813_2015}. Similar techniques revealed the presence of 
a clumpy gaseous \textcolor{black}{disc-like structure} around the young high-mass stellar object of the S255IR 
area~\citep{zinchenko_apj_810_2015} which is associated to a FU Orionis-like 
outburst~\citep{burns_mnras_460_2016,caratti_nature_2016}. An equivalent discovery around the 
young early massive star G11.92$−$0.61 MM1 has been reported using the velocity 
gradient of compact molecular line emission oriented perpendicularly to a 
bipolar molecular outflow in~\citet{ilee_mnras_462_2016} and the number of young massive 
candidates surrounded by a self-gravitating disc 
increases~\citep{forgan_mnras_463_2016}. Several ongoing 
observational campaigns currently aim at detecting signs of disc fragmentation, 
e.g. with the {\it ALMA} facility \textcolor{black}{(Cesaroni et
al. 2017, in press)}.

To appreciate our results in the context of observations, we \textcolor{black}{perform a} dust 
continuum radiative transfer calculation with the code 
{\sc radmc-3d}\footnote{ http://www.ita.uni-heidelberg.de/~dullemond/software/radmc-3d/}~\citep{
dullemond_2012} using a~\citet{laor_apj_402_1993} dust mixture based of 
Silicates crystals \textcolor{black}{with} density and temperature distributions are directly imported 
from our simulations.  
Fig.~\ref{fig:kj_discs} shows a synthetic 1.2mm dust continuum emission of our 
Run 1 at a time $\approx 40\, \rm kyr$, \textcolor{black}{viewed under an inclination angle of $30\degree$}. 
The protostellar mass is $\approx 25\, \rm M_{\odot}$
which corresponds to AFGL 4176's constrained mass, however, the disc mass is 
$\approx 30\, \rm M_{\odot}$ which is larger that the $12\, \rm M_{\odot}$ used to 
\textcolor{black}{fit} the properties of AFGL 4176. The radius of the disc of $\approx 2\, \rm kAU$ is 
consistent with the observations. We assumed \textcolor{black}{the} protostar to be a black body 
source of effective temperature directly given \textcolor{black}{by the} stellar evolution routine 
of the simulation $T_{\rm eff} \approx 37500\, \rm K$ and of bolometric 
luminosity $L \approx 85500\, \rm L_{\odot}$ from which photon 
packages are ray-traced through the accretion disc. The calculations are further 
post-processed with the {\sc simobserve} and {\sc clean} tasks of the {\it Common Astronomy 
Software Applications}\footnote{ https://casa.nrao.edu/}~\citep[{\it 
CASA,}][]{McMullin_aspc_376_2007} to produce synthetic {\it ALMA} observations 
of our \textcolor{black}{disc as if located at the coordinates of AFGL 4176}. 
\textcolor{black}{Telescope} parameters are choosen to be the configuration 4.9 for Cycle 4 {\it ALMA} 
observations, with a bandwidth of $1.8\, \rm  GHz$ with $\approx\, 3$ hours of integration time.  
\textcolor{black}{
The central frequency of the combined continuum emission is 249.827 GHz (1.2 mm) and
imaging was carried out using Briggs weighting with a robust parameter
of 0.5.
}
Under those assumptions and parameters, our {\it ALMA} rendering of Run~1 \textcolor{black}{that is shown in 
Fig.~\ref{fig:kj_discs}} has observable 
emission. It presents the typical structure 
of an accretion disc around with spiral arms and clumpy substructures in it. A more 
detailed analysis of the molecular surroundings of this model, including line emission, 
is \textcolor{black}{left} for future work (Johnston et al., in prep).

\begin{figure}
	\centering
	\begin{minipage}[b]{ 0.49\textwidth}
		\includegraphics[width=1.0\textwidth]{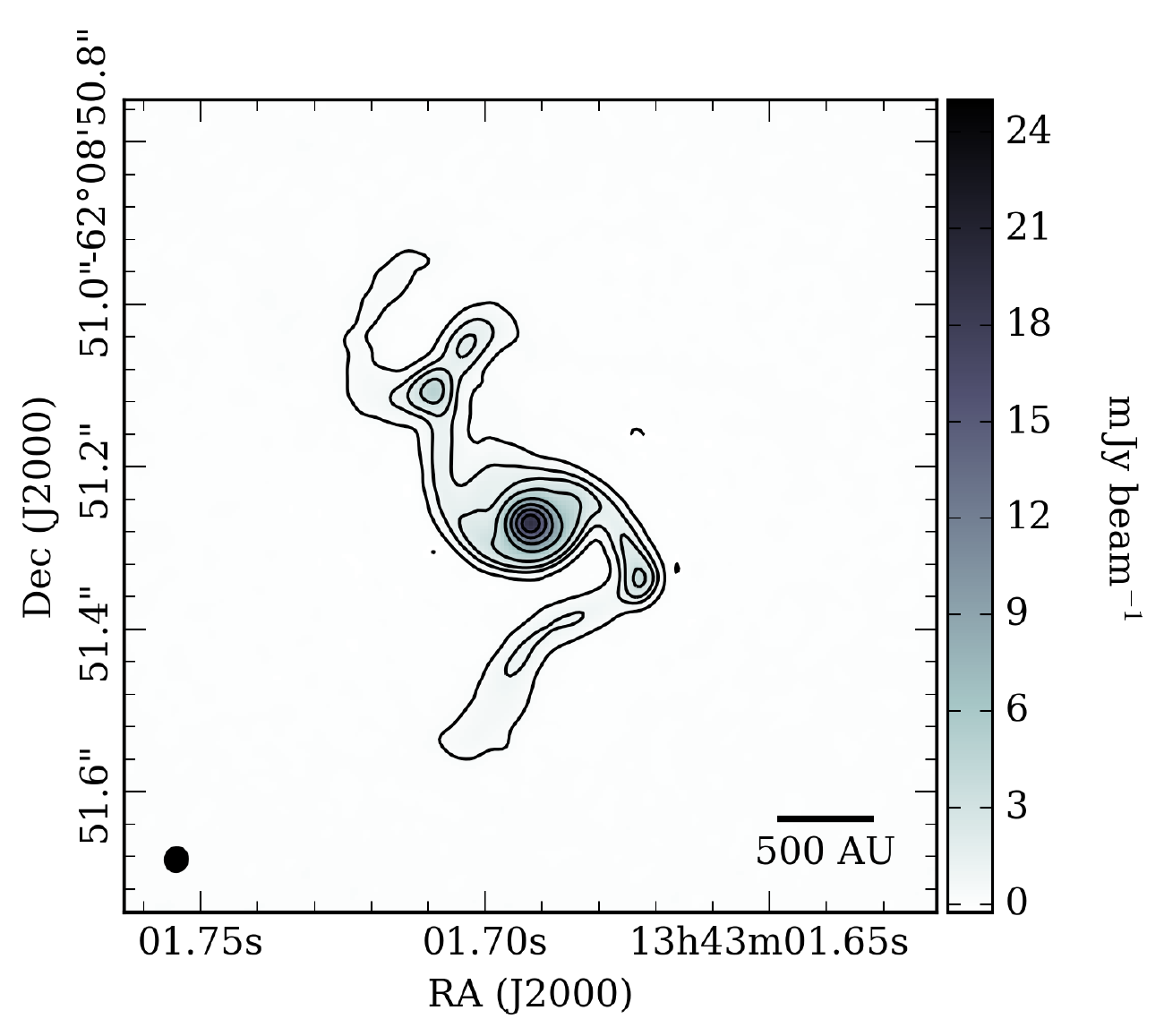}
	\end{minipage} 	
	\caption{ 
	         \textcolor{black}{
		 Dust continuum {\it ALMA } prediction of our Run 1 in the context of the young high-mass stars 
		 AFGL 4176~\citep{johnston_apj_813_2015}. We consider a snapshot from the model at a time $40.0\, \rm kyr$, 
		 when the protostar has reached $24.5\, \rm M_{\odot}$. The overall disc 
		 and its brightest circumstellar clumps have a radius $2\, \rm kAU$
		 and $\approx 0.6-1.0\, \rm kAU$ from the protostar, respectively. 
		 Image intensity is in mJy/beam. 
		 }
		 }	
	\label{fig:kj_discs}  
\end{figure}


\section{Conclusion}
\label{section:cc}

We have run three-dimensional \textcolor{black}{gravito-radiation-hydro} simulations of the collapse of 
$100\, \rm M_{\odot}$ pre-stellar cores rotating with a kinetic-to-gravitational 
energy ratio of $\beta=4\%$ and \textcolor{black}{with different} initial conditions in terms of 
initial angular momentum distribution, \textcolor{black}{in the context of a non-magnetized disc generated 
by an initially non-turbulent pre-stellar core}. 
We perform the simulations during at \textcolor{black}{least 
$\approx 35\, \rm kyr$} to generate protostars of $\approx 18$ to $35\, \rm 
M_{\odot}$. Our \textcolor{black}{state-of-the-art} treatment of the stellar radiation feedback, 
together with the sub-AU spatial 
resolution of the inner region of the accretion disc \textcolor{black}{allow} us to realistically 
follow the evolution of the circumstellar medium of \textcolor{black}{the} early high-mass stars.
\textcolor{black}{
We have investigated the stabilizing role of the direct stellar irradiation and performed a resolution 
study, ensuring that disc gravitational fragmentation accounts for the central protostellar feedback 
and/or artificially triggered. 
}
All our models have accretion discs \textcolor{black}{whose} spiral arms fragments by 
gravitational instability into a pattern of gaseous clumps, while continuously 
feeding the central protostars at highly variable accretion rates of $\sim 
10^{-4}$$-$$10^{-3}\, \rm M_{\odot}\, \rm yr^{-1}$. As in the 
other regimes of star formation~\citep[see][and references 
therein]{smith_mnras_424_2012,vorobyov_apj_768_2013,vorobyov_apj_805_2015}, those circumstellar clumps 
episodically generate luminous accretion-driven outbursts when migrating down onto the protostar. 
\textcolor{black}{
Our study shows that the mechanism introduced in~\citet{meyer_mnras_464_2017} also 
happens for the different rotation profiles investigated herein. 
}

Particularly, those flares are similar to the FU-Orionis-like episodic accretion-driven outbursts 
observed in the young high-mass stellar system S255IR-SMA1~\citep{burns_mnras_460_2016,caratti_nature_2016}. 
Some gaseous clumps migrate \textcolor{black}{to the close surroundings ($\le 10\, \rm AU$) of the central protostar} 
while simultaneously experiencing an increase of 
their central density and temperature that can reach up to $\approx 10^{-8}\, \rm g\, \rm cm^{-3}$ and $\ge 2000\, 
\rm K$, respectively. The clumps are sufficiently massive ($\simeq 0.5-$a few$\, 
\rm M_{\odot}$) and their \textcolor{black}{core is hot and dense} enough to be considered as being on the path to 
\textcolor{black}{the second gravitational collapse down to stellar densities. We show that such nascent 
companions can fast-migrate down to the central protostar, 
constituting a massive binary system, made of a massive component plus a low-mass companion in close Keplerian 
orbit around it. 
We conclude on the viability of the disc fragmentation channel to explain the formation of close 
massive proto-binaries made of a young high-mass component and at least a low mass companion, 
which are the progenitors of the future post-main-sequence spectroscopic massive 
binaries~\citep[see also][for an observational study of close-orbit, O-star-involving 
massive binary statistics]{sana_sci_337_2012,2013A&A...550A..27M,2014ApJS..213...34K}.
} 
We underline that disc fragmentation, high variations \textcolor{black}{of} the protostellar accretion rate, 
episodic accretion-driven outbursts and close binary formation are 
tightly correlated \textcolor{black}{mechanisms, and we predict that both phenomenons can} happen together. 
Consequently, luminous outbursts from young massive stars are the signature of the 
presence of its surrounding accretion disc, a tracer of the migration of circumstellar 
disc fragments but may also be at the same time a signature of the formation of close companions 
\textcolor{black}{that will become the spectroscopic companions of a future O-type star.} 

We test our accretion discs against several semi-analytical criteria 
characterising the fragmentation of self-gravitating discs. We find that 
our discs are consistent with the so-called Toomre, Gammie and Hill criteria, 
which is consistent with the analysis presented in~\citet{klassen_apj_823_2016}, which 
we can summurise as follows:
\textcolor{black}{
the Gammie criterion is fulfilled by our discs while the Toomre criterion alone allows us to discriminate 
fragmenting from non-fragmenting accretion discs~\citep[$Q\le 0.6$, see][]{takahashi_mnras_458_2016}. 
The Hill criterion, even applied to the early formation phase of our discs, predicts their subsequent evolution with respect to fragmentation. 
}
Radiative transfer calculations against dust opacity in the context of the 
Keplerian disc surrounding the young high-mass star AFGL-4176~\citep{johnston_apj_813_2015}
indicate that \textcolor{black}{clumps in our disc models are detectable with {\it ALMA}}. 
Such disc fragments should be searched with modern 
facilities such as {\it ALMA} or the future {\it European Extremely Large Telescope (E-ELT)} 
within high-mass star formation regions from which strong maser emission 
or evidence of accretion flows have been recorded, e.g. 
W51~\citep{keto_apj_678_2008,zapata_apj_698_2009} and 
W75~\citep{Carrasco_sci_348_2015}.

\textcolor{black}{
This work highlights the need for highly spatially resolved simulations of massive star formation 
simulations as an additional issue to numerical methods and physical processes.  
It further stresses the challenging character of numerical studies devoted 
to the surrounding of young hot stars, as well as the \textcolor{black}{similarities between} massive star formation 
mechanisms (accretion variability and disc fragmentation) with the other (\textcolor{black}{low- and primordial-}) 
mass regimes of star formation. 
\textcolor{black}{
Our work, showing a possible pathway for the formation of spectroscopic massive 
binaries, will be continued and expanded to obtain a deeper understanding of the 
circumstellar medium of young massive stars and the implications for massive 
stellar evolution.}
}
Follow-up studies will investigate the effects of several physical 
processes which we have so far neglected, such as the inertia of the protostar, 
its ionizing feedback or the magnetization of the pre-stellar core. This should 
allow us to pronounce more in detail the effects of disc fragmentation 
around early protostars as a function of the initial properties of their 
pre-stellar cores.


\section*{Acknowledgements}

The authors thank the anonymous referee for their useful advice and suggestions which greatly improved the manuscript.  
D.~M.-A. Meyer and E.~Vorobyov thank B.~Stecklum for his invitation at Th\" uringer Landessternwarte Tautenburg. 
This study was conducted within the Emmy Noether research group on "Accretion 
Flows and Feedback in Realistic Models of Massive Star Formation" funded by the 
German Research Foundation under grant no. KU 2849/3-1.   
E.~I.~Vorobyov acknowledges support
from the Austrian Science Fund (FWF) under 
research grant I2549-N27. 
The authors gratefully acknowledge the computing time provided on the bwGrid 
cluster T\" ubingen. 
and on the ForHLR~I cluster at the Steinbuch Center for Computing Karlsruhe. 
This research has made use of Astropy, a community-developed core Python package for 
Astronomy (Astropy Collaboration, 2013) and of SAOImage DS9.



\footnotesize{
\bibliography{grid}
}


\end{document}